\documentclass[
    twocolumn, 
    tighten, 
    twocolappendix
]{aastex631} 

\usepackage[utf8]{inputenc}
\usepackage[T1]{fontenc}

\usepackage{amsmath}
\usepackage{booktabs}
\usepackage{cleveref}
\usepackage{enumitem}
\usepackage{microtype}
\usepackage{pifont}
\usepackage{tabularx}
\usepackage{tikz}
\usepackage{xspace}

\usepackage{savesym}
\savesymbol{tablenum}

\usepackage{local-packages/siunitx/tex/latex/siunitx/siunitx}
\DeclareSIUnit\au{au}
\DeclareSIUnit\jupitermass{M\textsubscript{J}}
\DeclareSIUnit\solarmass{M\textsubscript{\ensuremath{\odot}}}
\DeclareSIUnit\lod{\ensuremath{\lambda / D}}
\DeclareSIUnit\magnitude{mag}
\DeclareSIUnit\mas{mas}
\DeclareSIUnit\pixel{pixel}
\DeclareSIUnit\year{yr}

\newcommand{\cmark}{\ding{51}}%
\newcommand{\xmark}{\quad}%
\newcolumntype{P}[1]{>{\centering\arraybackslash}p{#1}}

\newcommand{\naco}{VLT-NACO\xspace}
\newcommand{\sphere}{VLT-SPHERE\xspace}
\newcommand{\snr}{\ensuremath{\text{S/N}}\xspace}
\newcommand{\fours}{4S\xspace}
\newcommand{\rom}[1]{\uppercase\expandafter{\romannumeral #1\relax}}
\DeclareMathOperator*{\argmax}{arg\,max}
\newcommand*\circled[1]{%
    \tikz[baseline=(char.base)]{\node[shape=circle,draw,inner sep=1.2pt,font=\footnotesize\bfseries\sffamily] (char) {#1};}%
}

\makeatletter
\patchcmd\H@refstepcounter{\protected@edef}{\protected@xdef}{}{}
\makeatother

\makeatletter
\patchcmd{\ltx@foottext}{%
  .5\textwidth\advance\hsize-18pt}{%
  \linewidth\advance\hsize-1.8em%
}{}{}
\makeatother

\begin{document}

    \title{Use the \fours (Signal-Safe Speckle Subtraction): Explainable Machine Learning reveals the \mbox{Giant Exoplanet AF Lep b in High-Contrast Imaging Data from 2011}}
    
    \shortauthors{Bonse et al.}
    \correspondingauthor{Markus J. Bonse}
    \email{mbonse@phys.ethz.ch}
    
    \author[0000-0003-2202-1745]{Markus J. Bonse}
    \affiliation{ETH Zurich, Institute for Particle Physics \& Astrophysics, Wolfgang-Pauli-Str. 27, 8093 Zurich, Switzerland}
    \affiliation{Max Planck Institute for Intelligent Systems, Max-Planck-Ring 4, 72076 T\"ubingen, Germany}
    
    \author[0000-0001-9310-8579]{Timothy D. Gebhard}
    \affiliation{ETH Zurich, Institute for Particle Physics \& Astrophysics, Wolfgang-Pauli-Str. 27, 8093 Zurich, Switzerland}
    \affiliation{Max Planck Institute for Intelligent Systems, Max-Planck-Ring 4, 72076 T\"ubingen, Germany}
    \affiliation{Max Planck ETH Center for Learning Systems, Max-Planck-Ring 4, 72076 T\"ubingen, Germany}
    
    \author[0000-0002-5476-2663]{Felix A. Dannert}
    \affiliation{ETH Zurich, Institute for Particle Physics \& Astrophysics, Wolfgang-Pauli-Str. 27, 8093 Zurich, Switzerland}
    \affiliation{National Center of Competence in Research PlanetS}
    
    \author[0000-0002-4006-6237]{Olivier Absil}
    \affiliation{University of Li\`ege, STAR Institute, Allée du Six Août 19C, 4000 Li\`ege, Belgium}
    
    \author[0000-0002-3968-3780]{Faustine Cantalloube}
    \affiliation{Université Grenoble Alpes, CNRS, IPAG, 38000 Grenoble, France}
    
    \author[0000-0002-0101-8814]{Valentin Christiaens}
    \affiliation{University of Li\`ege, STAR Institute, Allée du Six Août 19C, 4000 Li\`ege, Belgium}
    \affiliation{KU Leuven, Institute for Astronomy, Celestijnenlaan 200D, Leuven, Belgium}
    
    \author[0000-0001-7255-3251]{Gabriele Cugno}
    \affiliation{ETH Zurich, Institute for Particle Physics \& Astrophysics, Wolfgang-Pauli-Str. 27, 8093 Zurich, Switzerland}
    \affiliation{University of Michigan, Department of Astronomy, Ann Arbor, MI 48109, USA}
    
    \author[0000-0003-2530-9330]{Emily O. Garvin}
    \affiliation{ETH Zurich, Institute for Particle Physics \& Astrophysics, Wolfgang-Pauli-Str. 27, 8093 Zurich, Switzerland}
    
    \author[0000-0003-3768-5712]{Jean Hayoz}
    \affiliation{ETH Zurich, Institute for Particle Physics \& Astrophysics, Wolfgang-Pauli-Str. 27, 8093 Zurich, Switzerland}
    
    \author[0000-0002-8425-6606]{Markus Kasper}
    \affiliation{European Southern Observatory, Garching bei M\"unchen, Germany}
    
    \author[0000-0003-0593-1560]{Elisabeth Matthews}
    \affiliation{Max Planck Institute for Astronomy, K\"onigstuhl 17, 69117 Heidelberg, Germany}
    
    \author[0000-0002-8177-0925]{Bernhard Sch\"olkopf}
    \affiliation{Max Planck Institute for Intelligent Systems, Max-Planck-Ring 4, 72076 T\"ubingen, Germany}
    \affiliation{ETH Zurich, Department of Computer Science, Universit\"atstrasse 6, 8092 Zurich, Switzerland}
    
    \author[0000-0003-3829-7412]{Sascha P. Quanz}
    \affiliation{ETH Zurich, Institute for Particle Physics \& Astrophysics, Wolfgang-Pauli-Str. 27, 8093 Zurich, Switzerland}
    \affiliation{National Center of Competence in Research PlanetS}
    \affiliation{ETH Zurich, Department of Earth and Planetary Sciences, Sonneggstrasse 5, 8092 Zurich, Switzerland}
    
    \begin{abstract}
        The main challenge of exoplanet high-contrast imaging (HCI) is to separate the signal of exoplanets from their host stars, which are many orders of magnitude brighter. HCI for ground-based observations is further exacerbated by speckle noise originating from perturbations in Earth’s atmosphere and imperfections in the telescope optics. Various data post-processing techniques are used to remove this speckle noise and reveal the faint planet signal. Often, however, a significant part of the planet signal is accidentally subtracted together with the noise.
        In the present work, we use explainable machine learning to investigate the reason for the loss of the planet signal for one of the most used post-processing methods: principal component analysis (PCA). 
        We find that PCA learns the shape of the telescope point spread function for high numbers of PCA components.
        This representation of the noise captures not only the speckle noise but also the characteristic shape of the planet signal. 
        Building on these insights, we develop a new post-processing method (\fours) that constrains the noise model to minimize this signal loss. 
        We apply our model to 11 archival HCI data sets from the \naco instrument in the L'-band and find that our model consistently outperforms PCA. The improvement is largest at close separations to the star ($\leq \qty{4}{\lod}$) providing up to 1.5 magnitudes deeper contrast. This enhancement enables us to detect the exoplanet AF Lep b in data from 2011, 11 years before its subsequent discovery. We present updated orbital parameters for this object.
    \end{abstract}
    

    \section{Introduction} 
\label{sec:intro}
The primary challenge of exoplanet high-contrast imaging (HCI) is to separate and detect the light of faint exoplanets from their host stars, which are many orders of magnitude brighter. During an observation, the light of the star gets scattered in the image plane, creating a halo of highly structured \textit{speckle noise} superimposed on the planet signal. Speckles tend to mimic the expected shape of the planet, which makes it difficult to distinguish them from a real planet signal.
There are two types of speckle noise \citep{malesMysteriousLivesSpeckles2021}: \textit{quasi-static speckles}, caused by nanometer scale imperfections in the telescope optics, and \textit{atmospheric speckles}, resulting from the turbulence of Earth's atmosphere.
%
Over the past decade, considerable efforts have been spent to push the limits of HCI observations and reduce speckle noise. In this context, one key aspect is the development of faster and more accurate \textit{eXtreme} Adaptive Optics (XAO) systems. Examples of such systems are the \textit{Gemini Planet Imager} (GPI), the \textit{Spectro-Polarimetric High-contrast Exoplanet REsearch instrument} (SPHERE), the \textit{Subaru Coronagraphic Extreme Adaptive Optics} (SCExAO) and MagAO-X \citep{macintoshFirstLightGemini2014,jovanovicSubaruCoronagraphicExtreme2015,malesMagAOXProjectStatus2018,beuzitSPHEREExoplanetImager2019}. XAO uses new detectors, advanced wavefront sensors, and deformable mirrors with higher actuator counts, along with faster computing hardware and control software. For HCI, they are often coupled with different types of coronagraphs to efficiently suppress the stellar point spread function (PSF) \citep{kenworthyFirstOnSkyHighContrast2007,mawetVECTORVORTEXCORONAGRAPH2009,snikVectorAPPBroadbandApodizing2012,ndiayeApodizedPupilLyot2015,ottenONSKYPERFORMANCEANALYSIS2017a}.

Equally crucial for achieving deeper contrast and for removing speckle noise is the development of post-processing techniques \citep[see ][for an overview]{cantalloubeExoplanetImagingData2021}. Post-processing techniques are combined with dedicated observing strategies like angular differential imaging \citep[ADI;][]{maroisAngularDifferentialImaging2006}, spectral differential imaging  \citep[SDI; ][]{racineSpeckleNoiseDetection1999,sparksImagingSpectroscopyExtrasolar2002}, reference star differential imaging \citep[RDI;][]{mawetDirectImagingExtrasolar2012,ruaneReferenceStarDifferential2019} or polarimetric differential imaging  \citep[PDI;][]{kuhnImagingPolarimetricObservations2001,quanzVERYLARGEESCOPE2011} which are designed to enable planet signals to be distinguished from speckle noise. A major breakthrough for the processing of HCI data was the introduction of principal component analysis \citep[PCA/KLIP;][]{amaraPYNPOINTImageProcessing2012,soummerDETECTIONCHARACTERIZATIONEXOPLANETS2012} to estimate the correlated speckle noise.
Since its adoption to HCI, PCA has formed the basis for the analysis of many HCI surveys \citep{nielsenGeminiPlanetImager2019,launhardtISPYNACOImagingSurvey2020,jansonBEASTBeginsSample2021,langloisSPHEREInfraredSurvey2021,cugnoISPYNACOImaging2023} and allowed the detection of several new planets \citep{macintoshDiscoverySpectroscopyYoung2015,chauvinDiscoveryWarmDusty2017}.
\looseness=-1
 
However, after more than a decade of HCI observations, only a relatively small number of objects have been found (see \citealt{currieDirectImagingSpectroscopy2023} for an overview and a table of substellar companions), and the results of several surveys have consistently shown that super-Jovian planets with masses between \qtyrange{5}{13}{\jupitermass} at large separations (\qtyrange{10}{100}{\au}) are extremely rare (between $\sim$ \qtyrange{0.3}{2.2}{\percent} depending on stellar type; see \citealt{viganSPHEREInfraredSurvey2021}). Only recently, taking advantage of the latest generation of instruments, has it been possible to search for less massive planets at closer separations. While the sample size is still small, new planets have been detected in this regime (e.g., 51~Eri~b, \citealt{macintoshDiscoverySpectroscopyYoung2015}; AF~Lep~b, \citealt{derosaDirectImagingDiscovery2023,fransonAstrometricAccelerationsDynamical2023,mesaAFLepLowestmass2023})
 hinting at an elevated occurrence rate towards smaller masses and separations. 
On the other hand, radial velocity (RV) studies have found that the occurrence rate of giant planets peaks around \qty{3.6(2.0:1.8)}{\au}, decreasing closer to the star and further out \citep{fultonCaliforniaLegacySurvey2021}. However, the decrease beyond \qtyrange{5}{8}{\au} has not been confirmed and is still the subject of ongoing research \citep{lagrangeRadialDistributionGiant2023}. It is therefore crucial to push down the limits of HCI observations close to the star to bridge the gap and create an overlap between RV and HCI surveys. This will provide empirical access to an important parameter space for exoplanet population and formation models and allow for direct imaging follow-up of exoplanets detected by RV surveys.
\looseness=-1

Unfortunately, close to the star, it becomes increasingly difficult to suppress speckle noise using data post-processing techniques. If we use methods like PCA, a significant fraction of the planet signal is accidentally subtracted together with the noise. While at large separations the loss of planet signal is moderate, close to the star more than \qty{95}{\percent} of the planet signal might get lost \citep[see ][]{gebhardHalfsiblingRegressionMeets2022,bonseComparingApplesApples2023}. 
This effect is caused by the fact that the speckle noise and the signal of the planet share the same morphology. Early work on the LOCI post-processing method attempted to overcome this problem by using different data masking strategies \citep{maroisExoplanetImagingLOCI2010,soummerORBITALMOTIONHR2011a}. Later, forward modelling approaches were introduced, leading to a better understanding of the effect, now known as planet over- and self-subtraction \citep{pueyoDETECTIONCHARACTERIZATIONEXOPLANETS2016}.
Despite these efforts, the loss of planet signal remains a major bottleneck in the post-processing of HCI data.

\paragraph{Contributions:} 
In this paper, we gain a deep understanding of the reason for the loss of planet signal in PCA and related methods and use it to develop a new post-processing algorithm. To this end, we revisit the mathematical background of PCA and give an overview of more recent post-processing methods in \Cref{sec:background}. By using methods from the field of explainable machine learning, we gain additional insights about planet over- and self-subtraction (\Cref{sec:signal_loss_explained}), which we use to develop our new post-processing algorithm \fours (\Cref{sec:s4}). 
Unlike previous works that mask the data, \fours uses a mask applied to the model parameters, as well as a new loss function and different regularization strategies.
A detailed analysis of the contrast performance of \fours on archival data from the Very Large Telescope (VLT) NACO instrument is given in \Cref{sec:ana}. We find that \fours reaches significantly deeper contrast at close separations to the star and learns a model of the noise that matches our understanding of the speckle noise. Using \fours we can reveal the faint exoplanet AF~Lep~b in archival \naco data from 2011 (\Cref{sec:af_lep}) which is not detectable with PCA. In \Cref{sec:future_work,sec:summary}, we outline directions for future work and provide concluding remarks.
\looseness=-1

\paragraph{Reproducibility:}
Along with this paper, we are releasing our algorithm \fours as a Python package called \texttt{fours}. The code is available on \href{https://github.com/markusbonse/fours}{GitHub} and comes with a documentation page on \href{https://fours.readthedocs.io/en/latest/}{ReadtheDocs}. All raw data \citep{bonse_raw_data2024} and intermediate results \citep{bonse_inter_res2024} to reproduce our plots are available on Zenodo.

    \begin{figure*}[t!]
	\includegraphics[width=\linewidth]{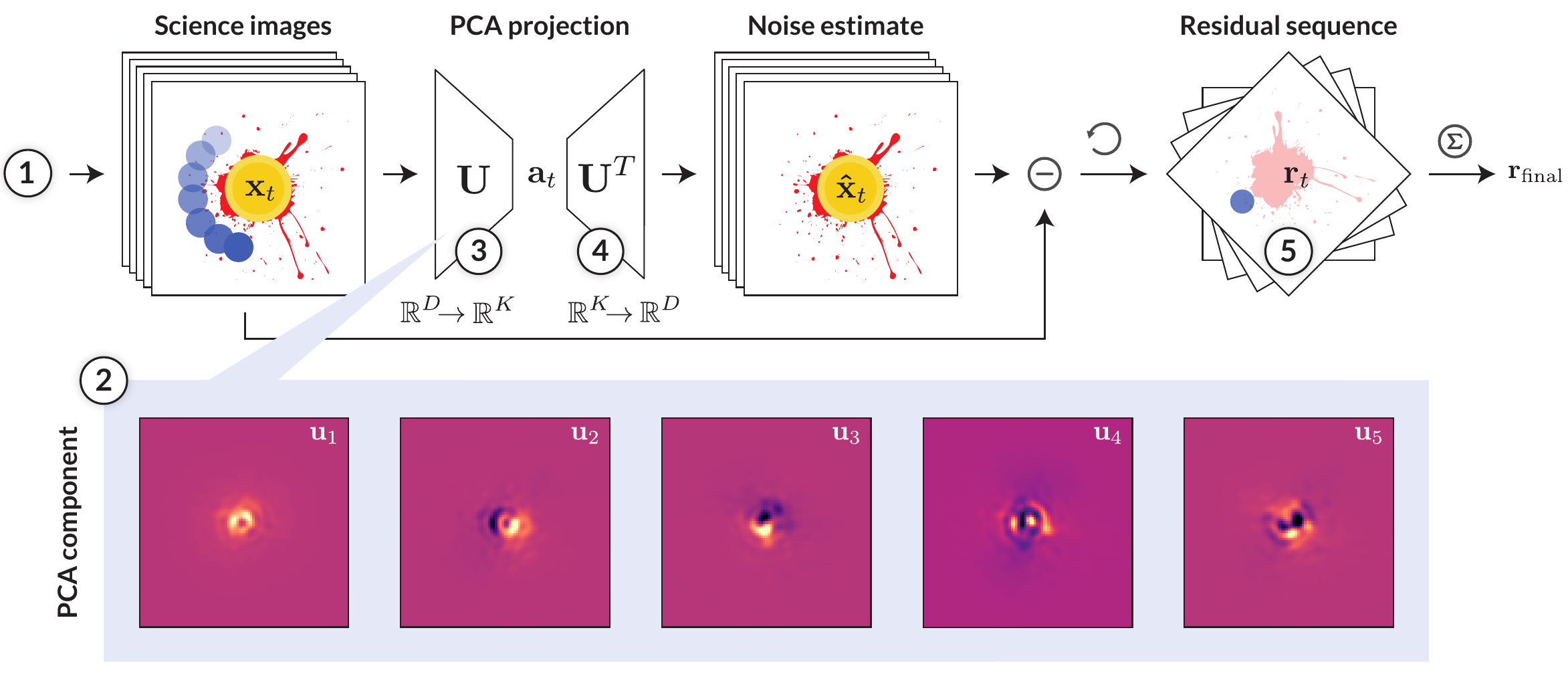}
	\caption{
        \emph{Top row:} illustration of how PCA is applied to ADI data in HCI. The planetary signal is shown in blue, while the stellar speckle noise is represented by the red splash. Details about the steps~\mbox{(\circled{1} to \circled{5})} are given in the text. 
        \emph{Bottom row:} examples of the first five principal components are shown, estimated for data set \#5~(HD~22049) in \Cref{tab:datasets}. Further details on the data are given in \Cref{sec:ana}. The first components of PCA are often interpretable. For example, $\mathbf{u}_1$ in the given example models the overall brightness variations of the post-coronagraphic PSF. The components $\mathbf{u}_2$ and $\mathbf{u}_3$ represent horizontal and vertical variations caused by small errors in the centering of the star behind the coronagraph.
    }
	\label{fig:02_pca_explained}
\end{figure*}
\newpage

\section{PCA Revisited} 
\label{sec:background}
The concept of using PCA (or \textit{Karhunen--Loève} transformation) in HCI is to estimate the stellar PSF and speckle noise to subtract it from the science data. 
For this purpose, PCA explores spatial covariances utilizing the time dimension and brightness of the speckles. 
A typical HCI data set consists of a sequence of individual exposures $\mathbf{x}_{t, \text{raw}} \in \mathbb{R}^D$, where $D = n_i \cdot n_j$ is the number of pixels per frame and $t = 1, \ldots, N$ is the temporal index of the frames. If we ignore background and detector noise, each image is composed of light from the star $\mathbf{s}_t$ and possibly light from the planet~$\mathbf{p}_t$:
\looseness=-1
\begin{equation}
	\mathbf{x}_{t, \text{raw}} = \mathbf{s}_t + \mathbf{p}_t \,.
\end{equation}
The stellar light $\mathbf{s}_t$ outshines the planet signal $\mathbf{p}_t$ by several orders of magnitude. As we go through the temporal stack of science images, the PSF and speckle pattern changes. This means that the strongest brightness variations in the data are due to the light of the star and not to the planet. This data characteristic is the foundation of speckle subtraction with PCA:
PCA is a linear method that tries to find a low-dimensional representation of the data with dimension $K \ll D$, which only captures the strong covariances of the data. If the planet signal is weak compared to the speckle noise, it should thus not be part of this low-dimensional representation. If we reconstruct the data from the low-dimensional space, we get an estimate of the speckle noise that can be subtracted from the science data. 
It is important to note that PCA only utilizes the time dimension to explore spatial covariances. It does not use the temporal evolution of the speckle noise. It is also possible to apply PCA along the time dimension as discussed by \cite{samlandTRAPTemporalSystematics2021} and  \cite{longImprovedCompanionMass2023}. More details can be found in the related work in \Cref{sec:related_work}.
Applying PCA to ADI data in HCI along the spatial dimension involves the following steps (see \Cref{fig:02_pca_explained}):
\begin{enumerate}[label=\protect\circled{\arabic*}]
	\item \textbf{Data preparation}: The temporal average of the science data $\overline{\mathbf{x}}$ is subtracted from each individual science frame:
	\begin{equation}
		\mathbf{x}_t = \mathbf{x}_{t, \text{raw}} - \overline{\mathbf{x}} \,.
	\end{equation}
	The mean-subtracted 2D science frames are then flattened into one-dimensional vectors and stacked into a data matrix $\mathbf{X} \in \mathbb{R}^{N\times D}$:
	\begin{equation}
		\mathbf{X} = 
		\begin{bmatrix}
           \mathbf{x}_1^T \\
           \mathbf{x}_2^T \\
           \vdots \\
           \mathbf{x}_N^T
         \end{bmatrix} = 
         \begin{bmatrix}
           x_{1,1} & x_{1,2} & \hdots & x_{1, D} \\
           x_{2,1} & x_{2,2} & \hdots & x_{2, D} \\
           \vdots \\
           x_{N,1} & x_{N,2} & \hdots & x_{N, D}
         \end{bmatrix} \,.
	\end{equation}
	Each row in $\mathbf{X}$ corresponds to one mean-subtracted science frame.%
    \footnote{
        We choose the notation to be consistent with \citet{bishopPatternRecognitionMachine2006} and our Python implementation. 
        All vectors in this paper (e.g., $\mathbf{x}, \mathbf{a}$ ,and $\mathbf{u}$) are column vectors.
        This notation might deviate from the conventions used in other papers \citep[e.g.,][]{lewisSpeckleSpaceTime2023a}.
        \looseness=-1
    }
	
	\item \textbf{Calculation of the projection}: We choose the dimensionality of the low-dimensional space $K$ and calculate the PCA basis (i.e., the component matrix $\mathbf{U} \in \mathbb{R}^{D\times K}$), which transforms our data into the low-dimensional space:
	\begin{equation}
		\mathbf{U} = \begin{bmatrix}
           \mathbf{u}_{1} & \hdots & \mathbf{u}_{K}
         \end{bmatrix} \,.
	\end{equation}
	The column vectors of $\mathbf{U}$ are the principal components $\mathbf{u}_1, \ldots, \mathbf{u}_K, \forall \mathbf{u} \in \mathbb{R}^D$. Details on how to calculate these components are given in \Cref{subsec:pca_projection}.
	
	\item \textbf{Dimensionality reduction}: We project our data into the low-dimensional space \mbox{$\mathbf{x}_t \rightarrow \mathbf{a}_t, \quad \mathbf{a}_t \in \mathbb{R}^{K}$}:
	\begin{equation}
		\mathbf{X} \cdot \mathbf{U} = 
		\begin{bmatrix}
           \mathbf{a}_1^T \\
           \mathbf{a}_2^T \\
           \vdots \\
           \mathbf{a}_N^T
         \end{bmatrix} = \mathbf{A} \in \mathbb{R}^{N \times K} \,.
	\end{equation}
	If the dimensionality $K$ is chosen carefully, the low-dimensional representation $\mathbf{A}$ should capture the variations caused by the star only, and not the signal of the planet.
	
	\item \textbf{Data reconstruction}: We reconstruct our data from the low-dimensional space $\mathbf{a}_t \rightarrow \hat{\mathbf{x}}_t$, $\hat{\mathbf{x}}_t \in \mathbb{R}^D$:
	\begin{equation}
		 \mathbf{A} \cdot \mathbf{U}^T= \mathbf{X} \cdot \mathbf{U} \mathbf{U}^T = 
		\begin{bmatrix}
           \mathbf{\hat{x}}_1^T \\
           \mathbf{\hat{x}}_2^T \\
           \vdots \\
           \mathbf{\hat{x}}_N^T
         \end{bmatrix}
		= \mathbf{\hat{X}} \,.
	\end{equation}
	Since $\mathbf{a}_t$ mostly captures the strong covariances of the data, we can use its reconstruction to approximate the speckle noise:
	\begin{equation}
		\hat{\mathbf{x}}_t \approx \mathbf{s}_t \,.
	\end{equation}
	
	\item \textbf{Calculation of the residual}: We subtract the noise estimate $\hat{\mathbf{x}}_t$ for each mean-subtracted science frame $\mathbf{x}_t$. The result is the so-called residual sequence $\mathbf{r}_t$:
	\begin{equation}
		\mathbf{r}_t = \phi_t(\mathbf{x}_t - \hat{\mathbf{x}}_t) \,,
	    \label{eq:residual_sequence}
	\end{equation}
	where $\phi_t$ is a de-rotation function that uses the parallactic angles to realign north in all images (ADI; \citealt{maroisAngularDifferentialImaging2006}).
	The final residual image $\mathbf{r}_{\text{final}}$ can then be obtained by averaging along the time axis:%
    \footnote{Some implementations take the median along time instead of the mean, which can be advantageous in the case of detector artifacts such as bad pixels.}
	\begin{equation}
		\mathbf{r}_{\text{final}} = \frac{1}{N} \sum_{t=1}^{N} \mathbf{r}_t \,.
	    \label{eq:residual_image}
	\end{equation}
\end{enumerate}

\begin{figure*}[t]
	\includegraphics[width=\linewidth]{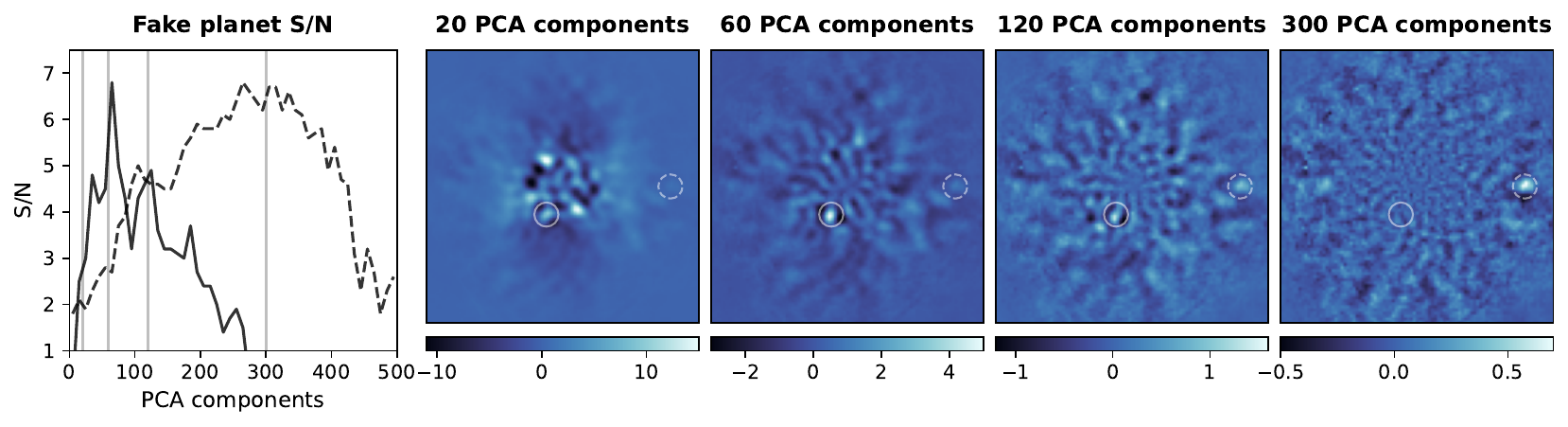}
	\caption{
	   Variation in planet visibility as a function of  principal components ($K=20$, $K=60$, $K=120$, $K=300$). The shown data set is data set \#5~(HD~22049) in \Cref{tab:datasets}. More details on data pre-processing are given in \Cref{sec:ana}. Two fake planets have been inserted into the data: one positioned at $\sim$ \qty{3}{\lod} separation from the star with a contrast brightness of \qty{9}{\magnitude} (white circle), and one at $\sim$ \qty{10}{\lod} with a contrast brightness of \qty{12.5}{\magnitude} (white dashed circle). The inner planet reaches the highest \snr at $\sim K=60$ components (left panel, solid line) and the outer planet at $\sim K=300$ components (left panel, dashed line). This example highlights the challenge of choosing the right number of principal components within a single data set.
    }
	\label{fig:02_pca_residuals}
\end{figure*}

\subsection{How to find the optimal projection with PCA?}
\label{subsec:pca_projection}
Depending on the type of the data set, the component matrix $\mathbf{U}$ can be calculated based on the science data directly \citep[e.g., in the case of ADI;][]{amaraPYNPOINTImageProcessing2012,soummerDETECTIONCHARACTERIZATIONEXOPLANETS2012} or based on reference stars \citep[RDI;][]{ruaneReferenceStarDifferential2019}. In this work, we focus on ADI, but the same concepts apply for RDI. As motivated in the previous section, we want to choose $\mathbf{U}$ such that the strongest variations of the data, that is, the speckle noise, are captured in the low-dimensional space. This objective is equivalent to the maximum variance derivation of PCA (see, e.g., chapter~12.1 of \citealt{bishopPatternRecognitionMachine2006} for details).

For simplicity, let us consider the projection onto the first component $\mathbf{u}_1$. We want to choose $\mathbf{u}_1$ such that the variance along the first axis of the low-dimensional space $\sigma^2(a_{t, 1})$ is large:
\begin{equation}
	\sigma^2(a_{t, 1}) = \mathbf{u}_1^T \mathbf{C} \mathbf{u}_1 \,,
\end{equation}
where $\mathbf{C} \in \mathbb{R}^{D \times D}$ is the covariance matrix of the data
\begin{equation}
	\mathbf{C} = \frac{1}{T} \mathbf{X}^T \mathbf{X} \,.
\end{equation}
We add another constraint $||\mathbf{u}_1|| = \mathbf{u}_1^T\mathbf{u}_1 = 1$ to avoid $||\mathbf{u}_1|| \rightarrow \infty$. To enforce this constraint, we introduce a Lagrange multiplier $\lambda_1$ which gives us the following optimization objective:
\begin{equation}
    \argmax_{\mathbf{u}_1} = \mathbf{u}_1^T \mathbf{C} \mathbf{u}_1 + \lambda_1 \left(1 - \mathbf{u}_1^T\mathbf{u}_1\right) \,.
\end{equation}
If we set the derivative with respect to $\mathbf{u}_1$ equal to zero, we get
\begin{equation}
	\mathbf{C}\mathbf{u}_1 = \lambda_1\mathbf{u}_1 \,.
\end{equation}
This means that $\mathbf{u}_1 $ should be chosen to be an eigenvector of $\mathbf{C}$. Using $\mathbf{u}_1^T\mathbf{u}_1 = 1$ we see that
\begin{equation}
	\sigma^2(a_{t,1}) = \mathbf{u}_1^T \mathbf{C} \mathbf{u}_1 = \lambda_1 \,.
\end{equation}
To maximize the variance of the projection, we should therefore choose the eigenvector $\mathbf{u}$ with the largest eigenvalue. We sort all eigenvectors of $\mathbf{C}$ by their eigenvalues:
\begin{align}
    \begin{array}{ccccccccccc}
	\lambda_1 &>& \lambda_2 &>& \hdots &>& \lambda_K &>& \hdots &>& \lambda_D \\
	\mathbf{u}_1 & & \mathbf{u}_2 & & \hdots & & \mathbf{u}_K & & \hdots & & \mathbf{u}_D
    \end{array} \,,
\end{align}
and select the first $K$ eigenvectors with the largest eigenvalues.
In practice, $\mathbf{U}$ can be obtained by computing a singular value decomposition (SVD) of the data matrix $\mathbf{X}$ \citep[see, e.g. implementations in VIP and PynPoint;][]{stolkerPynPointModularPipeline2019,christiaensVIPPythonPackage2023}. This approach improves numerical stability and often reduces computation time.
\looseness=-1

A big advantage of PCA is that the first components of the matrix $\mathbf{U}$ can be interpreted visually (see \Cref{fig:02_pca_explained}). For example, the first component $\mathbf{u}_1$ often models the general brightness variations of the stellar PSF. The data used to calculate the components in \Cref{fig:02_pca_explained} were taken with a vortex coronagraph \citep{mawetBandAGPMVector2013}. Thus, the first component shows the typical donut shape of the post-coronagraphic PSF. The second and third components show the response of the coronagraphic PSF to tip-tilt errors (i.e., the first modes of the AO).

\subsection{Limitations of PCA}

PCA is widely used by the exoplanet imaging community and available in most HCI packages such as VIP \citep{christiaensVIPPythonPackage2023}, PynPoint \citep{stolkerPynPointModularPipeline2019} or pyKLIP \citep{wangPyKLIPPSFSubtraction2015}.
Its appeal is mainly due to its simplicity and to the fact that it comes with only one central hyperparameter\footnote{In practice, there are additional geometry hyperparameters. These include masking strategies and temporal exclusion criteria and depend on the implementation \citep[see][]{wangPyKLIPPSFSubtraction2015,stolkerPynPointModularPipeline2019,christiaensVIPPythonPackage2023}}: the number of \mbox{principal components $K$}. However, this apparent simplicity hides a complex interplay of several factors, which makes it difficult to tune $K$ in practice. The optimal choice is highly data set specific, influenced by variables such as the instrument, observing mode, wavelength, and observing conditions. Even within a single data set, the optimal number of components can vary as a function of the separation from the star \citep{meshkatOPTIMIZEDPRINCIPALCOMPONENT2013}. This is to be expected, as low-order wavefront modes, which affect the closest separations to the star, vary the most. The farther away from the center, the higher the modal order and the larger the required $K$.
%
An example of this effect is shown in \Cref{fig:02_pca_residuals}. Two artificial planets are inserted into the same data set: one close to the star ($\sim$ \qty{3}{\lod}) and another farther out ($\sim$ \qty{10}{\lod}). The closer planet is best seen if 60 principal components are selected, whereas the farther one requires about 300 components to be fully visible. Notably, at 300 components, the signal of the inner planet is completely lost. This demonstrates a critical limitation of PCA: as the number of components increases, PCA not only models the noise but also starts to include parts of the planet signal, leading to what is known as planet self-subtraction and over-subtraction \citep{pueyoDETECTIONCHARACTERIZATIONEXOPLANETS2016}.
%
As discussed further in \Cref{sec:ana}, the choice of the number of components can affect the final contrast by several magnitudes. Hence, selecting the right number of components is crucial; it can be the difference between a detection and a nondetection. 

\subsection{Beyond PCA}
\label{sec:related_work}
%
The loss of planet signal in PCA has encouraged the development of more advanced post-processing techniques. Nevertheless, many of these methods either enhance PCA or incorporate its principles.
Annular PCA, for instance, applies PCA in concentric annular segments around the star \citep{absilSearchingCompanionsAU2013,gomezgonzalezVIPVortexImage2017a}. A specific exclusion criterion for ADI prevents the signal of the planet from entering the PCA basis. 
Similar exclusion criteria have been proposed for LOCI and its variations \citep{lafreniereNewAlgorithmPointSpread2007,maroisExoplanetImagingLOCI2010,soummerORBITALMOTIONHR2011a,maroisGPIPSFSubtraction2014,wahhajImprovingSignaltonoiseDirect2015,thompsonImprovedContrastImages2021}.
\citet{lewisSpeckleSpaceTime2023a} extend PCA to utilize the temporal and spatial axis simultaneously.
Conceptually similar to PCA are other subtraction-based techniques that replace the bottleneck of PCA. These methods include, for example, Nonnegative matrix factorization \citep[NMF; ][]{gomezgonzalezVIPVortexImage2017a,arcidiaconoApproximateNonnegativeMatrix2018, renNonnegativeMatrixFactorization2018}, Low-rank plus sparse decomposition \citep[LLSG][]{gomezgonzalezLowrankSparseDecomposition2016} or more recently the Half-sibling regression  \citep[HSR][]{gebhardHalfsiblingRegressionMeets2022}.
For example, LLSG adds an extra sparse term to the model with the goal of limiting the inclusion of the planet in the noise model. \cite{daglayanDirectExoplanetDetection2023} propose to change the objective function to an L1 loss to account for the heavy-tailed distribution of the speckle noise. The recently proposed ConStruct method uses nonlinear autoencoders to locally predict the speckle noise \citep{wolfDirectExoplanetDetection2023}.
Conventional PCA as discussed in \Cref{sec:background} does not explore the temporal evolution of the speckle noise. \cite{bonseWaveletBasedSpeckle2018},  \cite{samlandTRAPTemporalSystematics2021} and  \cite{longImprovedCompanionMass2023} propose methods that explicitly use the time dimension. 
Most of these methods discuss different exclusion criteria and masks to avoid subtraction of the planet signal.

After the speckle subtraction, the residual images are commonly analyzed with meta-techniques that quantify detections and nondetections, for example, using $t$-tests \citep{mawetFUNDAMENTALLIMITATIONSHIGH2014}, STIM maps \citep{pairetSTIMMapDetection2019} or performance maps \citep{jensen-clemNewStandardAssessing2017}. A~recent overview of these methods can be found in \citet{bonseComparingApplesApples2023}.
Beyond the analysis of a single residual, supervised machine learning techniques have been proposed \citep{gomezgonzalezSupervisedDetectionExoplanets2018,canteroNASODINNDeepLearning2023}. 
These methods use several PCA residuals with different numbers of components as input and train a neural network to detect the planet. The r﻿egime-switching model (RSM) map by \citet{dahlqvistRegimeswitchingModelDetection2020} and \citet{dahlqvistImprovingRSMMap2021} replaces the temporal averaging in the last step of PCA. It allows the combination of multiple PCA outputs alongside other post-processing techniques.

Complementary to subtraction-based techniques are inverse-problem-based methods. Instead of modeling and subtracting the noise, these methods rely on forward modeling to search for the expected signature of planets in the data. Examples are ANDROMEDA \citep{cantalloubeDirectExoplanetDetection2015}, FMMF \citep{pueyoDETECTIONCHARACTERIZATIONEXOPLANETS2016,ruffioImprovingAssessingPlanet2017}, PACO \citep{flasseurExoplanetDetectionAngular2018,flasseurDeepPACOCombining2024} and TRAP \citep{samlandTRAPTemporalSystematics2021}. In this context, FMMF is the direct extension of PCA to forward modeling. 

Although the concepts behind these modern post-processing techniques are diverse, they share the common idea of better using the knowledge about the shape and behavior of the planet signal. 
However, despite all these efforts, PCA still represents the main benchmark in practice: for example, \citet{fransonAstrometricAccelerationsDynamical2023} used it for the discovery of the exoplanet AF~Lep~b.
Further, none of these advanced algorithms have managed to completely avoid the problem of planet signal loss, which is the main bottleneck for achieving deeper detection limits.

    \section{Explaining Signal Loss in PCA}
\label{sec:signal_loss_explained}

In the previous section, we observed that with an increasing number of principal components, progressively more signal of the planet is lost.
A common explanation for this loss is that the principal components are not orthogonal to the signal of the planet, which is why they can fit and subtract the planet \mbox{(see Appendix~\ref{sec:masking_strategies})}. 
However, is there a specific component responsible for this effect?
While the first components of PCA can be interpreted visually (\Cref{fig:02_pca_explained}), it is difficult to understand the higher-order modes. To better understand why we lose signal in PCA, we propose to use methods from the field of eXplainable Artificial Intelligence (XAI).
With the recent boom in machine learning, there is a trend toward increasingly complex deep network architectures. Unfortunately, the decisions made by such models are hard to understand and therefore hard to trust. XAI seeks to gain a better understanding of \textit{why} and \textit{how} machine learning models make their decisions, thus giving us a glimpse into the black box. A large number of XAI methods have been proposed in the literature (see \citealt{aliExplainableArtificialIntelligence2023}, for a recent overview). Some of them are model specific (e.g., tailored to neural networks), while other methods are more generally applicable (e.g., LIME by \citealt{ribeiroWhyShouldTrust2016}). 
%
%
%
In the following paragraph, we will keep our explanation of the XAI method used in the present paper general, to provide the basis for a better understanding of other post-processing methods beyond PCA.\looseness=-1

\subsection{Saliency Maps}
A widely used approach to explain the predictions of machine learning methods is the use of input gradients. The most basic approach was proposed by \cite{simonyanDeepConvolutionalNetworks2013}. It is often referred to as \textit{Vanilla Gradient} and is straightforward to apply to subtraction-based post-processing methods in HCI.

Let us consider a single science image $\mathbf{x}_t$ at time $t$. Given a speckle-estimation method $S$, we calculate the speckle noise for this particular image:
\begin{equation}
	S(\mathbf{x}_t) = \hat{\mathbf{x}}_t.
\end{equation}
For each pixel position in the noise estimate, we can use input gradients to calculate so-called \emph{saliency maps}. Saliency maps are heatmaps that highlight areas in the input image $\mathbf{x}_t$ that contribute the most to the noise estimate at the selected pixel position. 
Given a pixel position $(i, j)$, we calculate the derivative of $S$ with respect to the input image $\mathbf{x}$ at the point $\mathbf{x}_t$:
\begin{equation}
\label{eq:gradient}
	\mathbf{w}_{l} = \frac{\partial S_{l}}{\partial \mathbf{x}} \bigg|_{\mathbf{x}_t} ,
\end{equation}
where $l = i \cdot n_j + j = 1, ..., D$ gives the index for the selected pixel position.
The magnitude of the derivative $\mathbf{w}$ points to pixels in $\mathbf{x}$ that cause the largest difference on the noise estimate at the position $(i, j)$. For each pixel position in the noise estimate, we get one saliency map $M_{l}$ by first calculating the input gradient using \Cref{eq:gradient}, then rearranging the one-dimensional vector $\mathbf{w}$ back into a two-dimensional image, and finally taking the absolute value:
\begin{equation}
	M_{l} = \left| \mathbf{w}_{l} \right|
\end{equation}
Saliency maps can be computed for any speckle estimation method $S$ for which we can calculate the gradient in \Cref{eq:gradient}. For nonlinear methods such as ConStruct \citep{wolfDirectExoplanetDetection2023} or SODINN \citep{gomezgonzalezSupervisedDetectionExoplanets2018}, more advanced methods such as SmoothGrad \citep{smilkovSmoothGradRemovingNoise2017} or Integrated Gradients \citep{sundararajanAxiomaticAttributionDeep2017} are likely to give better results than \textit{Vanilla Gradients}.

\subsection{Right for the Wrong Reasons}
\label{sec:right_for_wrong_reasons}

In the case of PCA, the noise estimation method $S$ is just the projection onto the low-dimensional space and its reconstruction:
\begin{equation}
	S_{\text{PCA}} (\mathbf{x}_t) = \mathbf{x}^T_t \mathbf{U} \mathbf{U}^T = \mathbf{\hat{x}}_t
    \label{eq:pca_bottleneck}
\end{equation}
Given the number of principal components $K$, we compute $\mathbf{U}$ using the procedure explained in \Cref{subsec:pca_projection}. Assuming that $\mathbf{U}$ is fixed, the saliency map of PCA is then given by
\begin{equation}
	M_{l, \text{PCA}} = \left|\mathbf{U} \mathbf{U}^T\right|_{:, l}
    \label{eq:saliency_pca}
\end{equation}
Each pixel value in the noise estimate of PCA is a weighted sum of the pixel values in the science frame. The weights of this sum are the column vectors of the projection matrix $\mathbf{U} \mathbf{U}^T$.
This means that \emph{each column} of $\mathbf{U} \mathbf{U}^T$ gives us an explanation for which information is used to estimate the noise at \emph{one output pixel position}. To get the saliency map for the pixel position $l$, we have to select the \mbox{$l$-th} column vector of $\mathbf{U} \mathbf{U}^T$ and take its absolute value.
Note that since PCA is a linear model, the saliency map for a given position is the same for all science frames along time.

\begin{figure}[t]
	\includegraphics[width=\linewidth]{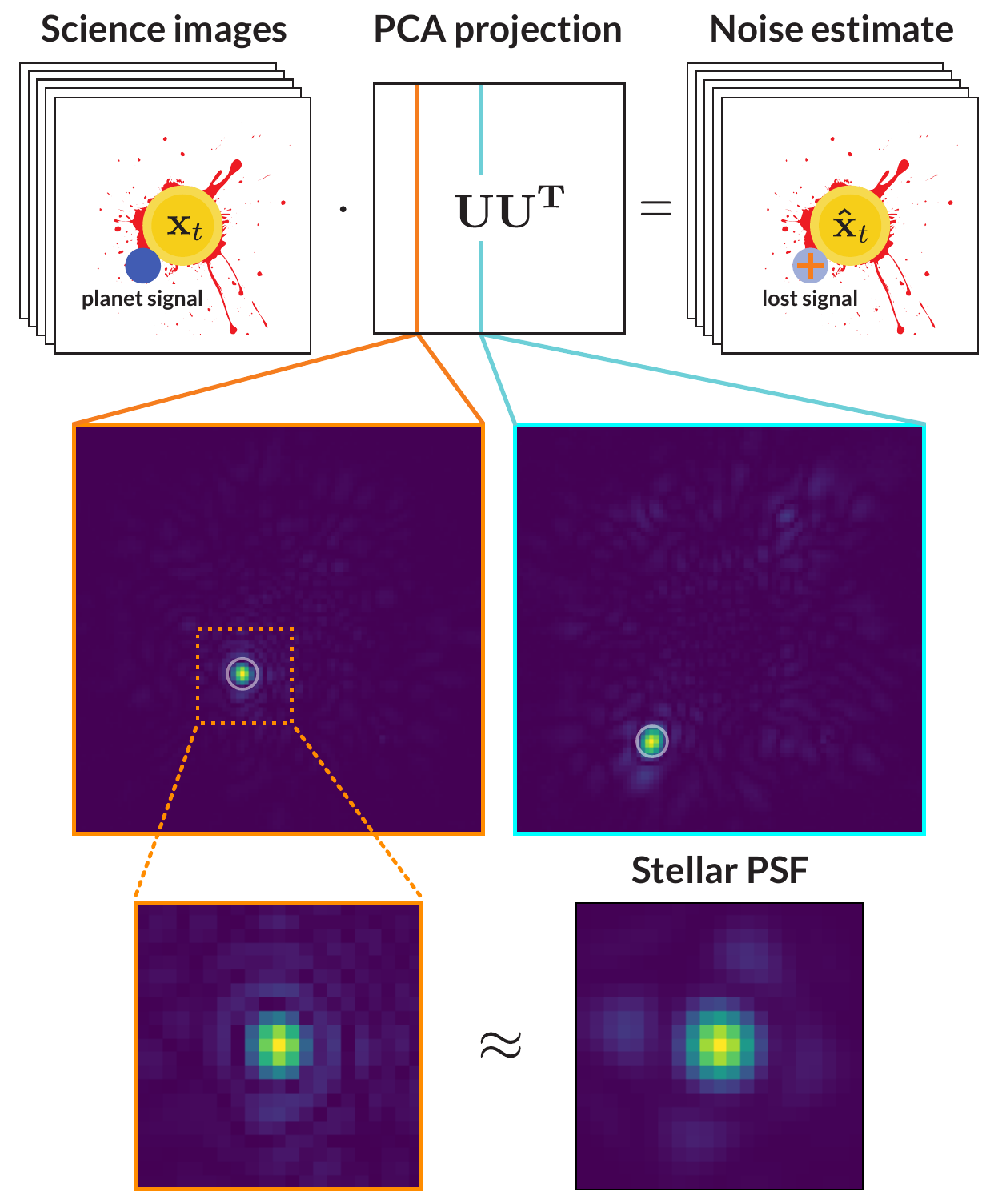}
	\caption{
        Illustration of how to compute saliency maps for PCA. For each pixel position in the noise estimate, we can compute one saliency map. Each saliency map highlights the areas in the science frame that are used to obtain the noise estimate at the respective position. Since the noise model of PCA is linear, every pixel in the noise estimate is a weighted sum of the pixels in the input image. The saliency map shows the absolute value of these weights. Two examples of such saliency maps are shown in the middle row (300 principal components). The example used in the text is marked in orange. For large numbers of principal components, the saliency map converges to the shape of the PSF. A gallery of saliency maps for different separations and numbers of principal components is given in \Cref{fig:0a3_gallery_saliency}. The saliency maps should not be confused with the component vectors $\mathbf{u}_1, ..., \mathbf{u}_K$ shown in \Cref{fig:02_pca_explained}.
    }
	\label{fig:03_pca_saliency_map}
\end{figure}

\Cref{fig:03_pca_saliency_map} illustrates the concept and provides examples of saliency maps for PCA based on data set \#4~(HD~22049; see \Cref{sec:ana} for more details on the data set). 
To better understand the loss of planet signal in PCA, let us assume that the planet signal in the science frame (\Cref{fig:03_pca_saliency_map}, top left panel) is at the position marked by the blue circle. If parts of the planet signal are lost in the subtraction step, it means that some signal leaks into the noise estimate at the same position (top right panel, orange~$+$). We can use \Cref{eq:saliency_pca} to compute the saliency map for this position (\Cref{fig:03_pca_saliency_map}, middle left panel). It turns out that most of the information used to estimate the noise at the planet's position is only taken from the input image at the planet's position. This means that PCA just copies the planet signal from the input image to the noise estimate. The same pattern can be observed for other pixel positions which do not contain the planet signal (\Cref{fig:03_pca_saliency_map}, middle right panel).
Even if the planet's position changes as a function of time (e.g., in the case of ADI), it is still subtracted.

It is interesting to note that the shape of the information used corresponds to the shape of the unsaturated PSF. We know that, to first order, the speckles follow the shape of the PSF (i.e., discarding wavelength-smearing effects). This means that PCA explains the speckle noise by its local morphology. 
However, since both the speckle noise and the planet signal follow the shape of the PSF, the planet signal is partially subtracted.
By using the shape of the PSF, PCA provides a good estimate of the noise (correct answer). However, the explanation is not limited to the noise, but also explains the planet signal. \emph{PCA gives the right answers for the noise, but for the wrong reasons.}

It should be noted that this problem occurs regardless of whether the planet is actually in the data. That is, even if we build the PCA basis using data from a different star (RDI), we still get a model that can subtract the planet signal. 
As the number of components increases, the projection $\mathbf{U} \mathbf{U}^T$ converges to a \textit{de facto} identity by learning the shape of the PSF. This convergence prevents us from learning more complex noise patterns.

\paragraph{Quantification of the effect}
The convergence rate of PCA to the shape of the PSF is very data set and instrument specific and depends on whether the noise is dominated by speckles. For reasons explained later in \Cref{sec:ana}, we confine our analysis to the \naco data taken in the $L'$ filter, leaving research on other instruments for future work.

\begin{figure}[t]
	\includegraphics[width=\linewidth]{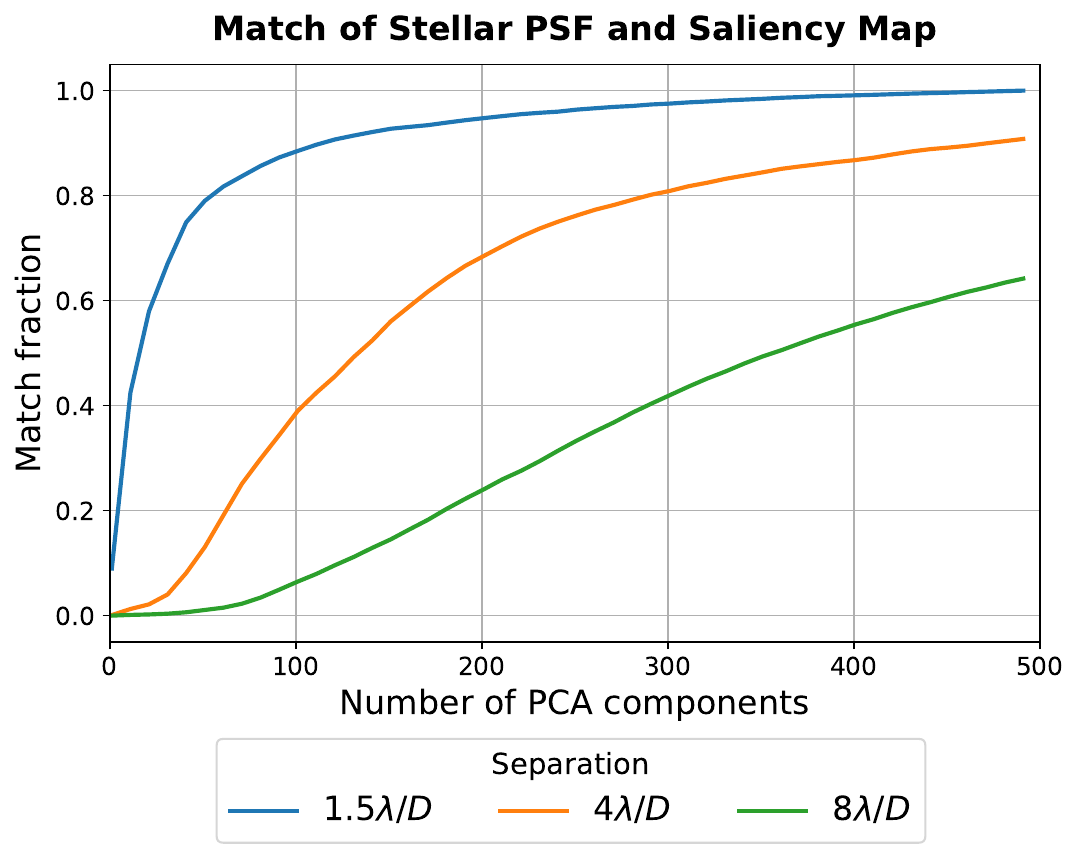}
	\caption{Convergence of the saliency map to the shape of the PSF as a function of the number of principal components used. At closer separations from the star, the convergence is faster, resulting in a greater loss of the planet signal with a smaller number of components.
    }
	\label{fig:03_match_saliency_map}
\end{figure}

To quantify how many components are needed to converge to the shape of the PSF, we compute $\mathbf{U} \mathbf{U}^T$ for different numbers of principal components $K \in [1, ..., 500]$. We calculate saliency maps for pixel positions at \qty{1.5}{\lod}, \qty{4}{\lod} and \qty{8}{\lod}, and estimate the \textit{match} between the saliency map and the unsaturated PSF template. More precisely, we compute a scalar product between the local saliency map and the unsaturated PSF and normalize it by the maximum value over all separations and components. The result of this experiment is shown in \Cref{fig:03_match_saliency_map}. Examples of saliency maps used for this analysis are shown in the \Cref{fig:0a3_gallery_saliency}.
For separations close to the star, PCA converges much faster to the shape of the PSF than farther out. This observation is in line with the fake planet experiment shown in \Cref{fig:02_pca_residuals}. While the signal of the fake planet close to the star is lost at 300 components, the planet farther out is still clearly visible. The speckle intensity gradually decreases with separation from the star. As PCA models the strongest covariances first, the convergence to the shape of the PSF is faster close to the star. As a result, the risk of losing parts of the planet signal is also higher close to the star.

\subsection{Misguided Loss Function}
\label{sec:Misguided_loss_function}
The origin of the planet signal loss described in the previous section occurs regardless of whether a planet signal is present in the data used to compute the PCA basis. It is therefore often referred to as planet over-subtraction. In the case of ADI data sets, in addition to over-subtraction, the planet signal can also directly affect the PCA basis. This second effect is known as planet self-subtraction and can lead to even greater signal loss.
In \Cref{subsec:pca_projection}, we explained how to derive PCA by maximizing the variance in the low-dimensional space. This derivation is well motivated for HCI, since the speckles are usually much brighter and more variable than the planet signal. An alternative derivation of PCA is the minimum reconstruction error formulation. 
We want to find a linear dimensionality reduction that minimizes the following reconstruction loss:
\begin{equation}
	\label{eq:loss_pca}
	\mathcal{L}_{\text{PCA}} = \frac{1}{N} \sum_{t=1}^N \underbrace{\left\| \mathbf{x}_t - \mathbf{\hat{x}}_t \right\|^2}_{\mathbf{s}_t + \mathbf{p}_t - \mathbf{\hat{x}}_t} .
\end{equation}
That is, we want our reconstruction $\mathbf{\hat{x}}_t$ to be close to our science data $\mathbf{x}_t$. 
Both the maximum variance derivation and the minimum reconstruction error formulation yield the same algorithm (see chapter 12.1.2 in \citealt{bishopPatternRecognitionMachine2006}).
In the context of HCI, minimizing \Cref{eq:loss_pca} can be problematic, as $\mathbf{x}_t$ contains both speckle noise and planet signal.
This means that $\mathcal{L}_{\text{PCA}}$ can only become small if the noise estimate $\mathbf{\hat{x}}_t$ also fits the signal of the planet $\mathbf{p}_t$. For large numbers of principal components and for bright companions, PCA tends to include the companion signal in its basis.

\begin{figure}[t]
	\includegraphics[width=\linewidth]{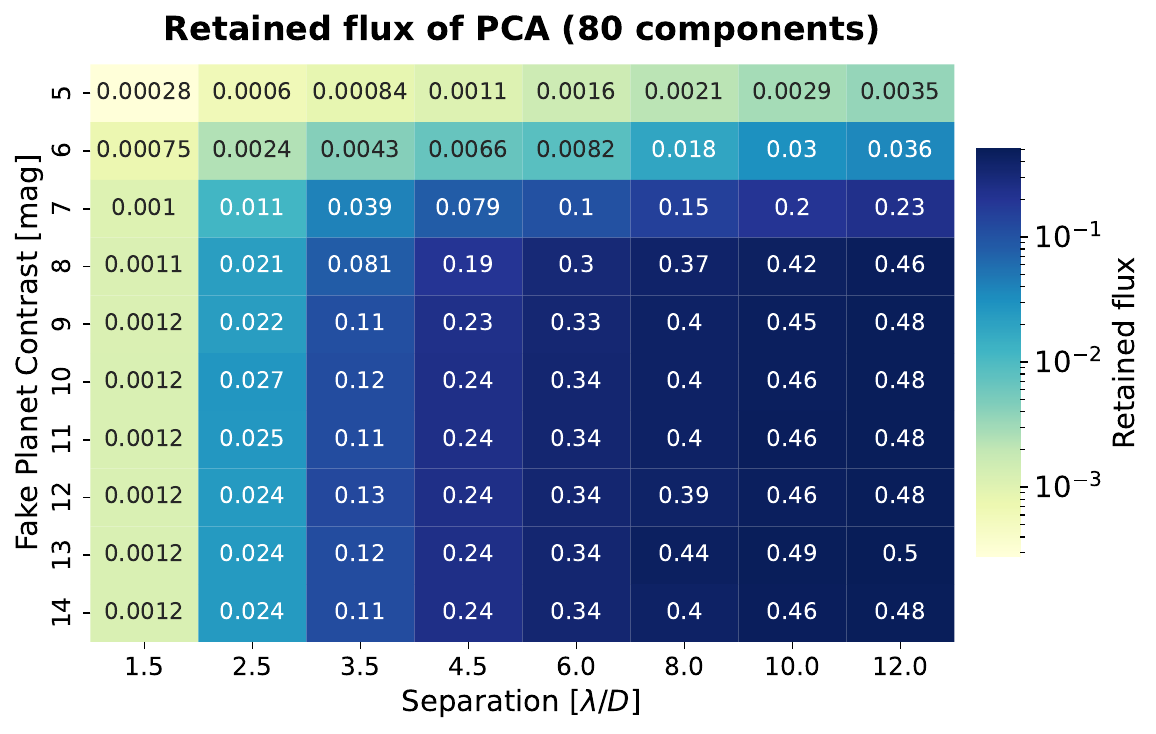}
	\caption{
        Retained flux after post-processing with PCA as a function of planet brightness and separation from the star. The grid was computed using \texttt{applefy}. Each value in the grid is the result of 3 fake planet experiments with planets inserted at different position angles. A significant fraction of the planet signal is lost, especially close to the star.
    }
	\label{fig:04_Throughput}
\end{figure}
To quantify this effect, we compute a throughput grid using the Python package \texttt{applefy} \citep{bonseComparingApplesApples2023}. A throughput grid illustrates the attenuation of the planet signal due to the data post-processing. The grid is computed by inserting fake companions with different known brightnesses and distances to the star and then measuring the fraction of the flux that remains after PSF subtraction. The result of this experiment is shown in \Cref{fig:04_Throughput}. 
We observe that for faint companions the retained flux is only a function of the separation and does not depend on the brightness of the companion. This gradual loss of planet signal closer to the star is caused by the convergence of PCA to the shape of the PSF (see \Cref{sec:right_for_wrong_reasons}). For bright companions, however, PCA begins to include parts of the signal in its basis, causing an even greater loss of planet signal. The brightness at which planets are affected depends on the number of principal components. If more components are used, fainter companions start to be affected.

%
    \section{\fours: Signal-Safe Speckle Subtraction} 
\label{sec:s4}
\begin{figure*}[t!]
	\includegraphics[width=\linewidth]{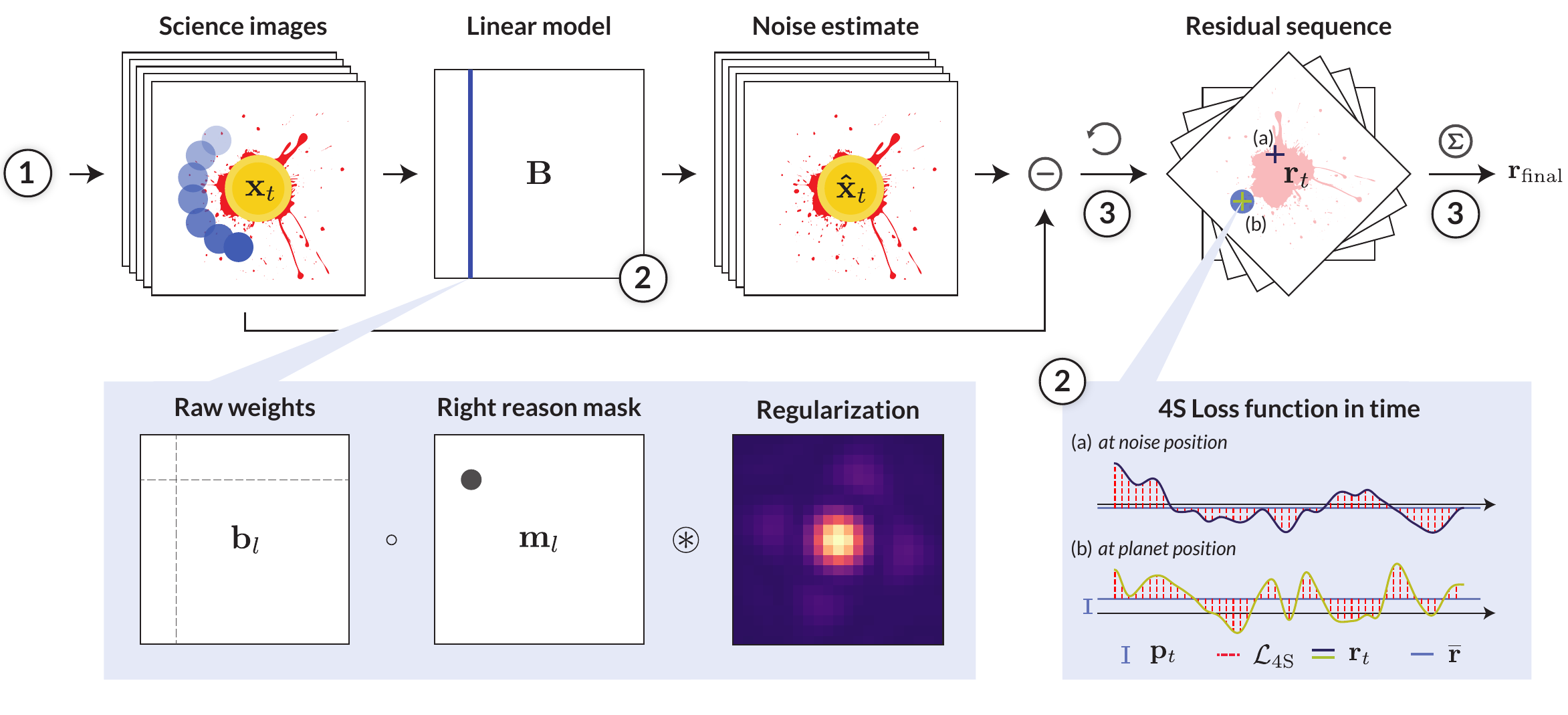}
	\caption{
        Illustration of our new post-processing technique \fours. To overcome the limitations of PCA, \fours makes use of three innovations: \rom{1}. a \textit{right reason} mask to prevent over-subtraction (bottom left box), \rom{2}. a loss function that is invariant to the signal of the planet (bottom right box), and \rom{3}. a regularization based on domain knowledge to prevent overfitting. The main steps \circled{1} to \circled{3} are explained in detail in \Cref{sec:s4_main_steps}.
    }
	\label{fig:04_s4_explained}
\end{figure*}

In the previous section, we identified two main reasons for the loss of planet signal in PCA: first, the convergence of the noise model to the telescope's PSF, and second, a problematic loss function that accidentally leads the noise estimate to fit the planet signal. Based on these findings, we propose a novel post-processing method that adopts the main steps of PCA but overcomes its limitations. We have named this method \fours, short for Signal-Safe Speckle Subtraction. \fours is characterized by three innovations: first, a linear noise model with a \emph{right reason} constraint; second, a new loss function that is invariant to the planet signal; and third, a domain-knowledge-informed regularization.

\subsection{The Right Reason Constraint}
\label{sec:right_reason_mask}
In \Cref{sec:right_for_wrong_reasons}, we found that the noise model of PCA converges to a \textit{de facto} identity that reproduces the shape of the telescope's PSF. This is problematic, as both the speckle noise and the planet signal follow this shape. Ideally, we would like our noise model to specifically recognize the speckle noise, that is, we want to find explanations for the speckle noise that are unique to the noise and do not describe the planet signal. 
In the context of XAI, \cite{rossRightRightReasons2017} proposed to add a \emph{right reason} penalty to guide the predictions of neural networks using domain knowledge. The domain knowledge is represented as a mask that penalizes \emph{wrong} explanations. The method has been successfully applied in practice \citep{schramowskiMakingDeepNeural2020}.
Inspired by this work, we propose a \emph{right reason} constraint to prevent the noise model of \fours from learning the shape of the telescope's PSF.\looseness=-1

Like PCA (see \Cref{eq:pca_bottleneck}), \fours uses a linear model, but the projection matrix $\mathbf{U}\mathbf{U}^T$ is replaced by a square matrix $\mathbf{B} \in \mathbb{R}^{D \times D}$:
\begin{equation}
	\text{4S}(\mathbf{x}_t) = \mathbf{x}_t^T \mathbf{B} = \mathbf{\hat{x}}_t \quad.
	\label{eq:s4_noise_model}
\end{equation}
It is important to note that $\mathbf{B}$ is much larger than the component matrix of PCA, $\mathbf{U}\in \mathbb{R}^{D \times K}$.
For PCA, each column of the projection matrix $\mathbf{U}\mathbf{U}^T$ gives us an explanation of which information of the input frame $\mathbf{x}_t$ is used to estimate the noise at one pixel position in~$\mathbf{\hat{x}}_{t}$. Similarly, the columns of $\mathbf{B}$ contain the explanations of \fours. 
Following the idea of \citet{rossRightRightReasons2017}, we want to limit the information \fours can use from the input frames to predict the speckle noise.
We achieve this goal by designing a \emph{right reason} mask that sets the values along and around the diagonal of $\mathbf{B}$ to zero. 
%
For each pixel position $l$, we have one parameter vector $\mathbf{b}_{l} \in \mathbb{R}^D$ as a column of $\mathbf{B}$ that is masked by one \emph{right reason} mask $\mathbf{m}_{l} \in \mathbb{R}^D$:
\begin{equation}
	\label{eq:B_first}
	\mathbf{B}_{:, l} = \mathbf{b}_{l} \circ \mathbf{m}_{l},
\end{equation}
where $\circ$ is the element-wise multiplication. In total we have $D$ \emph{right reason} masks, one for every column of $\mathbf{B}$.
Each mask $\mathbf{m}_{l}$ is designed such that values around the position $l = i\cdot n_j + j$ are set to zero:\footnote{Note: The mask does not have to be binary, which is beneficial for data sets with low resolution. Our implementation makes use of circular apertures based on the \texttt{photutils} package.}
\begin{equation}
	\mathbf{m}_{l}(l') = 
	\begin{cases}
    0,				& \text{if } \sqrt{(i-i')^2 + (j - j')^2} \leq d \\
    1,              & \text{otherwise}
	\end{cases},
\end{equation}
where $d$ sets the radius of the mask and $l' = i' \cdot n_{j} + j' = 1, ..., D$ gives the pixel position within the $l$-th mask (see bottom left box of \Cref{fig:04_s4_explained}). 
We choose $d$ relative to the size of the PSF: \mbox{$d = 0.75 \cdot \text{FWHM}$}. This mask size ensures that the core of the PSF is protected from the \textit{de facto} identity, while local information about the speckle noise is still preserved. A larger mask size allows us to protect even more signal, but it also weakens the noise model. The value of \mbox{$d = 0.75 \cdot \text{FWHM}$} is a compromise for which we have obtained good results in practice.

The right reason mask should not be confused with the exclusion criteria used by other post-processing techniques \citep[e.g.,][]{maroisExoplanetImagingLOCI2010,soummerORBITALMOTIONHR2011a,absilSearchingCompanionsAU2013,gomezgonzalezVIPVortexImage2017a,lafreniereNewAlgorithmPointSpread2007,wahhajImprovingSignaltonoiseDirect2015}. 
While the masks of these techniques also aim to reduce the loss of planet signal, the way they work is fundamentally different.
The masks proposed in the literature exclude PSF-sized regions in the \emph{science data} to prevent the planet signal from leaking into the noise estimate. 
They often have a temporal dependency to track potential trajectories of point-sources as a function of field rotation.
\fours, on the other hand, masks not the science data but the \emph{parameters} of the noise model during the optimization. 
The mask does not block the signal of the planet, but instead forces the model to explore other areas of the input image that explain the speckle noise. The right reason mask has no time dependency.
In Appendix~\ref{sec:masking_strategies} we compare the linear noise models of LOCI, PCA, and \fours and discuss how their masking strategies differ.

Our approach is similar to the post-processing methods based on the HSR \citep{samlandTRAPTemporalSystematics2021,gebhardHalfsiblingRegressionMeets2022}. However, while \cite{gebhardHalfsiblingRegressionMeets2022} select areas in the input images that promise to contain useful information about the speckle noise, we use the entire science frame but exclude areas that correspond to the shape of the telescope's PSF near the de facto identity.

\subsection{Signal Invariant Loss Function}
\label{sec:4s_loss_function}
A second limitation of PCA is its loss function (see \Cref{eq:loss_pca}), which can only become small if the noise estimate includes the planet signal. This is problematic in the case of ADI, for which the planet signal is part of the data set used to compute the PCA basis. Different types of exclusion regions \citep{lafreniereNewAlgorithmPointSpread2007,maroisExoplanetImagingLOCI2010,cantalloubeDirectExoplanetDetection2015,thompsonImprovedContrastImages2021} or explicit planet models \citep{samlandTRAPTemporalSystematics2021} have been proposed in the literature to protect the planet's signal from becoming part of the noise estimate. These methods, however, often come with additional hyperparameters that are hard to optimize in practice. Therefore, instead of minimizing the reconstruction loss, as done in \Cref{eq:loss_pca}, we propose to minimize the variance along the temporal axis in the de-rotated residual sequence: 
\begin{equation}
	\label{eq:loss_4s}
	\mathcal{L}_{\text{4S}} = \frac{1}{N} \sum_{t=1}^{N} \left\| \mathbf{r}_t - \overline{\mathbf{r}} \right\|^2, \quad \overline{\mathbf{r}} = \frac{1}{N} \sum_{t=1}^{N} \mathbf{r}_t \,.
\end{equation}
An illustration of why $\mathcal{L}_{\text{4S}}$ is sensitive to the speckle noise but does not affect the planet signal is shown in the bottom right box of \Cref{fig:04_s4_explained}. In the de-rotated residual sequence $\mathbf{r}_t$, the signal of the planet $\mathbf{p}_t$ is always at the same position and has approximately constant brightness. Consequently, it only contributes to the temporal mean $\overline{\mathbf{r}}$ but does not influence the temporal variance. Since the temporal mean $\overline{\mathbf{r}}$ is subtracted in \Cref{eq:loss_4s}, the loss function becomes invariant to the presence of the planet signal.
Quasi-static and atmospheric speckles, however, vary with time, which is why they affect the loss function (red dashed lines in \Cref{fig:04_s4_explained}). Note that the noise model of \fours is still applied to the science images before de-rotation (see \Cref{eq:s4_noise_model}). Only the loss function is evaluated on the de-rotated residual sequence\footnote{We do not account for changing conditions during the night. However, if information about the variability of the planet signal as a function of time is available (e.g., for data taken with a vAPP coronagraph such as in \citealt{ottenONSKYPERFORMANCEANALYSIS2017a}), one could normalize for these variations as part of the pre-processing.}. Details on how to minimize this loss function are given in \Cref{sec:s4_main_steps}.

\subsection{Regularization}
So far, the number of free model parameters of \fours is huge: given a frame resolution of $100 \times 100$ pixels, $\mathbf{B}$ contains about 100 million parameters. To mitigate the risk of overfitting, we use two types of regularization to reduce the expressiveness of the model.
First, we use a classical L2 penalty term on the model parameters (similar to Tikhonov/ridge regression; see chapter~3.1.4 of \citealt{bishopPatternRecognitionMachine2006}), which we add to the loss function:
\begin{equation}
	\mathcal{L}_{\text{4S}} = \underbrace{\frac{1}{N} \sum_{t=1}^{N} \left\| \mathbf{r}_t - \overline{\mathbf{r}} \right\|^2}_{\text{temporal variance}} + \underbrace{\lambda \cdot \sum_{l=1}^{D} \left\| \mathbf{b}_l \right\|^2 \vphantom{\sum_{t=1}^{N}}}_{\text{L2 penalty}} \,,
	\label{eq:loss_s4}
\end{equation}
where $\lambda$ is the only hyperparameter of \fours. Larger values of $\lambda$ force the parameters $\mathbf{b}$ to be closer to zero, reducing the expressiveness of the model. This means that a smaller value of $\lambda$ results in a stronger reduction of the speckle noise while coming at the risk of overfitting (i.e., memorizing the data). In contrast to PCA, we notice that the best choice of $\lambda$ does not depend on the separation from the star. Instead, it depends on the temporal resolution and the number of pixels in the science frames. If only a few frames are available in time, a stronger regularization (i.e., larger values of $\lambda$) is needed and the reduction of speckle noise is less effective. However, a detailed analysis of how to tune $\lambda$ is beyond the scope of this paper.
\looseness=-1

In addition to the L2 regularization, we also convolve the parameters of our model with the unsaturated PSF. For this purpose we replace \Cref{eq:B_first} by
\begin{equation}
	\label{eq:B_final}
	\mathbf{B}_{:, l} = \left(\mathbf{b}_{l} \circ \mathbf{m}_{l} \right) \circledast \text{PSF} \,.
\end{equation}
The convolution ties spatially adjacent parameters together, reducing the effective number of model parameters. From our domain knowledge, we know that speckles approximately follow the shape of the PSF. All features in the data that are smaller are not due to the speckle noise but are caused, for example, by detector noise. By convolving the parameters $\mathbf{b}$ with the PSF, we prevent our model from overfitting pixel-level artifacts. The convolution is an integral part of the optimization, and the parameters $\mathbf{b}$ are optimized taking into account both the right reason mask and the convolution.
As explained in step \circled{3} in the next section, we use automatic differentiation to back-propagate through the convolution.

\subsection{Main Steps of \fours and Optimization}
\label{sec:s4_main_steps}
Applying \fours to ADI data in HCI consists of the following steps, which we illustrate in \Cref{fig:04_s4_explained}:
\begin{enumerate}[label=\protect\circled{\arabic*}]
	\item \textbf{Data preparation}: Similar to PCA, \fours works on normalized data. We first subtract the temporal mean $\overline{\mathbf{x}}$ and divide by the standard deviation along the time axis:
	\begin{equation}
		\mathbf{x}_t = \left(\mathbf{x}_{t, \text{raw}} - \overline{\mathbf{x}} \right) / \text{std}(\mathbf{x}_{t, \text{raw}})
	\end{equation}
	The additional division by $\text{std}(\mathbf{x}_{t, \text{raw}})$ is required to ensure that the L2 regularization affects all areas in the image equally. The 2D science frames are flattened into one-dimensional vectors and stacked into the data matrix $\mathbf{X} \in \mathbb{R}^{N \times D}$ (see the PCA procedure in \Cref{sec:background}).
	
	\item \textbf{Initialization of the model}: We initialize all parameters $\mathbf{b}_{l}$ with zeros and create the \textit{right reason} masks $\mathbf{m}_{l}$ for all output pixel positions $l = 1, ..., D$. 
	
	\item \textbf{Optimization of the model parameters}: We optimize the parameters of the noise model $\mathbf{b}_l$ to minimize the loss function in \Cref{eq:loss_s4} using the Python package \texttt{PyTorch} \citep{paszkePyTorchImperativeStyle2019}. 
	We chose the L-BFGS optimizer \citep{liuLimitedMemoryBFGS1989}, which converged faster for our application than other optimizers such as SGD with momentum \citep{sutskeverImportanceInitializationMomentum2013} or Adam \citep{kingmaAdamMethodStochastic2017}.
	The L-BFGS optimizer is a quasi-Newton method that iteratively minimizes the loss function. 
	The optimization is carried out in an end-to-end fashion taking into account the \emph{right reason mask}, the convolution with the PSF (see \Cref{eq:B_final}), and the field rotation due to ADI. 
	We use the automatic differentiation framework \citep[see][for a tutorial on the topic]{baydinAutomaticDifferentiationMachine2018} of \texttt{Pytorch} to compute gradients of the loss function~$\mathcal{L}_{\text{4S}}$ with respect to the parameters $\mathbf{b}$. 
	Since the loss function in \Cref{eq:loss_s4} is defined with respect to the de-rotated residuals sequence, this step requires us to back-propagate through the image rotation~$\phi_t$.
	For this purpose, we take advantage of sampler grids originally developed for spatial transformer networks  \citep{jaderbergSpatialTransformerNetworks2016}.
	The L-BFGS optimizer takes the first-order gradients and approximates second-order derivatives to update the parameters~$\mathbf{b}_l$. 
	The iterative optimization is finished once the loss plateaus. This is usually the case after 100--500 iterations depending on the data set and the value of $\lambda$. Larger values of $\lambda$ (stronger regularization) lead to faster convergence compared to smaller values of $\lambda$. 
	We find that the use of minibatches, as often used in neural networks, leads to significantly worse results. This is probably due to the values of $\overline{\mathbf{r}}$, which contain the signal of the planet and become noisy when using minibatches. Therefore, the entire data set is used in each step of the optimization.
	
	\item \textbf{Calculation of the residual}: After the optimization is completed, we calculate the residual sequence using \Cref{eq:residual_sequence}. As for PCA, the final residual is obtained by averaging along the time dimension (\Cref{eq:residual_image}).
\end{enumerate}
We note that the subtraction of the temporal average in step \circled{1} is not free from planet self-subtraction. More advanced ideas to overcome this step should be explored in future work. Our \texttt{Python} implementation takes advantage of modern NVIDIA A100 and H100 GPUs to speed up the optimization. Depending on the spatial and temporal resolution of the data set, the optimization converges in about 10 to 30 minutes using one GPU. For large data sets that do not fit into GPU memory, gradient accumulation can be used.
It is possible to calculate results for a range of $\lambda_1 > \lambda_2 > \lambda_3 > ... $ without significant computational overheads. For this we first optimize the parameters starting with the strongest regularization $\lambda_1$. 
For the second setup $\lambda_2$, we repeat steps \circled{2} to \circled{4}. 
But instead of initializing the model parameters $\mathbf{b}_l$ with zeros in step \circled{2}, we use the $\mathbf{b}_l$ computed for $\lambda_1$. This strategy significantly speeds up the convergence of the optimization for $\lambda_2$. 
    
\begin{table*}[t]
	\caption{
        \naco Data Sets Used for Performance Evaluation.
    }
    \label{tab:datasets}
    { 
    \hbadness=10000
    \begin{tabular*}{\linewidth}{@{\hskip\tabcolsep} @{\extracolsep{\fill}} c l l c S[table-format=1.2] S[table-format=3.1] S[table-format=1.2] S[table-format=4] S[table-format=5] c @{\hskip\tabcolsep}}
    	\toprule
     	\# & 
     	Target &
     	Program ID &
     	Observing Date &
     	\multicolumn{1}{c}{$r_0$ (\unit{\arcsecond})} &
     	\multicolumn{1}{c}{Field Rotation (\unit{\degree})} &
     	\multicolumn{1}{c}{DIT (\unit{\second})} &
     	\multicolumn{1}{c}{Total (\unit{\second})} &
     	\multicolumn{1}{c}{\# Frames} &
     	AGPM \\
     	\midrule
     	1 & HD 2262 & 199.C-0065 (C)$^\textit{a}$ & 2017-11-01 & 0.56  & 73.7 & 0.35 & 5043 &  12711 & \cmark \\
     	2 & HD 7570 & 1101.C-0092 (C)$^\textit{c}$ & 2018-11-27 & 0.65  & 68.9 & 0.35 & 6895 & 14515 &\cmark \\
     	3 & HD 11171 & 1101.C-0092 (C)$^\textit{c}$ & 2018-11-28 & 0.63  & 108.7 & 0.35 &  7350 & 18572 & \cmark  \\
     	4 & HD 22049 & 096.C-0679 (A)$^\textit{c}$ & 2015-12-17 & 0.57  & 90.1 & 0.08 & 4608 &  48300 &\cmark \\
     	5 & HD 22049 & 199.C-0065 (C)$^\textit{a}$ & 2017-10-30 & 0.61  & 104.7 & 0.08 & 5304 &  57966 &\cmark \\
    	6 & HD 38678 & 084.C-0396 (A)$^\textit{b}$ & 2009-11-27 & 1.47  & 39.2 & 0.2 & 2460 &   11185 &\xmark \\
    	7 & HD 40136 & 1101.C-0092(C)$^\textit{c}$ & 2018-11-29 &  0.87 & 114.7 & 0.3 & 7500 &   20265 &\cmark \\
    	8 & HD 115892 & 1101.C-0092 (E)$^\textit{c}$ & 2019-05-23 & 0.48  & 95.2 & 0.35 & 6895 &  17401 & \cmark \\
    	9 & HD 169022 & 1101.C-0092 (E)$^\textit{c}$ & 2019-05-20 & 1.28  & 113.8 & 0.35 & 6195 &   9935 &\cmark \\
    	10 & HD 177724 & 091.C-0654 (A)$^\textit{c}$ & 2013-08-19 & 0.65  & 24.9 & 0.2 & 3200 &  12720 & \xmark \\
    	11 & HD 209952 & 089.C-0149 (A)$^\textit{b}$ & 2012-07-14 & 1.12  & 40.4 & 0.2 & 2880 &  12647 & \xmark \\
      	\midrule
      	12 & HD 35850 & 088.C-0085(A)$^\textit{b}$ & 2011-10-21 & 1.1  & 69.6 & 0.2 & 3120 & 13809  &  \xmark \\
      	\bottomrule
    \end{tabular*}
	\tablecomments{
        Data sets \#1 through \#11 are used for the quantitative analysis; data set \#12 (AF Lep) is used for the scientifc demonstration in \Cref{sec:af_lep}. The number of frames provided is the number of frames after removing bad frames. All data sets are pre-processed with \texttt{PynPoint} and publicly available on Zenodo \citep{bonse_raw_data2024}.
    }
	\tablerefs{
        \quad 
		\mbox{\textit{a}: \citet{launhardtISPYNACOImagingSurvey2020},}
		\mbox{\textit{b}: \citet{rameauSurveyYoungNearby2013},}
	 	\mbox{\textit{c}: no reference or unpublished}
	}
    }
\end{table*}

\section{Quantitative Analysis} 
\label{sec:ana}

Comparing the contrast performance of post-processing techniques can be difficult for two main reasons.
First, different post-processing methods follow fundamentally different concepts. PCA and \fours are both subtraction-based techniques. They return a residual image, which is analyzed in a second step with statistical methods like t-tests or STIM maps \citep{mawetFUNDAMENTALLIMITATIONSHIGH2014,pairetSTIMMapDetection2019}. Methods like FMMF, PACO, or TRAP \citep{ruffioImprovingAssessingPlanet2017,flasseurExoplanetDetectionAngular2018,samlandTRAPTemporalSystematics2021} belong to the family of inverse-problem approaches. They produce detection maps that show the models' belief in the presence of a planet. The detection process for subtraction-based and inverse-problem approaches relies on very different statistics. Therefore, their results are not directly comparable. 
In addition, the residual noise characteristics are method specific and usually unknown. This makes a fair comparison even more difficult, especially if we want to compare different methods given the same false-positive fraction (FPF). 
For this reason, methods should only be compared within the same category \citep[see, e.g.,][]{cantalloubeExoplanetImagingData2021}.

Secondly, the contrast performance is influenced by the instrument and the observing mode. Depending on the wavelength, different noise sources dominate, such that the best choice of post-processing method is likely to be instrument dependent. To keep our analysis consistent, we decided to focus on archival data from the \naco instrument in the $L'$-band ($\lambda = \qty{3.8}{\micro\meter}$). The exposure times of these data are typically very short, which gives us a high temporal resolution. This is crucial for \fours to minimize the risk of overfitting. It also allows us to correct not only for quasi-static speckles but also for atmospheric speckles on shorter time scales. To ensure the robustness of our results under changing observing conditions and targets, we base our analysis on 11 archival data sets (see \Cref{tab:datasets}). This guarantees that any observed improvements are not random, but consistent across multiple data sets. All targets are bright ($L' < \qty{4}{\magnitude}$) and were chosen such that no obvious companions or disk signals are present. We ensure that each data set has at least 30 minutes of observations with seeing no worse than \qty{1.5}{\arcsecond}. The final list is a random selection of data sets in the archive that satisfy these conditions.
\looseness=-1

We process all our data sets with the state-of-the-art pipeline \texttt{PynPoint}. The preprocessing steps include: \mbox{1. a simple} dark and flat calibration, \mbox{2. an} interpolation of bad pixels, \mbox{3. a simple} background subtraction using sky frames, \mbox{4. a centering} of the star using cross-correlations, and \mbox{5. a removal} of bad frames (e.g. frames with open AO loop or bad centering of the star behind the coronagraph). To keep the computation time manageable, we combine every 5 frames of the science sequence and crop the frames to a field of view of radius \qty{1.2}{\arcsecond} (equivalent to \qty{12}{\lod}). Even after temporal binning, we still have between \num{2400} and \num{13000} frames along the time dimension. The final pre-processed data cubes are publicly available on Zenodo \citep{bonse_raw_data2024}.

\subsection{Right for the Right Reasons}

In \Cref{sec:right_for_wrong_reasons}, we used saliency maps to better understand which information is used by PCA to estimate the noise at a given position. This approach revealed that PCA converges to the telescope's PSF for high numbers of components, causing significant signal loss. \fours is designed to overcome this limitation and forces the model to explore other explanations for the speckle noise using a \textit{right reason constraint}. But is the noise estimate of \fours \textit{right for the right reasons}?

\begin{figure}[t]
	\includegraphics[width=\linewidth]{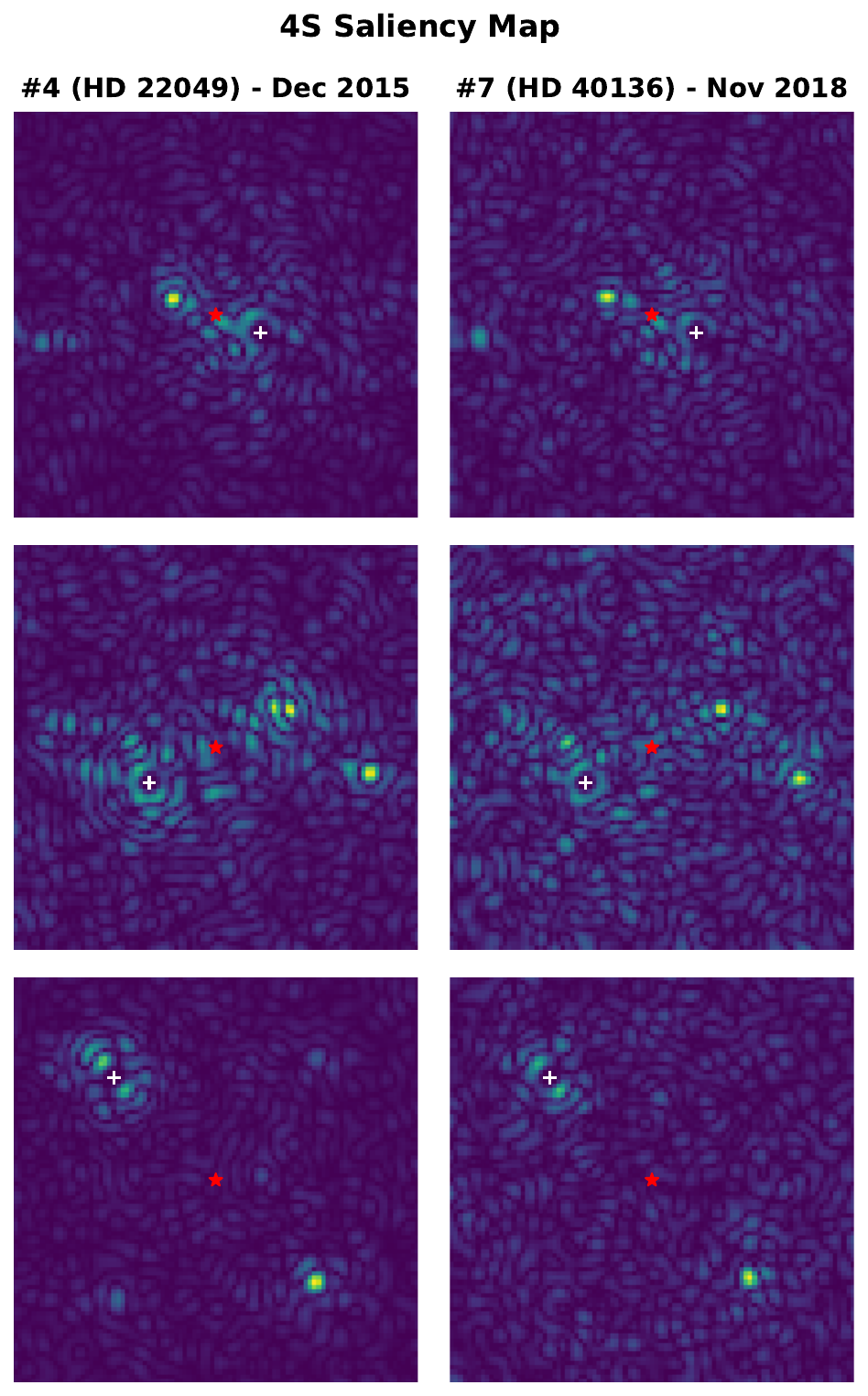}
	\caption{
        Saliency maps for \fours, computed for data sets \#4 (HD 22049; left column) and \#7 (HD 40136; right column). The saliency maps show which information from the science frames $\mathbf{x}_t$ is used by \fours  to estimate the noise at the position marked by the white cross. In contrast to PCA (see \Cref{fig:03_pca_saliency_map}), \fours successfully learns the point-symmetric signatures of the speckle noise. The red star marks the center of the frame.
    }
	\label{fig:04_s4_sailency_map}
\end{figure}

Similar to PCA, we can compute saliency maps for \fours that are just the column vectors of our noise model:
\begin{equation}
	M_{l, \text{4S}} = \left| \mathbf{B} \right|_{:, l}.
\end{equation}
We run \fours on data sets \#4~(HD~22049) and \#7~(HD~40136) with a regularization strength \mbox{$\lambda = 10^3$} and compute saliency maps for three different positions. 
The two data sets were selected randomly, with about three years between observations.
The output saliency maps are shown in \Cref{fig:04_s4_sailency_map}.
We notice that \fours no longer learns the shape of the telescope's PSF, but instead explores other areas in the science frames. The most dominant features in the saliency maps of \fours are point symmetries. This is a strong indication that \fours indeed learns the characteristic behavior of speckles, which are known to exhibit symmetric or antisymmetric correlations \citep{bloemhofBehaviorRemnantSpeckles2001,perrinStructureHighStrehl2003,bloemhofFeasibilitySymmetrybasedSpeckle2007,ribakFainterCloserFinding2008}. Another pattern, characterized by four points of equal distance from the star, can be seen in the lower left saliency map of \Cref{fig:04_s4_sailency_map}. This pattern is due to waffle mode, a distortion of the wavefront that is not visible to the wavefront sensor. 
It is an artifact in the data that we want to remove in the data post-processing. 
Due to the width of the $L'$-band filter, the speckles are slightly elongated. \fours finds this characteristic of the speckle noise and uses information on a line pointing toward the center of the frame (bottom two panels of \Cref{fig:04_s4_sailency_map}). 

The two data sets, \#4 (HD~22049) and \#7 (HD~40136), were acquired with the same observing mode. However, there is a significant gap of three years between the two observations. Despite this time difference, it is remarkable to observe the similarity in the saliency maps produced by \fours. This suggests that \fours can learn signatures of the noise that are linked to the instrument and not the individual observation. This result opens up several possibilities for future work, which we discuss in \Cref{sec:future_work}.
\looseness=-1

\subsection{Fake Planet Experiments}
\begin{figure*}[t]
    \centering
	\includegraphics[width=\linewidth]{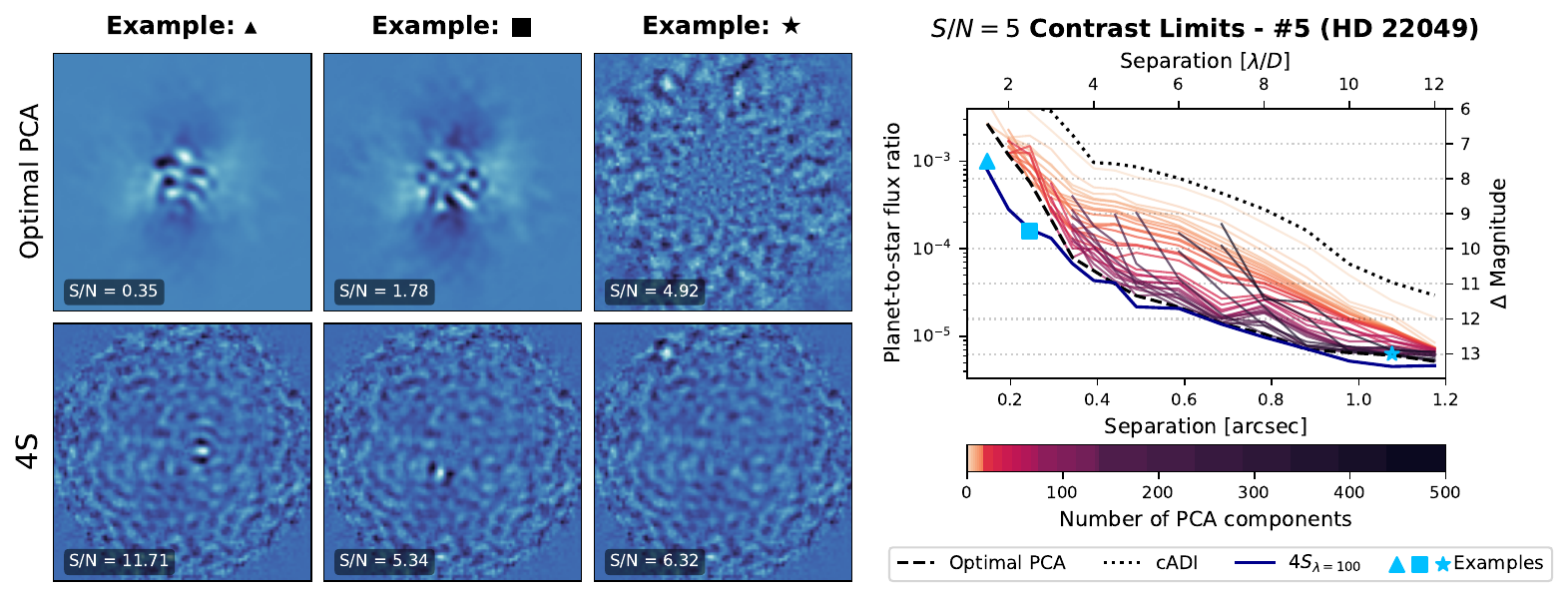}
	\caption{
        Results of our fake planet experiments for data set \mbox{\#5 (HD~22049; see \Cref{tab:datasets})}. On the left, example residuals are shown for fake planets at \mbox{$\blacktriangle = (\qty{1.5}{\lod}, \qty{7.5}{\magnitude})$}, \mbox{$\blacksquare = (\qty{2.5}{\lod}, \qty{9.5}{\magnitude})$} and \mbox{$\bigstar = (\qty{11}{\lod}, \qty{13}{\magnitude})$}. In the case of PCA, we tune the number of principal components to get the deepest contrast (\emph{optimal PCA}). For \fours, we use the same setup with $\lambda = 100$ for all separations. An interactive version of the plot is available on our \href{https://fours.readthedocs.io/en/latest/}{documentation page}. The plot on the right shows contrast curves of \fours and PCA with different numbers of principal components. Each contrast curve was obtained by computing a contrast grid using the Python package \texttt{applefy} and is based on 945 fake planet experiments. The dotted black line gives the cADI contrast, which is equivalent to a simple subtraction of the temporal median. Note that for large numbers of components and small separations, none of the inserted fake planets reach an $\snr > 5$.
    }
	\label{fig:05_residuals_PCA_S4}
\end{figure*}
%
To reliably calculate the contrast performance of \fours, we perform extensive fake planet experiments. We use \textit{contrast grids} implemented in the \texttt{python} package \texttt{applefy}  \citep{bonseComparingApplesApples2023}. 
Contrast grids systematically calculate the signal-to-noise ratio (\snr) of inserted fake companions as a function of distance from the star and planet brightness. 
We insert fake planets every \mbox{0.5 mag} for contrasts \mbox{$[5, 15]$ mag}, and for separations $[1.5, 12.5]$ $\lambda /D$ every $0.5 \lambda /D$.
For each combination of planet brightness and separation, three planets are inserted at different position angles. We insert only one fake planet at a time, resulting in 945 fake planet experiments per data set. 
Each data set with an inserted fake planet is processed with PCA and \fours to obtain a residual image. Afterward, the \snr of the companion is calculated using the method proposed by \cite{mawetFUNDAMENTALLIMITATIONSHIGH2014}. Since our limits are in the speckle-dominated regime, we use spaced pixels instead of apertures to extract the photometry, as suggested by \cite{bonseComparingApplesApples2023}. The final contrast curves are obtained by thresholding the contrast grid using all 945 fake planet experiments.

The results of PCA depend strongly on the number of components used.
Therefore, 33 different setups with $K \in [1, 500]$ were calculated for each fake planet experiment. For \fours, the regularization strength was set to $\lambda \in [10^2, 10^3, 10^4, 10^5]$.
We compute contrast grids for each data set (see \Cref{tab:datasets}) and all algorithm setups. 
In total, we compute more than \num{300000} PCA and more than \num{30000} \fours residuals. 
This experiment required substantial compute and was thus realized on a compute cluster, allowing us to use up to 200 NVIDIA A100 / H100 GPUs simultaneously. 
To further speed up the computations, we reimplemented the \texttt{PynPoint} version of PCA using \texttt{PyTorch} to enable GPU computations. Our new GPU implementation of PCA uses the fast SVD from \cite{halkoFindingStructureRandomness2011}, resulting in a 50-fold speedup on one NVIDIA H100 GPU compared to 64 CPUs. The code for our new implementation is available along with our Python package \texttt{fours}. 
Example residuals and contrast curves for one of our data sets, \#5~(HD~22049) are shown in \Cref{fig:05_residuals_PCA_S4}.  
As shown in the plot, \fours reaches significantly deeper contrast compared to PCA, especially at separations close to the star. The two close fake companions at $1.5 \lambda / D$ and $2.5 \lambda / D$ are undetectable with PCA, regardless of the number of principal components. For separations $<$ \qty{2.5}{\lod}, we reach about \qty{1.4}{\magnitude} deeper contrast with \fours compared to PCA.
Even at larger separations, a decent improvement of about $\sim 0.3$ mag can be achieved. 
Although \fours is designed to prevent planet self-subtraction, we still observe negative \emph{wings} next to the planet's signal at close separations. These negative wings are caused by the subtraction of the temporal average in the first step of \fours. If we were able to subtract the temporal mean without the planet signal, the negative wings would disappear and even better results could be achieved. This path should be explored in future work within the context of RDI.

To achieve the deepest contrast of PCA, we need to carefully tune the number of components for each separation. While 16 components give the best results at $\sim$~\qty{3}{\lod}, 400 components are required at $\sim$~\qty{12}{\lod}.  
For \fours, on the other hand, \mbox{$\lambda = 100$} consistently outperforms all PCA setups. Since tuning the number of principal components is difficult in practice and could suffer from human bias, this is a clear advantage of \fours.

\begin{figure}[t]
	\includegraphics[width=\linewidth]{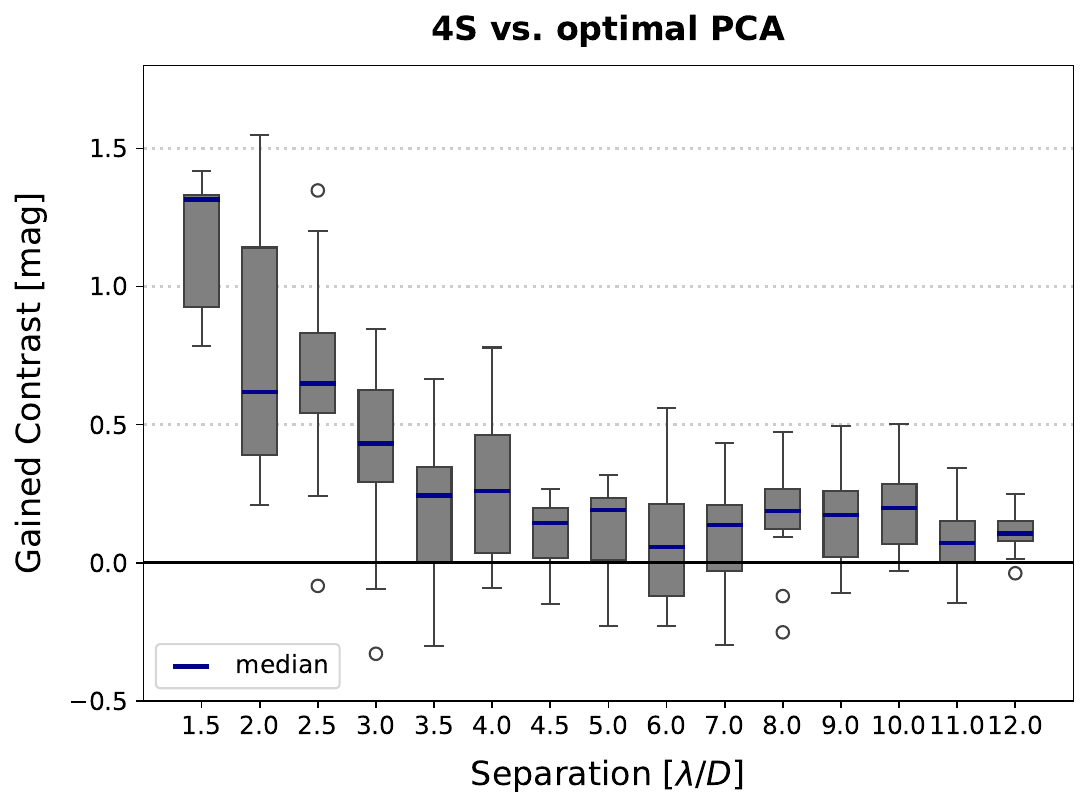}
	\caption{
        Relative improvement in contrast of \fours over PCA, accumulated over the 11 data sets shown in \Cref{tab:datasets}. The blue line shows the median improvement, the gray boxes give the interquartile range (IQR), and the whiskers extend to points that lie within 1.5~IQR. The points marked as circles are outliers.
	}
	\label{fig:05_relative_improvement}
\end{figure}

We repeat the same experiment shown in \Cref{fig:05_residuals_PCA_S4} for all data sets summarized in \Cref{tab:datasets}. Depending on the data set, we select the best $\lambda \in [10^2, 10^3, 10^4, 10^5]$ for \fours and compare it to the \textit{optimal} PCA limits. For data sets \#10 and \#11 we had to exclude the result for the innermost \qty{2.5}{\lod} due to strong saturation from the host star. The relative improvement of \fours over PCA accumulated over all data sets is shown in \Cref{fig:05_relative_improvement}.

The results are consistent with the improvement achieved on data set \#5, with about \qty{0.5}{\magnitude} variation depending on the data set. 
The improvement is largest at close separations to the star. We know that the loss of the planet signal in PCA is at its maximum at these separations (see \Cref{sec:signal_loss_explained}). \fours successfully mitigates this limitation, allowing it to learn a better representation of the speckle noise.
\looseness=-1

The improvement of \fours over PCA is probably larger in practice, mainly for three reasons:
\begin{itemize}
    \item If we go through the stack of PCA residuals for different numbers of components, the speckle pattern changes considerably. For \fours the speckle pattern in the residual images is very static for different values of $\lambda$. Since we try 33 different setups for PCA, the chance of a speckle co-aligning with the inserted fake companion increases. This artificially increases the \snr in the PCA residuals.
    
    \item It is likely that the \textit{optimal PCA} limit is not reached in practice. This is because it is difficult to find the best number of principal components. For data set \mbox{\#4 (HD 22049)}, \cite{launhardtISPYNACOImagingSurvey2020} report a contrast of \qty{10.5}{\magnitude} at \qty{0.5}{\arcsecond} and \qty{11.6}{\magnitude} at \qty{0.75}{\arcsecond}. Our \textit{optimal PCA} limits reach down to \qty{11.5}{\magnitude} at \qty{0.5}{\arcsecond} and \qty{12.3}{\magnitude} at \qty{0.75}{\arcsecond}, which is almost one magnitude deeper. This difference is likely due to the large number of setups tried in our PCA analysis. In addition, better pre-processing of the data with \texttt{PynPoint} could explain part of this improvement.
	
    \item The noise distribution of the PCA residuals often deviates from Gaussian noise \citep{pairetSTIMMapDetection2019,bonseComparingApplesApples2023}. This is especially true for small numbers of principal components. Heavy-tailed non-Gaussian residual noise increases the risk of false positives. In Appendix~\ref{sec:a_residual_noise_dist}, we study the residual noise distribution of \fours and PCA. We find that the noise of \fours is not perfectly Gaussian either, but it is more Gaussian than the residual noise of PCA. 
	This means that the chance of observing an $\snr > 5$ that is caused by noise is greater for PCA than for \fours. However, since the true noise distribution is unknown for both methods, the limits cannot be calculated at the same FPF level.
\end{itemize}

Over all data sets, the median contrast performance of \fours is consistently better than limits reached with \emph{optimal} PCA. This improvement opens a new discovery space on archival data and for future observations.

\section{Scientific Demonstration} 
\label{sec:af_lep}

\begin{figure*}[t]
	\includegraphics[width=\linewidth]{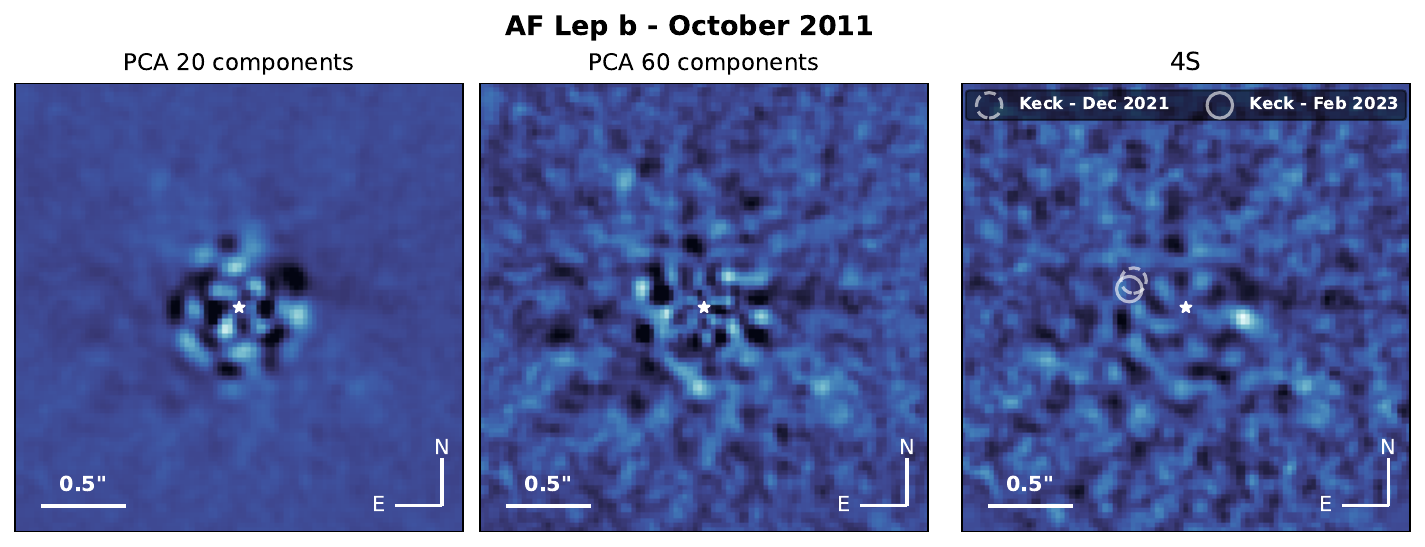}
	\caption{
        Residual images of the AF~Lep data set \mbox{(\#12 in \Cref{tab:datasets})} taken with \naco in 2011. The left and middle panels show the residuals obtained with PCA with $K=20$ and $K=60$ components, respectively. The right panel shows the residual of \fours, which reveals the giant exoplanet AF~Lep~b at $\snr = 6.8$, 11 years before its discovery. For better visibility, we filter the residual images with a Gaussian kernel ($\sigma = \qty{0.8}{\pixel}$). The \snr is calculated on the raw residuals before applying the Gaussian filter. An interactive version of the plot for different setups of PCA and \fours is available in our \href{https://fours.readthedocs.io/en/latest/}{online documentation}. A gallery of all PCA and \fours residuals is given in \Cref{fig:0a4_af_lep}.
    }
	\label{fig:05_af_lep}
\end{figure*}

To demonstrate that \fours is not only capable of improving the contrast performance for artificial companions, we have searched the ESO archive for companions that have been found with \sphere but not with \naco. \sphere has found several new companions, but unfortunately most of them have never been observed with \naco.

\begin{figure}[t]
	\includegraphics[width=\linewidth]{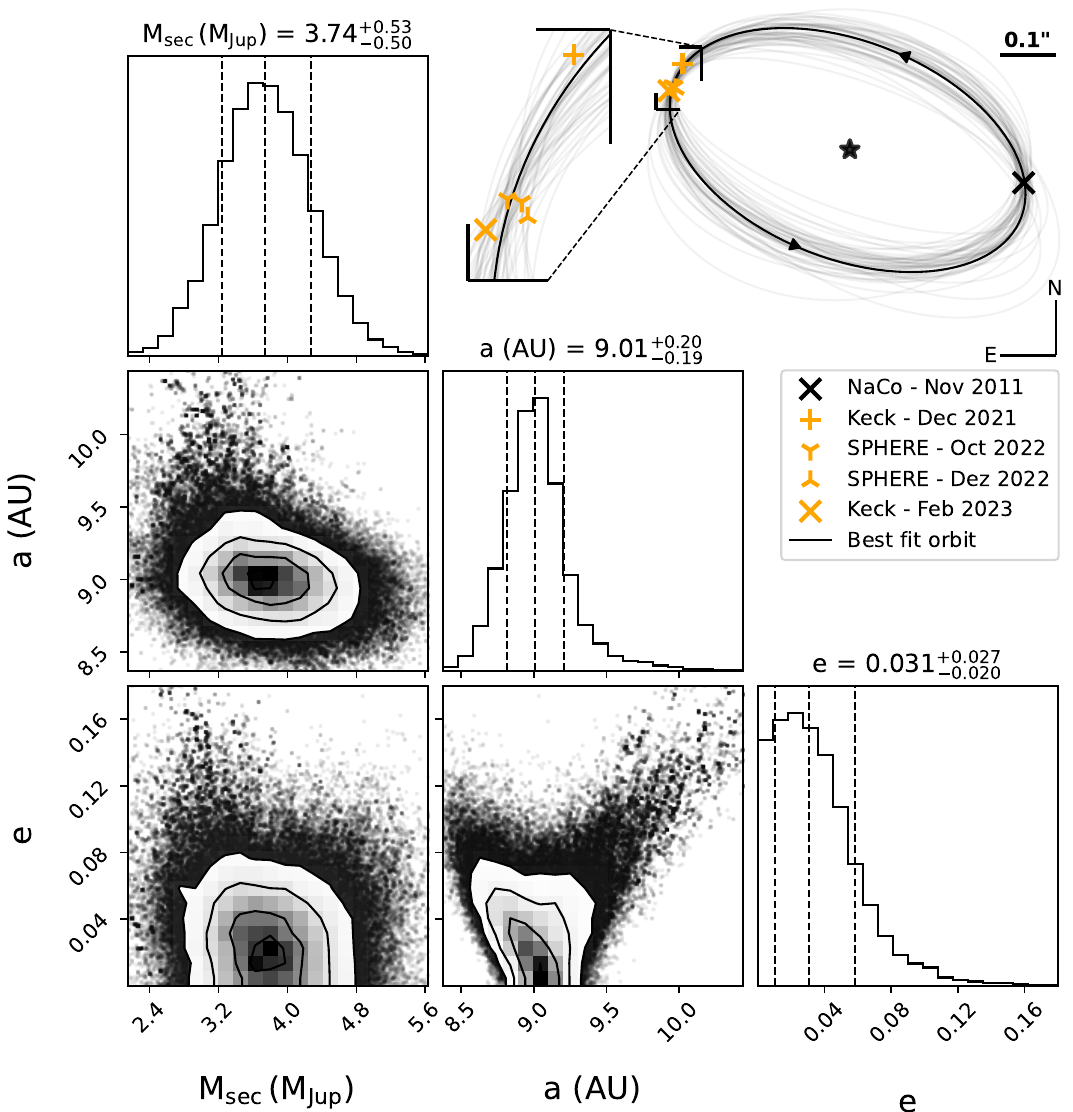}
	\caption{
        Results from the updated orbit fit of AF~Lep~b. The corner diagram reports the posterior distribution of an MCMC-based orbit fit using \texttt{orvara}. The median with the \qty{16}{\percent} and \qty{84}{\percent} quantiles is reported. The top-right inset shows all available relative astrometry measurements with the best orbital fit. To illustrate the confidence of the fit, 100 orbits are drawn from the fit posterior (gray). For a summary of all results, see \Cref{tab:af_lep_prop}.}
	\label{fig:05_af_lep_orbit}
\end{figure}

AF Leporis is a young ($\sim \SI{28}{\mega\year}$ based on moving group age estimate; \citealt{bellSelfconsistentAbsoluteIsochronal2015}) F8V-type star \citep{grayContributionsNearbyStars2006} belonging to the $\beta$ Pictoris moving group \citep{zuniga-fernandezSearchAssociationsContaining2021}. Since \mbox{AF Lep} presents a significant astrometric \textsc{Hipparcos} - \textsc{Gaia} acceleration, it was observed by multiple direct imaging surveys targeted at accelerating stars. Three direct detections of the companion AF~Lep~b obtained with \sphere and Keck-NICR2 were claimed almost simultaneously in 2023 \citep{derosaDirectImagingDiscovery2023,mesaAFLepLowestmass2023,fransonAstrometricAccelerationsDynamical2023}. While these observations firmly place AF~Lep~b as a faint L-type dwarf in the planetary regime, estimations of its dynamical mass via the astrometric acceleration are conflicting between instruments and with age-luminosity model-based estimates. While the latter disagreement is often observed in recent analyses \citep{cheethamDirectImagingUltracool2018a,brandtImprovedDynamicalMasses2021,tobinDirectimagingDiscoverySubstellar2024}, the former is likely due to the very short time baseline on which the orbit of AF~Lep~b could be sampled (2021 December to 2023 February; see \Cref{fig:05_af_lep_orbit}).

\cite{fransonAstrometricAccelerationsDynamical2023} discovered the planet in the $L'$-band at a separation of $\rho_{2023.090342} = \qty{342 \pm 8}{\mas}$ and at a contrast of $\Delta L' = \qty{9.94 \pm 0.14}{\magnitude}$. This is precisely the regime where we expect an improvement with \fours over PCA.
The star AF~Lep was imaged with \naco in 2011 at $L'$-band using the normal imaging mode (see \Cref{tab:datasets}). The data set was taken in medium observing conditions (seeing \qty{1.1}{\arcsecond}), but provides $\qty{69.6}{\degree}$ of field rotation. 
We process this archival data set with \texttt{PynPoint} \citep{stolkerPynPointModularPipeline2019} using the same preprocessing routine explained in \Cref{sec:ana}, and then subtract the stellar PSF using PCA and \fours. 
Our \fours reduction reveals the giant exoplanet AF~Lep~b on the opposite side of the star, 11 years before its discovery in 2022 (see \Cref{fig:05_af_lep}). The signal of the planet is also present in the residuals of PCA, but it is indistinguishable from the speckle noise.
%

This precovery of \mbox{AF Lep b} presents a unique opportunity to extend the time baseline of its orbit sampling from about $\SI{1}{\year}$ to more than $\SI{11}{\year}$. Furthermore, it enables an independent confirmation of the existing $L'$-band magnitude estimate \citep{fransonAstrometricAccelerationsDynamical2023}.

\paragraph{Astrometry and Photometry} 
\begin{table}[t]
	\caption{Properties of AF Lep b.}
    \label{tab:af_lep_prop}
    { 
    \hbadness=10000
    \begin{tabular*}{\linewidth}{@{\hskip\tabcolsep} @{\extracolsep{\fill}} l r @{\hskip\tabcolsep}}
    	\toprule
     	$\rho_{2011.800}$ (\unit{\mas}) & ${323.24}_{-6.44}^{+6.71}$ \\
    	$\theta_{2011.800}$ (\unit{\degree}) & ${258.81}_{-0.59}^{+0.53}$ \\
    	$\Delta L'$ (\unit{\magnitude}) & ${10.03}_{-0.12}^{+0.13}$ \\
    	$L'$ (\unit{\magnitude}) & ${14.96}_{-0.13}^{+0.14}$ \\
    	$M_{L'}$ (\unit{\magnitude}) & ${12.81}_{-0.13}^{+0.14}$ \\
    	Primary mass (\unit{\solarmass}) & ${1.219}_{-0.054}^{+0.052}$\\
        Secondary mass  (\unit{\jupitermass})& ${3.74}_{-0.50}^{+0.53}$\\
        Semi-major axis (\unit{\au})& ${9.01}_{-0.19}^{+0.20}$\\
        Inclination (\unit{\degree})& ${55.8}_{-7.2}^{+6.2}$\\
        Ascending node (\unit{\degree})& ${69.7}_{-5.4}^{+5.6}$\\
        Mean longitude at 2010.0 (\unit{\degree})& ${170}_{-13}^{+11}$\\
        Eccentricity & ${0.031}_{-0.020}^{+0.027}$\\
        Argument of periastron (\unit{\degree}) & ${-10}_{-52}^{+66}$\\
      	\bottomrule
    \end{tabular*}
    \tablecomments{
        Reported uncertainties are the median with the \qty{16}{\percent} and \qty{84}{\percent} quantiles of the MCMC samples after burn-in.
    }
    }
\end{table}

Because of the attenuation of the planet signal due to the post-processing, we cannot extract the photometry and astrometry from the residual images directly. 
Note that, due to the normalization of the data in the first step of \fours and to the finite size of the right reason mask, \fours still loses some planet signal.
In the case of PCA, it is common practice to insert a negative fake planet into the raw data at the position of the detection. The photometry and astrometry, as well as their uncertainties, are then calculated using Markov Chain Monte Carlo \citep[MCMC; see, e.g.,][]{stolkerPynPointModularPipeline2019}. 
Each step in the MCMC requires a complete recalculation of the PCA basis and the residual. A full MCMC often relies on several thousand rereductions, which becomes computationally expensive.
For \fours, a reduction of the AF~Lep data set takes about 10 minutes, resulting in a total computation time of about one month. 
It is likely possible to significantly reduce the computation time required for an MCMC with \fours. This could be achieved by avoiding the recalculation of the model matrix $\mathbf{B}$ for each iteration. Instead, $\mathbf{B}$ could be calculated once and then updated continuously during the MCMC optimization. However, to verify the accuracy of such an approach, further tests are needed that are beyond the scope of this paper. Further details for future work in this direction are discussed in \Cref{sec:future_work}.
Instead, we decided to estimate the astrometry and photometry based on the PCA residuals using the well-established and tested MCMC code presented in \cite{stolkerPynPointModularPipeline2019}.
For this, we first calculated PCA residuals for $K\in [1, 200]$ every 10 components. The peak $\snr = 4.7$ is reached at 65 components\footnote{It is important to note that the number of principal components that gives the highest $\snr$ can only be found given both the exact separation and the position angle. If only the separation is known, the optimal number of components may vary by a few components as a function of the position angle.}. We fix $K=65$ and use the MCMC procedure in \texttt{PynPoint} to estimate the astrometry and photometry. As in \citet{stolkerPynPointModularPipeline2019}, we first use a simplex minimization to find good initialization parameters. Afterward, we run the MCMC for 500 steps using 100 walkers. The minimization criterion is based on the Hessian \citep{stolkerPynPointModularPipeline2019,christiaensVIPPythonPackage2023}.
Based on the injection of many fake planets, the MCMC procedure allows us to retrieve both the astrometry and photometry of the companion, as well as the corresponding error bars.

Our results are summarized in \Cref{tab:af_lep_prop}. The estimated photometry of $\Delta L' = \qty{10.03(13:12)}{\magnitude}$ is in perfect agreement (i.e., within one standard deviation) with the results of \cite{fransonAstrometricAccelerationsDynamical2023}, who reported $\Delta L' = \qty{9.94 \pm 0.14}{\magnitude}$. In Appendix~\ref{sec:af_lep_fake_planets}, we discuss an additional fake planet experiment that verifies that the signal fit with PCA is the source of the detection in \fours.


\paragraph{Orbit Fit}
We combine the new relative astrometry data point obtained in this work with the relative astrometry points from the three AF~Lep~b detection papers \citep{derosaDirectImagingDiscovery2023,mesaAFLepLowestmass2023,fransonAstrometricAccelerationsDynamical2023}, \textsc{Hipparcos} \citep[see][]{leeuwenValidationNewHipparcos2007} and \textsc{Gaia} \citep{prustiGaiaMission2016, vallenariGaiaDataRelease2023} acceleration, and archival RV data published by \cite{butlerLCESHIRESKeck2017}. To combine the data into a single orbit fit, we use the MCMC-based orbit fitting package \texttt{orvara} \citep{brandtOrvaraEfficientCode2021}. We fit for all orbital parameters and the primary and secondary mass, setting the default priors as defined in Table~4 of \citet{brandtOrvaraEfficientCode2021}. To stay comparable to the results in \cite{fransonAstrometricAccelerationsDynamical2023}, we also set a highly constrained Gaussian prior on the mass of AF~Lep: $M = \qty{1.2 \pm 0.06}{\solarmass}$ \citep{kervellaStellarSubstellarCompanions2022}. We run the Monte Carlo chain to \num{500000} samples over \num{100} walkers, discarding the first \qty{20}{\percent} of each chain as burn-in.
The parameter distributions and resulting fit values are reported in \Cref{fig:05_af_lep_orbit} and \Cref{tab:af_lep_prop}. We report a dynamical mass estimate of $M = \qty{3.74(53:50)}{\jupitermass}$, confirming its status well within the planetary-mass regime. This estimate offers a significant improvement over and is compatible with the masses reported in \cite{fransonAstrometricAccelerationsDynamical2023} and \cite{derosaDirectImagingDiscovery2023}.

As the additional \naco data point is almost precisely opposite to the more recent observations, it offers an ideal opportunity to constrain the orbit's eccentricity. Indeed, we find AF~Lep~b to move on a circular orbit ($\epsilon = \qty{0.031(27:20)}{}$). Considering that, due to the low mass of the companion, it likely formed situ in the protoplanetary disk, our measurement agrees much better with tentative evidence of planet-mass companions occurring preferentially on circular orbits \citep{bowlerPopulationlevelEccentricityDistributions2020}.
We note that, due to the relative astrometry point being opposite, we do not improve the fit of the inclination. This circular orbit agrees with recent interferometric observations of the planet with VLTI-GRAVITY \citep{balmer2024}.
\looseness=-1

In addition to the 2011 observation, \mbox{AF Lep} was observed a second time with \naco in 2015 (prog. ID 096.C-0679(A); \citealt{launhardtISPYNACOImagingSurvey2020}). Our reanalysis of this data set did not yield a detection. In 2015, the projected separation of the planet from the star is much smaller ($\lesssim$ \qty{2}{\lod}) than in 2011, resulting in higher-contrast requirements. 
Therefore, the nondetection in 2015 is consistent with our contrast limits and orbit fit.

    \section{Future Work}
\label{sec:future_work}

The detection of \mbox{AF Lep b} in archival \naco data demonstrates the potential of \fours to find previously missed companions.
The archive of \naco observations comprises more than 800 data sets from more than 450 stars (considering only $L'$-band data taken in pupil tracking mode). It is possible that a reanalysis of these data sets will reveal additional companions. 
Future work should consider a reanalysis not only for \naco data but also for other instruments. In this way, a better understanding of the contrast performance of \fours for different observing modes and instruments can be obtained.
The use of saliency maps together with \fours may help us to better understand the optics and speckle pattern.
If some difficult-to-interpret patterns are systematically visible in the saliency maps of an instrument, they may indicate overlooked effects in the instrument optics.
This could open the door to new developments on the hardware side.
\looseness=-1

The primary drawback of \fours is its computational cost. Processing one data set takes about 10--30 minutes on a single NVIDIA A100/H100 GPU. While this computation time is unproblematic for detection purposes, it quickly becomes infeasible for scientific analysis. Extracting astrometry and photometry, or calculating detection limits, requires reprocessing the same data set hundreds of times.
In fact, the most computationally expensive step in \fours is the optimization of the model matrix in step~2 (see \Cref{fig:04_s4_explained}). We recompute this matrix from scratch for all our fake planet experiments. This might not be necessary.
Future work should explore strategies to compute the matrix once and update it as the scientific analysis progresses. Fast GPU-based MCMC implementations compatible with our code are available, for example, in \texttt{pyro} \citep{binghamPyroDeepUniversal2019}. 
The L-BFGS optimizer used in this work requires a lot of GPU memory, especially for large science frames. This sets the maximum resolution that can still be processed with one NVIDIA A100/H100 GPU to about $150 \times 150$ pixels, depending on the temporal resolution of the data set. We recommend cropping the science frames to benefit from the better contrast performance of \fours at close separations. Unfortunately, this memory requirement currently prevents us from processing the data sets of the exoplanet imaging data challenge \citep{cantalloubeDirectExoplanetDetection2015}. 
Future work should focus on making our implementation more memory efficient, for example, by using gradient accumulation during  optimization. 
\looseness=-1

\paragraph{Further development of the method}
The noise model of \fours is a simple linear model. Especially at separations close to the star and in data taken with a coronagraph, nonlinear artifacts can occur. 
For this purpose, the linear model of \fours in \Cref{eq:s4_noise_model} could be replaced by a neural network. Neural networks have already been proposed for speckle subtraction in HCI (see \citealt{wolfDirectExoplanetDetection2023}). 
In \Cref{sec:ana}, we observed that a substantial part of the information used by \fours to explain the speckle noise are symmetries across the entire frame.
Therefore, neural networks should not be used locally, as in \cite{wolfDirectExoplanetDetection2023}, but globally. A possible direction for this could be masked autoencoders \citep{heMaskedAutoencodersAre2022}.

A number of new post-processing methods based on PCA have been proposed in the literature. One example is the RSM map of \cite{dahlqvistRegimeswitchingModelDetection2020}, which replaces the last step in PCA, the averaging along time, and allows combining multiple post-processing algorithms. Another example is the combination of PCA with forward modeling \mbox{(FMMF; \citealt{ruffioImprovingAssessingPlanet2017})}.
The basic steps of \fours are identical to those of PCA, which is why methods such as RSM and FMMF can also be used with \fours. For this reason, \fours should be seen as a building block for new post-processing methods.

Our analysis of the saliency maps in \Cref{sec:ana} has shown that the noise model learned by \fours hardly changes over years.
This is a strong indicator that \fours could also be used with RDI. For RDI, the noise model of \fours needs to be learned only once and can then be applied to many data sets. This procedure also avoids the high computational costs of \fours for ADI. 
The parameters of \fours could be conditioned on observation conditions, as suggested by \cite{gebhardHalfsiblingRegressionMeets2022} for half-sibling regression. In combination with RDI, we might be able to obtain a universal model of the speckle noise that utilizes all previously collected data. 

\section{Summary and Conclusions} 
\label{sec:summary}

In this paper, we have gained a deep understanding for the loss of planet signal in PCA-based post-processing.  
Using saliency maps, we identified two main reasons for this loss: first, the convergence of the PCA noise model to the telescope PSF; and second, a problematic loss function that misleads the noise estimate to fit the planet signal. 
Building on these insights, we developed a new post-processing algorithm that makes a step toward overcoming the limitations of PCA. We named this method \fours. \fours is characterized by three innovations: first, a linear noise model with a right reason constraint, second, a new loss function that is invariant to the planet signal, and third, a domain-knowledge-informed regularization. 

We applied \fours to archival data from the \naco instrument to compare its contrast performance to PCA. On a sample of 11 data sets, we observed a contrast improvement of up to 1.5 magnitudes, especially at close separations to the star. In addition to achieving deeper contrast, \fours produces residual noise that is more Gaussian compared to PCA (see Appendix~\ref{sec:a_residual_noise_dist}), reducing the risk of false positives. 
The main reason for this improvement is the ability of \fours to learn a better representation of the speckle noise. 
A detailed analysis revealed that \fours can recover the speckle pattern that we would expect from the theory. 
The increase in contrast performance allowed us to detect the giant exoplanet AF~Lep~b in archival data from 2011, 11 years before its discovery. Using this new astrometric point, we significantly improve the constraints on the orbital parameters of the planet. 

The introduction of \fours marks a leap for the post-processing of HCI data. Looking ahead, our findings have the potential to boost exoplanet detection on archival data and for future observations. 
We believe that the findings from our work will provide inspiration for the development of new post-processing methods beyond \fours.

    \section{Acknowledgments}
We thank the anonymous referee for a critical and constructive review of the original manuscript, which helped us improve the clarity and quality of the paper significantly.
This work was supported by an ETH Zurich Research Grant. M.J.B. and S.P.Q. gratefully acknowledge the financial support from ETH Zurich. Parts of this work have been carried out within the framework of the National Centre of Competence in Research PlanetS supported by the Swiss National Science Foundation (SNSF). J.H. acknowledges the financial support from the Swiss National Science Foundation (SNSF) under project grant number 200020\_200399. This project has received funding from the European Research Council (ERC) under the European Union’s Horizon 2020 research and innovation program (grant agreement No. 819155). 

\textbf{Author Contributions:}
M.J.B. carried out the main analyses, developed the \fours algorithm, and wrote the manuscript. He programmed the publicly available python package \texttt{fours} and wrote the documentation page. T.D.G. helped with the analysis and the development of 4S. His previous work on half-sibling regression algorithm has set the basis for this paper. F.A.D. carried out the orbit fits of AF~Lep~b and helped write the corresponding section. S.P.Q. and B.S. provided access to the compute resources needed for this project. All authors discussed the algorithm, analysis, and results. All authors commented on the manuscript.

\textbf{Used software:}
This work has made use of a number of open-source Python packages, including
\texttt{astropy} \citep{astropy_2013, astropy_2018, astropy_2022},
\texttt{matplotlib} \citep{matplotlib},
\texttt{numpy} \citep{numpy},
\texttt{pandas} \citep{pandas},
\texttt{photutils} \citep{photutils},
\texttt{scikit-image} \citep{skimage},
\texttt{scikit-learn} \citep{sklearn},
\texttt{scipy} \citep{scipy},
\texttt{seaborn} \citep{seaborn},
\texttt{torch} \citep{paszkePyTorchImperativeStyle2019}, and
\texttt{tqdm} \citep{tqdm}.
    
    \bibliography{s4_bib_zotero,software,datasets}{}

\begin{thebibliography}{}
\expandafter\ifx\csname natexlab\endcsname\relax\def\natexlab#1{#1}\fi
\providecommand{\url}[1]{\href{#1}{#1}}
\providecommand{\dodoi}[1]{doi:~\href{http://doi.org/#1}{\nolinkurl{#1}}}
\providecommand{\doeprint}[1]{\href{http://ascl.net/#1}{\nolinkurl{http://ascl.net/#1}}}
\providecommand{\doarXiv}[1]{\href{https://arxiv.org/abs/#1}{\nolinkurl{https://arxiv.org/abs/#1}}}

\bibitem[{Absil {et~al.}(2013)Absil, Milli, Mawet, Lagrange, Girard, Chauvin, Boccaletti, Delacroix, \& Surdej}]{absilSearchingCompanionsAU2013}
Absil, O., Milli, J., Mawet, D., {et~al.} 2013, A\&A, 559, L12, \dodoi{10.1051/0004-6361/201322748}

\bibitem[{Ali {et~al.}(2023)Ali, Abuhmed, {El-Sappagh}, Muhammad, {Alonso-Moral}, Confalonieri, Guidotti, Del~Ser, {D{\'i}az-Rodr{\'i}guez}, \& Herrera}]{aliExplainableArtificialIntelligence2023}
Ali, S., Abuhmed, T., {El-Sappagh}, S., {et~al.} 2023, Information Fusion, 99, 101805, \dodoi{10.1016/j.inffus.2023.101805}

\bibitem[{Amara \& Quanz(2012)}]{amaraPYNPOINTImageProcessing2012}
Amara, A., \& Quanz, S.~P. 2012, MNRAS, 427, 948, \dodoi{10.1111/j.1365-2966.2012.21918.x}

\bibitem[{Arcidiacono \& Simoncini(2018)}]{arcidiaconoApproximateNonnegativeMatrix2018}
Arcidiacono, C., \& Simoncini, V. 2018, in Adaptive {{Optics Systems VI}}, Vol. 10703 ({SPIE}), 1070331, \dodoi{10.1117/12.2311681}

\bibitem[{{Balmer} {et~al.}(2025){Balmer}, {Franson}, {Chomez}, {Pueyo}, {Stolker}, {Lacour}, {Nowak}, {Nasedkin}, {Bonse}, {Thorngren}, {Palma-Bifani}, {Molli{\`e}re}, {Wang}, {Zhang}, {Chavez}, {Kammerer}, {Blunt}, {Bowler}, {Bonnefoy}, {Brandner}, {Charnay}, {Chauvin}, {Henning}, {Lagrange}, {Pourr{\'e}}, {Rickman}, {De Rosa}, {Vigan}, \& {Winterhalder}}]{balmer2024}
{Balmer}, W.~O., {Franson}, K., {Chomez}, A., {et~al.} 2025, \aj, 169, 30, \dodoi{10.3847/1538-3881/ad9265}

\bibitem[{Baydin {et~al.}(2018)Baydin, Pearlmutter, Radul, \& Siskind}]{baydinAutomaticDifferentiationMachine2018}
Baydin, A.~G., Pearlmutter, B.~A., Radul, A.~A., \& Siskind, J.~M. 2018, Journal of Machine Learning Research, 18, 1.
\newblock \url{https://arxiv.org/abs/1502.05767}

\bibitem[{Bell {et~al.}(2015)Bell, Mamajek, \& Naylor}]{bellSelfconsistentAbsoluteIsochronal2015}
Bell, C. P.~M., Mamajek, E.~E., \& Naylor, T. 2015, MNRAS, 454, 593, \dodoi{10.1093/mnras/stv1981}

\bibitem[{Beuzit {et~al.}(2019)Beuzit, Vigan, Mouillet, Dohlen, Gratton, Boccaletti, Sauvage, Schmid, Langlois, Petit, Baruffolo, Feldt, Milli, Wahhaj, Abe, Anselmi, Antichi, Barette, Baudrand, Baudoz, Bazzon, Bernardi, Blanchard, Brast, Bruno, Buey, Carbillet, Carle, Cascone, Chapron, Charton, Chauvin, Claudi, Costille, De~Caprio, {de Boer}, Delboulb{\'e}, Desidera, Dominik, Downing, Dupuis, Fabron, Fantinel, Farisato, Feautrier, Fedrigo, Fusco, Gigan, Ginski, Girard, Giro, Gisler, Gluck, Gry, Henning, Hubin, Hugot, Incorvaia, Jaquet, Kasper, Lagadec, Lagrange, Le~Coroller, Le~Mignant, Le~Ruyet, Lessio, Lizon, Llored, Lundin, Madec, Magnard, Marteaud, Martinez, Maurel, M{\'e}nard, Mesa, {M{\"o}ller-Nilsson}, Moulin, Moutou, Orign{\'e}, Parisot, Pavlov, Perret, Pragt, Puget, Rabou, Ramos, Reess, Rigal, Rochat, Roelfsema, Rousset, Roux, Saisse, Salasnich, Santambrogio, Scuderi, Segransan, Sevin, Siebenmorgen, Soenke, Stadler, Suarez, Tiph{\`e}ne, Turatto, Udry, Vakili, Waters, Weber, Wildi, Zins, \&
  Zurlo}]{beuzitSPHEREExoplanetImager2019}
Beuzit, J.~L., Vigan, A., Mouillet, D., {et~al.} 2019, A\&A, 631, A155, \dodoi{10.1051/0004-6361/201935251}

\bibitem[{Bingham {et~al.}(2019)Bingham, Chen, Jankowiak, Obermeyer, Pradhan, Karaletsos, Singh, Szerlip, Horsfall, \& Goodman}]{binghamPyroDeepUniversal2019}
Bingham, E., Chen, J.~P., Jankowiak, M., {et~al.} 2019, Journal of Machine Learning Research, 20, 1.
\newblock \url{https://arxiv.org/abs/1810.09538}

\bibitem[{Bishop(2006)}]{bishopPatternRecognitionMachine2006}
Bishop, C.~M. 2006, Pattern Recognition and Machine Learning, Information Science and Statistics (New York: Springer)

\bibitem[{Bloemhof(2007)}]{bloemhofFeasibilitySymmetrybasedSpeckle2007}
Bloemhof, E.~E. 2007, Optics Express, 15, 4705, \dodoi{10.1364/OE.15.004705}

\bibitem[{Bloemhof {et~al.}(2001)Bloemhof, Dekany, Troy, \& Oppenheimer}]{bloemhofBehaviorRemnantSpeckles2001}
Bloemhof, E.~E., Dekany, R.~G., Troy, M., \& Oppenheimer, B.~R. 2001, ApJ, 558, L71, \dodoi{10.1086/323494}

\bibitem[{Bonse(2024{\natexlab{a}})}]{bonse_raw_data2024}
Bonse, M.~J. 2024{\natexlab{a}}, {Raw data for: Use the 4S (Signal-Safe Speckle Subtraction)}, 0.1,  Zenodo, \dodoi{10.5281/zenodo.11456704}

\bibitem[{Bonse(2024{\natexlab{b}})}]{bonse_inter_res2024}
---. 2024{\natexlab{b}}, {Intermediate results for: Use the 4S (Signal-Safe Speckle Subtraction)}, 0.1,  Zenodo, \dodoi{10.5281/zenodo.11457071}

\bibitem[{Bonse {et~al.}(2018)Bonse, Quanz, \& Amara}]{bonseWaveletBasedSpeckle2018}
Bonse, M.~J., Quanz, S.~P., \& Amara, A. 2018, arXiv e-prints, \dodoi{10.48550/arXiv.1804.05063}

\bibitem[{Bonse {et~al.}(2023)Bonse, Garvin, Gebhard, Dannert, Cantalloube, Cugno, Absil, Hayoz, Milli, Kasper, \& Quanz}]{bonseComparingApplesApples2023}
Bonse, M.~J., Garvin, E.~O., Gebhard, T.~D., {et~al.} 2023, AJ, 166, 71, \dodoi{10.3847/1538-3881/acc93c}

\bibitem[{Bowler {et~al.}(2020)Bowler, Blunt, \& Nielsen}]{bowlerPopulationlevelEccentricityDistributions2020}
Bowler, B.~P., Blunt, S.~C., \& Nielsen, E.~L. 2020, AJ, 159, 63, \dodoi{10.3847/1538-3881/ab5b11}

\bibitem[{{Bradley} {et~al.}(2024){Bradley}, {Sip\H{o}cz}, {Robitaille}, {Tollerud}, {Vinícius}, {Deil}, {Barbary}, {Wilson}, {Busko}, {Donath}, {Günther}, {Cara}, {Lim}, {Meßlinger}, {Burnett}, {Conseil}, {Droettboom}, {Bostroem}, {Bray}, {Andersen Bratholm}, {Jamieson}, {Ginsburg}, {Barentsen}, {Craig}, {Pascual}, {Rathi}, {Perrin}, {Morris}, \& {Perren}}]{photutils}
{Bradley}, L., {Sip\H{o}cz}, B., {Robitaille}, T., {et~al.} 2024, {astropy/photutils: 1.12.0},  Zenodo, \dodoi{10.5281/zenodo.10967176}

\bibitem[{Brandt {et~al.}(2021{\natexlab{a}})Brandt, Dupuy, Li, Chen, Brandt, Wong, Currie, Bowler, Liu, Best, \& Phillips}]{brandtImprovedDynamicalMasses2021}
Brandt, G.~M., Dupuy, T.~J., Li, Y., {et~al.} 2021{\natexlab{a}}, AJ, 162, 301, \dodoi{10.3847/1538-3881/ac273e}

\bibitem[{Brandt {et~al.}(2021{\natexlab{b}})Brandt, Dupuy, Li, Brandt, Zeng, Michalik, Gagliuffi, \& {Raposo-Pulido}}]{brandtOrvaraEfficientCode2021}
Brandt, T.~D., Dupuy, T.~J., Li, Y., {et~al.} 2021{\natexlab{b}}, AJ, 162, 186, \dodoi{10.3847/1538-3881/ac042e}

\bibitem[{Butler {et~al.}(2017)Butler, Vogt, Laughlin, Burt, Rivera, Tuomi, Teske, Arriagada, Diaz, Holden, \& Keiser}]{butlerLCESHIRESKeck2017}
Butler, R.~P., Vogt, S.~S., Laughlin, G., {et~al.} 2017, AJ, 153, 208, \dodoi{10.3847/1538-3881/aa66ca}

\bibitem[{Cantalloube {et~al.}(2015)Cantalloube, Mouillet, Mugnier, Milli, Absil, Gomez~Gonzalez, Chauvin, Beuzit, \& Cornia}]{cantalloubeDirectExoplanetDetection2015}
Cantalloube, F., Mouillet, D., Mugnier, L.~M., {et~al.} 2015, A\&A, 582, A89, \dodoi{10.1051/0004-6361/201425571}

\bibitem[{Cantalloube {et~al.}(2020)Cantalloube, Gomez-Gonzalez, Absil, Cantero, Bacher, Bonse, Bottom, Dahlqvist, Desgrange, Flasseur, {et~al.}}]{cantalloubeExoplanetImagingData2021}
Cantalloube, F., Gomez-Gonzalez, C., Absil, O., {et~al.} 2020, in Adaptive Optics Systems VII, Vol. 11448 (SPIE), \dodoi{10.1117/12.2574803}

\bibitem[{Cantero {et~al.}(2023)Cantero, Absil, Dahlqvist, \& Van~Droogenbroeck}]{canteroNASODINNDeepLearning2023}
Cantero, C., Absil, O., Dahlqvist, C.~H., \& Van~Droogenbroeck, M. 2023, A\&A, 680, A86, \dodoi{10.1051/0004-6361/202346085}

\bibitem[{Chauvin {et~al.}(2017)Chauvin, Desidera, Lagrange, Vigan, Gratton, Langlois, Bonnefoy, Beuzit, Feldt, Mouillet, Meyer, Cheetham, Biller, Boccaletti, D'Orazi, Galicher, Hagelberg, Maire, Mesa, Olofsson, Samland, Schmidt, Sissa, Bonavita, Charnay, Cudel, Daemgen, Delorme, {Janin-Potiron}, Janson, Keppler, Le~Coroller, Ligi, Marleau, Messina, Molli{\`e}re, Mordasini, M{\"u}ller, Peretti, Perrot, Rodet, Rouan, Zurlo, Dominik, Henning, Menard, Schmid, Turatto, Udry, Vakili, Abe, Antichi, Baruffolo, Baudoz, Baudrand, Blanchard, Bazzon, Buey, Carbillet, Carle, Charton, Cascone, Claudi, Costille, Deboulbe, De~Caprio, Dohlen, Fantinel, Feautrier, Fusco, Gigan, Giro, Gisler, Gluck, Hubin, Hugot, Jaquet, Kasper, Madec, Magnard, Martinez, Maurel, Le~Mignant, {M{\"o}ller-Nilsson}, Llored, Moulin, Orign{\'e}, Pavlov, Perret, Petit, Pragt, Puget, Rabou, Ramos, Rigal, Rochat, Roelfsema, Rousset, Roux, Salasnich, Sauvage, Sevin, Soenke, Stadler, Suarez, Weber, Wildi, Antoniucci, Augereau, Baudino, Brandner, Engler,
  Girard, Gry, Kral, Kopytova, Lagadec, Milli, Moutou, Schlieder, Szul{\'a}gyi, Thalmann, \& Wahhaj}]{chauvinDiscoveryWarmDusty2017}
Chauvin, G., Desidera, S., Lagrange, A.~M., {et~al.} 2017, A\&A, 605, L9, \dodoi{10.1051/0004-6361/201731152}

\bibitem[{Cheetham {et~al.}(2018)Cheetham, S{\'e}gransan, Peretti, Delisle, Hagelberg, Beuzit, Forveille, Marmier, Udry, \& Wildi}]{cheethamDirectImagingUltracool2018a}
Cheetham, A., S{\'e}gransan, D., Peretti, S., {et~al.} 2018, A\&A, 614, A16, \dodoi{10.1051/0004-6361/201630136}

\bibitem[{Christiaens {et~al.}(2023)Christiaens, Gonzalez, Farkas, Dahlqvist, Nasedkin, Milli, Absil, Ngo, Cantero, Rainot, Hammond, Bonse, Cantalloube, Vigan, Kompella, \& Hancock}]{christiaensVIPPythonPackage2023}
Christiaens, V., Gonzalez, C. A.~G., Farkas, R., {et~al.} 2023, Journal of Open Source Software, 8, 4774, \dodoi{10.21105/joss.04774}

\bibitem[{Cugno {et~al.}(2023)Cugno, Pearce, Launhardt, Bonse, Ma, Henning, Quirrenbach, S{\'e}gransan, Matthews, Quanz, Kennedy, M{\"u}ller, Reffert, \& Rickman}]{cugnoISPYNACOImaging2023}
Cugno, G., Pearce, T.~D., Launhardt, R., {et~al.} 2023, A\&A, 669, A145, \dodoi{10.1051/0004-6361/202244891}

\bibitem[{Currie {et~al.}(2023)Currie, Biller, Lagrange, Marois, Guyon, Nielsen, Bonnefoy, \& De~Rosa}]{currieDirectImagingSpectroscopy2023}
Currie, T., Biller, B., Lagrange, A.-M., {et~al.} 2023, Direct {{Imaging}} and {{Spectroscopy}} of {{Extrasolar Planets}},  arXiv.
\newblock \doarXiv{2205.05696}

\bibitem[{{da Costa-Luis}(2019)}]{tqdm}
{da Costa-Luis}, C.~O. 2019, JOSS, 4, 1277, \dodoi{10.21105/joss.01277}

\bibitem[{Daglayan {et~al.}(2023)Daglayan, Vary, Leplat, Gillis, \& Absil}]{daglayanDirectExoplanetDetection2023}
Daglayan, H., Vary, S., Leplat, V., Gillis, N., \& Absil, P.-A. 2023, in Proceedings of BNAIC/BeNeLearn.
\newblock \doarXiv{2304.03619}

\bibitem[{Dahlqvist \& Absil(2021)}]{dahlqvistImprovingRSMMap2021}
Dahlqvist, C.-H., \& Absil, O. 2021, A\&A, 646, A49, \dodoi{10.1051/0004-6361/202039597}

\bibitem[{Dahlqvist {et~al.}(2020)Dahlqvist, Cantalloube, \& Absil}]{dahlqvistRegimeswitchingModelDetection2020}
Dahlqvist, C.-H., Cantalloube, F., \& Absil, O. 2020, A\&A, 633, A95, \dodoi{10.1051/0004-6361/201936421}

\bibitem[{De~Rosa {et~al.}(2023)De~Rosa, Nielsen, Wahhaj, Ruffio, Kalas, Peck, Hirsch, \& Roberson}]{derosaDirectImagingDiscovery2023}
De~Rosa, R.~J., Nielsen, E.~L., Wahhaj, Z., {et~al.} 2023, A\&A, 672, A94, \dodoi{10.1051/0004-6361/202345877}

\bibitem[{Flasseur {et~al.}(2024)Flasseur, Bodrito, Mairal, Ponce, Langlois, \& Lagrange}]{flasseurDeepPACOCombining2024}
Flasseur, O., Bodrito, T., Mairal, J., {et~al.} 2024, MNRAS, 527, 1534, \dodoi{10.1093/mnras/stad3143}

\bibitem[{Flasseur {et~al.}(2018)Flasseur, Denis, Thi{\'e}baut, \& Langlois}]{flasseurExoplanetDetectionAngular2018}
Flasseur, O., Denis, L., Thi{\'e}baut, {\'E}., \& Langlois, M. 2018, A\&A, 618, A138, \dodoi{10.1051/0004-6361/201832745}

\bibitem[{Franson {et~al.}(2023)Franson, Bowler, Zhou, Pearce, Bardalez~Gagliuffi, Biddle, Brandt, Crepp, Dupuy, Faherty, {Jensen-Clem}, Morgan, Sanghi, Theissen, Tran, \& Wolf}]{fransonAstrometricAccelerationsDynamical2023}
Franson, K., Bowler, B.~P., Zhou, Y., {et~al.} 2023, ApJ, 950, L19, \dodoi{10.3847/2041-8213/acd6f6}

\bibitem[{Fulton {et~al.}(2021)Fulton, Rosenthal, Hirsch, Isaacson, Howard, Dedrick, Sherstyuk, Blunt, Petigura, Knutson, Behmard, Chontos, Crepp, Crossfield, Dalba, Fischer, Henry, Kane, Kosiarek, Marcy, Rubenzahl, Weiss, \& Wright}]{fultonCaliforniaLegacySurvey2021}
Fulton, B.~J., Rosenthal, L.~J., Hirsch, L.~A., {et~al.} 2021, ApJ Supplement Series, 255, 14, \dodoi{10.3847/1538-4365/abfcc1}

\bibitem[{Gebhard {et~al.}(2022)Gebhard, Bonse, Quanz, \& Sch{\"o}lkopf}]{gebhardHalfsiblingRegressionMeets2022}
Gebhard, T.~D., Bonse, M.~J., Quanz, S.~P., \& Sch{\"o}lkopf, B. 2022, A\&A, 666, A9, \dodoi{10.1051/0004-6361/202142529}

\bibitem[{Gomez~Gonzalez {et~al.}(2016)Gomez~Gonzalez, Absil, Absil, Droogenbroeck, Mawet, \& Surdej}]{gomezgonzalezLowrankSparseDecomposition2016}
Gomez~Gonzalez, C.~A., Absil, O., Absil, P.-A., {et~al.} 2016, A\&A, 589, A54, \dodoi{10.1051/0004-6361/201527387}

\bibitem[{Gomez~Gonzalez {et~al.}(2018)Gomez~Gonzalez, Absil, \& Van~Droogenbroeck}]{gomezgonzalezSupervisedDetectionExoplanets2018}
Gomez~Gonzalez, C.~A., Absil, O., \& Van~Droogenbroeck, M. 2018, A\&A, 613, A71, \dodoi{10.1051/0004-6361/201731961}

\bibitem[{Gomez~Gonzalez {et~al.}(2017)Gomez~Gonzalez, Wertz, Absil, Christiaens, Defr{\`e}re, Mawet, Milli, Absil, Droogenbroeck, Cantalloube, Hinz, Skemer, Karlsson, \& Surdej}]{gomezgonzalezVIPVortexImage2017a}
Gomez~Gonzalez, C.~A., Wertz, O., Absil, O., {et~al.} 2017, AJ, 154, 7, \dodoi{10.3847/1538-3881/aa73d7}

\bibitem[{Gray {et~al.}(2006)Gray, Corbally, Garrison, McFadden, Bubar, McGahee, O'Donoghue, \& Knox}]{grayContributionsNearbyStars2006}
Gray, R.~O., Corbally, C.~J., Garrison, R.~F., {et~al.} 2006, AJ, 132, 161, \dodoi{10.1086/504637}

\bibitem[{Halko {et~al.}(2011)Halko, Martinsson, \& Tropp}]{halkoFindingStructureRandomness2011}
Halko, N., Martinsson, P.~G., \& Tropp, J.~A. 2011, SIAM Review, 53, 217, \dodoi{10.1137/090771806}

\bibitem[{{Harris} {et~al.}(2020){Harris}, {Millman}, {van der Walt}, {et~al.}}]{numpy}
{Harris}, C.~R., {Millman}, K.~J., {van der Walt}, S.~J., {et~al.} 2020, \nat, 585, 357, \dodoi{10.1038/s41586-020-2649-2}

\bibitem[{He {et~al.}(2022)He, Chen, Xie, Li, Dollar, \& Girshick}]{heMaskedAutoencodersAre2022}
He, K., Chen, X., Xie, S., {et~al.} 2022, in 2022 {{IEEE}}/{{CVF Conference}} on {{Computer Vision}} and {{Pattern Recognition}} ({{CVPR}}) (New Orleans, LA, USA: IEEE), 15979--15988, \dodoi{10.1109/CVPR52688.2022.01553}

\bibitem[{Hunter(2007)}]{matplotlib}
Hunter, J.~D. 2007, Computing in Science \& Engineering, 9, 90, \dodoi{10.1109/mcse.2007.55}

\bibitem[{Jaderberg {et~al.}(2016)Jaderberg, Simonyan, Zisserman, \& Kavukcuoglu}]{jaderbergSpatialTransformerNetworks2016}
Jaderberg, M., Simonyan, K., Zisserman, A., \& Kavukcuoglu, K. 2016, in NeurIPS, \dodoi{10.48550/arXiv.1506.02025}

\bibitem[{Janson {et~al.}(2021)Janson, Squicciarini, Delorme, Gratton, Bonnefoy, Reffert, Mamajek, Eriksson, Vigan, Langlois, Engler, Chauvin, Desidera, Mayer, Marleau, Bohn, Samland, Meyer, {d'Orazi}, Henning, Quanz, Kenworthy, \& Carson}]{jansonBEASTBeginsSample2021}
Janson, M., Squicciarini, V., Delorme, P., {et~al.} 2021, A\&A, 646, A164, \dodoi{10.1051/0004-6361/202039683}

\bibitem[{{Jensen-Clem} {et~al.}(2017){Jensen-Clem}, Mawet, Gomez~Gonzalez, Absil, Belikov, Currie, Kenworthy, Marois, Mazoyer, Ruane, Tanner, \& Cantalloube}]{jensen-clemNewStandardAssessing2017}
{Jensen-Clem}, R., Mawet, D., Gomez~Gonzalez, C.~A., {et~al.} 2017, AJ, 155, 19, \dodoi{10.3847/1538-3881/aa97e4}

\bibitem[{Jovanovic {et~al.}(2015)Jovanovic, Martinache, Guyon, Clergeon, Singh, Kudo, Garrel, Newman, Doughty, Lozi, Males, Minowa, Hayano, Takato, Morino, Kuhn, Serabyn, Norris, Tuthill, Schworer, Stewart, Close, Huby, Perrin, Lacour, Gauchet, Vievard, Murakami, Oshiyama, Baba, Matsuo, Nishikawa, Tamura, Lai, Marchis, Duchene, Kotani, \& Woillez}]{jovanovicSubaruCoronagraphicExtreme2015}
Jovanovic, N., Martinache, F., Guyon, O., {et~al.} 2015, PASP, 127, 890, \dodoi{10.1086/682989}

\bibitem[{Kenworthy {et~al.}(2007)Kenworthy, Codona, Hinz, Angel, Heinze, \& Sivanandam}]{kenworthyFirstOnSkyHighContrast2007}
Kenworthy, M.~A., Codona, J.~L., Hinz, P.~M., {et~al.} 2007, ApJ, 660, 762, \dodoi{10.1086/513596}

\bibitem[{Kervella {et~al.}(2022)Kervella, Arenou, \& Th{\'e}venin}]{kervellaStellarSubstellarCompanions2022}
Kervella, P., Arenou, F., \& Th{\'e}venin, F. 2022, A\&A, 657, A7, \dodoi{10.1051/0004-6361/202142146}

\bibitem[{Kingma \& Ba(2015)}]{kingmaAdamMethodStochastic2017}
Kingma, D.~P., \& Ba, J. 2015, in ICLR, \dodoi{10.48550/arXiv.1412.6980}

\bibitem[{Kuhn {et~al.}(2001)Kuhn, Potter, \& Parise}]{kuhnImagingPolarimetricObservations2001}
Kuhn, J.~R., Potter, D., \& Parise, B. 2001, ApJ, 553, L189, \dodoi{10.1086/320686}

\bibitem[{Lafreni{\`e}re {et~al.}(2007)Lafreni{\`e}re, Marois, Doyon, Nadeau, \& Artigau}]{lafreniereNewAlgorithmPointSpread2007}
Lafreni{\`e}re, D., Marois, C., Doyon, R., Nadeau, D., \& Artigau, {\'E}. 2007, ApJ, 660, 770, \dodoi{10.1086/513180}

\bibitem[{Lagrange {et~al.}(2023)Lagrange, Philipot, Rubini, Meunier, Kiefer, Kervella, Delorme, \& Beust}]{lagrangeRadialDistributionGiant2023}
Lagrange, A.~M., Philipot, F., Rubini, P., {et~al.} 2023, A\&A, 677, A71, \dodoi{10.1051/0004-6361/202346165}

\bibitem[{{Langlois} {et~al.}(2021){Langlois}, {Gratton}, {Lagrange}, {Delorme}, {Boccaletti}, {Bonnefoy}, {Maire}, {Mesa}, {Chauvin}, {Desidera}, {Vigan}, {Cheetham}, {Hagelberg}, {Feldt}, {Meyer}, {Rubini}, {Le Coroller}, {Cantalloube}, {Biller}, {Bonavita}, {Bhowmik}, {Brandner}, {Daemgen}, {D'Orazi}, {Flasseur}, {Fontanive}, {Galicher}, {Girard}, {Janin-Potiron}, {Janson}, {Keppler}, {Kopytova}, {Lagadec}, {Lannier}, {Lazzoni}, {Ligi}, {Meunier}, {Perreti}, {Perrot}, {Rodet}, {Romero}, {Rouan}, {Samland}, {Salter}, {Sissa}, {Schmidt}, {Zurlo}, {Mouillet}, {Denis}, {Thi{\'e}baut}, {Milli}, {Wahhaj}, {Beuzit}, {Dominik}, {Henning}, {M{\'e}nard}, {M{\"u}ller}, {Schmid}, {Turatto}, {Udry}, {Abe}, {Antichi}, {Allard}, {Baruffolo}, {Baudoz}, {Baudrand}, {Bazzon}, {Blanchard}, {Carbillet}, {Carle}, {Cascone}, {Charton}, {Claudi}, {Costille}, {De Caprio}, {Delboulb{\'e}}, {Dohlen}, {Fantinel}, {Feautrier}, {Fusco}, {Gigan}, {Giro}, {Gisler}, {Gluck}, {Gry}, {Hubin}, {Hugot}, {Jaquet}, {Kasper}, {Le Mignant},
  {Llored}, {Madec}, {Magnard}, {Martinez}, {Maurel}, {Messina}, {M{\"o}ller-Nilsson}, {Mugnier}, {Moulin}, {Orign{\'e}}, {Pavlov}, {Perret}, {Petit}, {Pragt}, {Puget}, {Rabou}, {Ramos}, {Rigal}, {Rochat}, {Roelfsema}, {Rousset}, {Roux}, {Salasnich}, {Sauvage}, {Sevin}, {Soenke}, {Stadler}, {Suarez}, {Weber}, {Wildi}, \& {Rickman}}]{langloisSPHEREInfraredSurvey2021}
{Langlois}, M., {Gratton}, R., {Lagrange}, A.~M., {et~al.} 2021, \aap, 651, A71, \dodoi{10.1051/0004-6361/202039753}

\bibitem[{Launhardt {et~al.}(2020)Launhardt, Henning, Quirrenbach, S{\'e}gransan, Avenhaus, {van Boekel}, Brems, Cheetham, Cugno, Girard, Godoy, Kennedy, Maire, Metchev, M{\"u}ller, Musso~Barcucci, Olofsson, Pepe, Quanz, Queloz, Reffert, Rickman, Ruh, \& Samland}]{launhardtISPYNACOImagingSurvey2020}
Launhardt, R., Henning, {\relax Th}., Quirrenbach, A., {et~al.} 2020, A\&A, 635, A162, \dodoi{10.1051/0004-6361/201937000}

\bibitem[{Lewis {et~al.}(2023)Lewis, Fitzgerald, Dodkins, Davis, \& Lin}]{lewisSpeckleSpaceTime2023a}
Lewis, B., Fitzgerald, M.~P., Dodkins, R.~H., Davis, K.~K., \& Lin, J. 2023, AJ, 165, 59, \dodoi{10.3847/1538-3881/aca761}

\bibitem[{Liu \& Nocedal(1989)}]{liuLimitedMemoryBFGS1989}
Liu, D.~C., \& Nocedal, J. 1989, Mathematical Programming, 45, 503, \dodoi{10.1007/BF01589116}

\bibitem[{Long {et~al.}(2023)Long, Males, Haffert, Pearce, Marley, Morzinski, Close, Otten, Snik, Kenworthy, Keller, Hinz, Monnier, Weinberger, \& Tolls}]{longImprovedCompanionMass2023}
Long, J.~D., Males, J.~R., Haffert, S.~Y., {et~al.} 2023, AJ, 165, 216, \dodoi{10.3847/1538-3881/acbd4b}

\bibitem[{Macintosh {et~al.}(2014)Macintosh, Graham, Ingraham, Konopacky, Marois, Perrin, Poyneer, Bauman, Barman, Burrows, Cardwell, Chilcote, De~Rosa, Dillon, Doyon, Dunn, Erikson, Fitzgerald, Gavel, Goodsell, Hartung, Hibon, Kalas, Larkin, Maire, Marchis, Marley, McBride, {Millar-Blanchaer}, Morzinski, Norton, Oppenheimer, Palmer, Patience, Pueyo, Rantakyro, Sadakuni, Saddlemyer, Savransky, Serio, Soummer, Sivaramakrishnan, Song, Thomas, Wallace, Wiktorowicz, \& Wolff}]{macintoshFirstLightGemini2014}
Macintosh, B., Graham, J.~R., Ingraham, P., {et~al.} 2014, PNAS, 111, 12661, \dodoi{10.1073/pnas.1304215111}

\bibitem[{Macintosh {et~al.}(2015)Macintosh, Graham, Barman, De~Rosa, Konopacky, Marley, Marois, Nielsen, Pueyo, Rajan, Rameau, Saumon, Wang, Patience, Ammons, Arriaga, Artigau, Beckwith, Brewster, Bruzzone, Bulger, Burningham, Burrows, Chen, Chiang, Chilcote, Dawson, Dong, Doyon, Draper, Duch{\^e}ne, Esposito, Fabrycky, Fitzgerald, Follette, Fortney, Gerard, Goodsell, Greenbaum, Hibon, Hinkley, Cotten, Hung, Ingraham, {Johnson-Groh}, Kalas, Lafreniere, Larkin, Lee, Line, Long, Maire, Marchis, Matthews, Max, Metchev, {Millar-Blanchaer}, Mittal, Morley, Morzinski, {Murray-Clay}, Oppenheimer, Palmer, Patel, Perrin, Poyneer, Rafikov, Rantakyr{\"o}, Rice, Rojo, Rudy, Ruffio, Ruiz, Sadakuni, Saddlemyer, Salama, Savransky, Schneider, Sivaramakrishnan, Song, Soummer, Thomas, Vasisht, Wallace, {Ward-Duong}, Wiktorowicz, Wolff, \& Zuckerman}]{macintoshDiscoverySpectroscopyYoung2015}
Macintosh, B., Graham, J.~R., Barman, T., {et~al.} 2015, Science, 350, 64, \dodoi{10.1126/science.aac5891}

\bibitem[{{Males} {et~al.}(2021){Males}, {Fitzgerald}, {Belikov}, \& {Guyon}}]{malesMysteriousLivesSpeckles2021}
{Males}, J.~R., {Fitzgerald}, M.~P., {Belikov}, R., \& {Guyon}, O. 2021, \pasp, 133, 104504, \dodoi{10.1088/1538-3873/ac0f0c}

\bibitem[{Males {et~al.}(2018)Males, Close, Miller, Schatz, Doelman, Lumbres, Snik, Rodack, Knight, Gorkom, Long, Hedglen, Kautz, Jovanovic, Morzinski, Guyon, Douglas, Follette, Lozi, Bohlman, Durney, Gasho, Hinz, Ireland, Jean, Keller, Kenworthy, Mazin, Noenickx, Alfred, Perez, Sanchez, Sauve, Weinberger, \& Conrad}]{malesMagAOXProjectStatus2018}
Males, J.~R., Close, L.~M., Miller, K., {et~al.} 2018, in Adaptive {{Optics Systems VI}}, Vol. 10703 (SPIE), 76--89, \dodoi{10.1117/12.2312992}

\bibitem[{{Marois} {et~al.}(2014){Marois}, {Correia}, {Galicher}, {Ingraham}, {Macintosh}, {Currie}, \& {De Rosa}}]{maroisGPIPSFSubtraction2014}
{Marois}, C., {Correia}, C., {Galicher}, R., {et~al.} 2014, in Adaptive Optics Systems IV, Vol. 9148 (SPIE), \dodoi{10.1117/12.2055245}

\bibitem[{Marois {et~al.}(2006)Marois, Lafreniere, Doyon, Macintosh, \& Nadeau}]{maroisAngularDifferentialImaging2006}
Marois, C., Lafreniere, D., Doyon, R., Macintosh, B., \& Nadeau, D. 2006, ApJ, 641, 556, \dodoi{10.1086/500401}

\bibitem[{{Marois} {et~al.}(2010){Marois}, {Macintosh}, \& {V{\'e}ran}}]{maroisExoplanetImagingLOCI2010}
{Marois}, C., {Macintosh}, B., \& {V{\'e}ran}, J.-P. 2010, in Adaptive Optics Systems II, Vol. 7736, 77361J, \dodoi{10.1117/12.857225}

\bibitem[{Mawet {et~al.}(2009)Mawet, Serabyn, Liewer, Burruss, Hickey, \& Shemo}]{mawetVECTORVORTEXCORONAGRAPH2009}
Mawet, D., Serabyn, E., Liewer, K., {et~al.} 2009, ApJ, 709, 53, \dodoi{10.1088/0004-637X/709/1/53}

\bibitem[{Mawet {et~al.}(2012)Mawet, Absil, Montagnier, Riaud, Surdej, Ducourant, Augereau, R{\"o}ttinger, Girard, Krist, \& Stapelfeldt}]{mawetDirectImagingExtrasolar2012}
Mawet, D., Absil, O., Montagnier, G., {et~al.} 2012, A\&A, 544, A131, \dodoi{10.1051/0004-6361/201219662}

\bibitem[{Mawet {et~al.}(2013)Mawet, Absil, Delacroix, Girard, Milli, O'Neal, Baudoz, Boccaletti, Bourget, Christiaens, Forsberg, Gonte, Habraken, Hanot, Karlsson, Kasper, Lizon, Muzic, Olivier, Pe{\~n}a, Slusarenko, {Tacconi-Garman}, \& Surdej}]{mawetBandAGPMVector2013}
Mawet, D., Absil, O., Delacroix, C., {et~al.} 2013, A\&A, 552, L13, \dodoi{10.1051/0004-6361/201321315}

\bibitem[{Mawet {et~al.}(2014)Mawet, Milli, Wahhaj, Pelat, Absil, Delacroix, Boccaletti, Kasper, Kenworthy, Marois, Mennesson, \& Pueyo}]{mawetFUNDAMENTALLIMITATIONSHIGH2014}
Mawet, D., Milli, J., Wahhaj, Z., {et~al.} 2014, ApJ, 792, 97, \dodoi{10.1088/0004-637X/792/2/97}

\bibitem[{{McKinney}(2010)}]{pandas}
{McKinney}, W. 2010, in {P}roceedings of the 9th {P}ython in {S}cience {C}onference, ed. S.~{van der Walt} \& J.~{Millman}, 56--61, \dodoi{10.25080/Majora-92bf1922-00a}

\bibitem[{Mesa {et~al.}(2023)Mesa, Gratton, Kervella, Bonavita, Desidera, D'Orazi, Marino, Zurlo, \& Rigliaco}]{mesaAFLepLowestmass2023}
Mesa, D., Gratton, R., Kervella, P., {et~al.} 2023, A\&A, 672, A93, \dodoi{10.1051/0004-6361/202345865}

\bibitem[{Meshkat {et~al.}(2013)Meshkat, Kenworthy, Quanz, \& Amara}]{meshkatOPTIMIZEDPRINCIPALCOMPONENT2013}
Meshkat, T., Kenworthy, M.~A., Quanz, S.~P., \& Amara, A. 2013, ApJ, 780, 17, \dodoi{10.1088/0004-637X/780/1/17}

\bibitem[{N'Diaye {et~al.}(2015)N'Diaye, Pueyo, \& Soummer}]{ndiayeApodizedPupilLyot2015}
N'Diaye, M., Pueyo, L., \& Soummer, R. 2015, ApJ, 799, 225, \dodoi{10.1088/0004-637X/799/2/225}

\bibitem[{Nielsen {et~al.}(2019)Nielsen, De~Rosa, Macintosh, Wang, Ruffio, Chiang, Marley, Saumon, Savransky, Mark~Ammons, Bailey, Barman, Blain, Bulger, Burrows, Chilcote, Cotten, Czekala, Doyon, Duch{\^e}ne, Esposito, Fabrycky, Fitzgerald, Follette, Fortney, Gerard, Goodsell, Graham, Greenbaum, Hibon, Hinkley, Hirsch, Hom, Hung, Ilene~Dawson, Ingraham, Kalas, Konopacky, Larkin, Lee, Lin, Maire, Marchis, Marois, Metchev, {Millar-Blanchaer}, Morzinski, Oppenheimer, Palmer, Patience, Perrin, Poyneer, Pueyo, Rafikov, Rajan, Rameau, Rantakyr{\"o}, Ren, Schneider, Sivaramakrishnan, Song, Soummer, Tallis, Thomas, {Ward-Duong}, \& Wolff}]{nielsenGeminiPlanetImager2019}
Nielsen, E.~L., De~Rosa, R.~J., Macintosh, B., {et~al.} 2019, AJ, 158, 13, \dodoi{10.3847/1538-3881/ab16e9}

\bibitem[{Otten {et~al.}(2017)Otten, Snik, Kenworthy, Keller, Males, Morzinski, Close, Codona, Hinz, Hornburg, Brickson, \& Escuti}]{ottenONSKYPERFORMANCEANALYSIS2017a}
Otten, G. P. P.~L., Snik, F., Kenworthy, M.~A., {et~al.} 2017, ApJ, 834, 175, \dodoi{10.3847/1538-4357/834/2/175}

\bibitem[{Pairet {et~al.}(2019)Pairet, Cantalloube, Gomez~Gonzalez, Absil, \& Jacques}]{pairetSTIMMapDetection2019}
Pairet, B., Cantalloube, F., Gomez~Gonzalez, C.~A., Absil, O., \& Jacques, L. 2019, MNRAS, 487, 2262, \dodoi{10.1093/mnras/stz1350}

\bibitem[{Paszke {et~al.}(2019)Paszke, Gross, Massa, Lerer, Bradbury, Chanan, Killeen, Lin, Gimelshein, Antiga, Desmaison, Kopf, Yang, DeVito, Raison, Tejani, Chilamkurthy, Steiner, Fang, Bai, \& Chintala}]{paszkePyTorchImperativeStyle2019}
Paszke, A., Gross, S., Massa, F., {et~al.} 2019, in NeurIPS, Vol.~32, \dodoi{10.48550/arXiv.1912.01703}

\bibitem[{Pedregosa {et~al.}(2011)Pedregosa, Varoquaux, Gramfort, Michel, Thirion, Grisel, Blondel, Prettenhofer, Weiss, Dubourg, Vanderplas, Passos, Cournapeau, Brucher, Perrot, \& Duchesnay}]{sklearn}
Pedregosa, F., Varoquaux, G., Gramfort, A., {et~al.} 2011, Journal of Machine Learning Research, 12, 2825.
\newblock \url{https://arxiv.org/abs/1201.0490}

\bibitem[{Perrin {et~al.}(2003)Perrin, Sivaramakrishnan, Makidon, Oppenheimer, \& Graham}]{perrinStructureHighStrehl2003}
Perrin, M.~D., Sivaramakrishnan, A., Makidon, R.~B., Oppenheimer, B.~R., \& Graham, J.~R. 2003, ApJ, 596, 702, \dodoi{10.1086/377689}

\bibitem[{Prusti {et~al.}(2016)Prusti, de~Bruijne, Brown, Vallenari, Babusiaux, {Bailer-Jones}, Bastian, Biermann, Evans, Eyer, Jansen, Jordi, Klioner, Lammers, Lindegren, Luri, Mignard, Milligan, Panem, Poinsignon, Pourbaix, Randich, Sarri, Sartoretti, Siddiqui, Soubiran, Valette, van Leeuwen, Walton, Aerts, Arenou, Cropper, Drimmel, H{\o}g, Katz, Lattanzi, O'Mullane, Grebel, Holland, Huc, Passot, Bramante, Cacciari, Casta{\~n}eda, Chaoul, Cheek, Angeli, Fabricius, Guerra, Hern{\'a}ndez, {Jean-Antoine-Piccolo}, Masana, Messineo, Mowlavi, Nienartowicz, {Ord{\'o}{\~n}ez-Blanco}, Panuzzo, Portell, Richards, Riello, Seabroke, Tanga, Th{\'e}venin, Torra, Els, {Gracia-Abril}, Comoretto, {Garcia-Reinaldos}, Lock, Mercier, Altmann, Andrae, Astraatmadja, {Bellas-Velidis}, Benson, Berthier, Blomme, Busso, Carry, Cellino, Clementini, Cowell, Creevey, Cuypers, Davidson, Ridder, de~Torres, Delchambre, Dell'Oro, Ducourant, Fr{\'e}mat, {Garc{\'i}a-Torres}, Gosset, Halbwachs, Hambly, Harrison, Hauser, Hestroffer, Hodgkin,
  Huckle, Hutton, Jasniewicz, Jordan, Kontizas, Korn, Lanzafame, Manteiga, Moitinho, Muinonen, Osinde, Pancino, Pauwels, Petit, {Recio-Blanco}, Robin, Sarro, Siopis, Smith, Smith, Sozzetti, Thuillot, van Reeven, Viala, Abbas, Aramburu, Accart, Aguado, Allan, Allasia, Altavilla, {\'A}lvarez, Alves, Anderson, Andrei, Varela, Antiche, Antoja, Ant{\'o}n, Arcay, Atzei, Ayache, Bach, Baker, {Balaguer-N{\'u}{\~n}ez}, Barache, Barata, Barbier, Barblan, Baroni, y~Navascu{\'e}s, Barros, Barstow, Becciani, Bellazzini, Bellei, Garc{\'i}a, Belokurov, Bendjoya, Berihuete, Bianchi, Bienaym{\'e}, Billebaud, Blagorodnova, {Blanco-Cuaresma}, Boch, Bombrun, Borrachero, Bouquillon, Bourda, Bouy, Bragaglia, Breddels, Brouillet, Br{\"u}semeister, Bucciarelli, Budnik, Burgess, Burgon, Burlacu, Busonero, Buzzi, Caffau, Cambras, Campbell, Cancelliere, {Cantat-Gaudin}, Carlucci, Carrasco, Castellani, Charlot, Charnas, Charvet, Chassat, Chiavassa, Clotet, Cocozza, Collins, Collins, Costigan, Crifo, Cross, Crosta, Crowley, Dafonte,
  Damerdji, Dapergolas, David, David, Cat, de~Felice, de~Laverny, Luise, March, de~Martino, de~Souza, Debosscher, del Pozo, Delbo, Delgado, Delgado, di~Marco, Matteo, Diakite, Distefano, Dolding, Anjos, Drazinos, Dur{\'a}n, Dzigan, Ecale, Edvardsson, Enke, Erdmann, Escolar, Espina, Evans, Bontemps, Fabre, Fabrizio, Faigler, Falc{\~a}o, Casas, Faye, Federici, Fedorets, {Fern{\'a}ndez-Hern{\'a}ndez}, Fernique, Fienga, Figueras, Filippi, Findeisen, Fonti, Fouesneau, Fraile, Fraser, Fuchs, Furnell, Gai, Galleti, Galluccio, Garabato, {Garc{\'i}a-Sedano}, Gar{\'e}, Garofalo, Garralda, Gavras, Gerssen, Geyer, Gilmore, Girona, Giuffrida, Gomes, {Gonz{\'a}lez-Marcos}, {Gonz{\'a}lez-N{\'u}{\~n}ez}, {Gonz{\'a}lez-Vidal}, Granvik, Guerrier, Guillout, Guiraud, G{\'u}rpide, {Guti{\'e}rrez-S{\'a}nchez}, Guy, Haigron, Hatzidimitriou, Haywood, Heiter, Helmi, Hobbs, Hofmann, Holl, Holland, Hunt, Hypki, Icardi, Irwin, de~Fombelle, Jofr{\'e}, Jonker, Jorissen, Julbe, Karampelas, Kochoska, Kohley, Kolenberg, Kontizas, Koposov,
  Kordopatis, Koubsky, Kowalczyk, {Krone-Martins}, Kudryashova, Kull, Bachchan, {Lacoste-Seris}, Lanza, Lavigne, {Poncin-Lafitte}, Lebreton, Lebzelter, Leccia, Leclerc, {Lecoeur-Taibi}, Lemaitre, Lenhardt, Leroux, Liao, Licata, Lindstr{\o}m, Lister, Livanou, Lobel, L{\"o}ffler, L{\'o}pez, {Lopez-Lozano}, Lorenz, Loureiro, MacDonald, Fernandes, Managau, Mann, Mantelet, Marchal, Marchant, Marconi, Marie, Marinoni, Marrese, Marschalk{\'o}, Marshall, {Mart{\'i}n-Fleitas}, Martino, Mary, Matijevi{\v c}, Mazeh, McMillan, Messina, Mestre, Michalik, Millar, Miranda, Molina, Molinaro, Molinaro, Moln{\'a}r, Moniez, Montegriffo, Monteiro, Mor, Mora, Morbidelli, Morel, Morgenthaler, Morley, Morris, Mulone, Muraveva, Musella, Narbonne, Nelemans, Nicastro, Noval, Ord{\'e}novic, {Ordieres-Mer{\'e}}, Osborne, Pagani, Pagano, Pailler, Palacin, Palaversa, Parsons, Paulsen, Pecoraro, Pedrosa, Pentik{\"a}inen, Pereira, Pichon, Piersimoni, Pineau, Plachy, Plum, Poujoulet, Pr{\v s}a, Pulone, Ragaini, Rago, Rambaux, {Ramos-Lerate},
  Ranalli, Rauw, Read, Regibo, Renk, Reyl{\'e}, Ribeiro, Rimoldini, Ripepi, Riva, Rixon, Roelens, {Romero-G{\'o}mez}, Rowell, Royer, Rudolph, {Ruiz-Dern}, Sadowski, Sell{\'e}s, Sahlmann, Salgado, Salguero, Sarasso, Savietto, Schnorhk, Schultheis, Sciacca, Segol, Segovia, Segransan, Serpell, Shih, Smareglia, Smart, Smith, Solano, Solitro, Sordo, Nieto, Souchay, Spagna, Spoto, Stampa, Steele, Steidelm{\"u}ller, Stephenson, Stoev, Suess, S{\"u}veges, Surdej, Szabados, {Szegedi-Elek}, Tapiador, Taris, Tauran, Taylor, Teixeira, Terrett, Tingley, Trager, Turon, Ulla, Utrilla, Valentini, van Elteren, Hemelryck, van Leeuwen, Varadi, Vecchiato, Veljanoski, Via, Vicente, Vogt, Voss, Votruba, Voutsinas, Walmsley, Weiler, Weingrill, Werner, Wevers, Whitehead, Wyrzykowski, Yoldas, {\v Z}erjal, Zucker, Zurbach, Zwitter, Alecu, Allen, Prieto, Amorim, {Anglada-Escud{\'e}}, Arsenijevic, Azaz, Balm, Beck, Bernstein, Bigot, Bijaoui, Blasco, Bonfigli, Bono, Boudreault, Bressan, Brown, Brunet, Bunclark, Buonanno, Butkevich,
  Carret, Carrion, Chemin, Ch{\'e}reau, Corcione, Darmigny, de~Boer, de~Teodoro, de~Zeeuw, Luche, Domingues, Dubath, Fodor, Fr{\'e}zouls, Fries, Fustes, Fyfe, Gallardo, Gallegos, Gardiol, Gebran, Gomboc, G{\'o}mez, Grux, Gueguen, Heyrovsky, Hoar, Iannicola, Parache, Janotto, Joliet, Jonckheere, Keil, Kim, Klagyivik, Klar, Knude, Kochukhov, Kolka, Kos, Kutka, Lainey, LeBouquin, Liu, Loreggia, Makarov, Marseille, Martayan, {Martinez-Rubi}, Massart, Meynadier, Mignot, Munari, Nguyen, Nordlander, Ocvirk, O'Flaherty, Sanz, Ortiz, Osorio, Oszkiewicz, Ouzounis, Palmer, Park, Pasquato, Peltzer, Peralta, P{\'e}turaud, Pieniluoma, Pigozzi, Poels, Prat, Prod'homme, Raison, Rebordao, Risquez, {Rocca-Volmerange}, Rosen, {Ruiz-Fuertes}, Russo, Sembay, Vizcaino, Short, Siebert, Silva, Sinachopoulos, Slezak, Soffel, Sosnowska, Strai{\v z}ys, ter Linden, Terrell, Theil, Tiede, Troisi, Tsalmantza, Tur, Vaccari, Vachier, Valles, Hamme, Veltz, Virtanen, Wallut, Wichmann, Wilkinson, Ziaeepour, \& Zschocke}]{prustiGaiaMission2016}
Prusti, T., de~Bruijne, J. H.~J., Brown, A. G.~A., {et~al.} 2016, A\&A, 595, A1, \dodoi{10.1051/0004-6361/201629272}

\bibitem[{{Pueyo}(2016)}]{pueyoDETECTIONCHARACTERIZATIONEXOPLANETS2016}
{Pueyo}, L. 2016, \apj, 824, 117, \dodoi{10.3847/0004-637X/824/2/117}

\bibitem[{Quanz {et~al.}(2011)Quanz, Schmid, Geissler, Meyer, Henning, Brandner, \& Wolf}]{quanzVERYLARGEESCOPE2011}
Quanz, S.~P., Schmid, H.~M., Geissler, K., {et~al.} 2011, ApJ, 738, 23, \dodoi{10.1088/0004-637X/738/1/23}

\bibitem[{Racine {et~al.}(1999)Racine, Walker, Nadeau, Doyon, \& Marois}]{racineSpeckleNoiseDetection1999}
Racine, R., Walker, G. A.~H., Nadeau, D., Doyon, R., \& Marois, C. 1999, PASP, 111, 587, \dodoi{10.1086/316367}

\bibitem[{Rameau {et~al.}(2013)Rameau, Chauvin, Lagrange, Klahr, Bonnefoy, Mordasini, Bonavita, Desidera, Dumas, \& Girard}]{rameauSurveyYoungNearby2013}
Rameau, J., Chauvin, G., Lagrange, A.-M., {et~al.} 2013, A\&A, 553, A60, \dodoi{10.1051/0004-6361/201220984}

\bibitem[{Ren {et~al.}(2018)Ren, Pueyo, Zhu, Debes, \& Duch{\^e}ne}]{renNonnegativeMatrixFactorization2018}
Ren, B., Pueyo, L., Zhu, G.~B., Debes, J., \& Duch{\^e}ne, G. 2018, ApJ, 852, 104, \dodoi{10.3847/1538-4357/aaa1f2}

\bibitem[{Ribak \& Gladysz(2008)}]{ribakFainterCloserFinding2008}
Ribak, E.~N., \& Gladysz, S. 2008, Optics Express, 16, 15553, \dodoi{10.1364/OE.16.015553}

\bibitem[{Ribeiro {et~al.}(2016)Ribeiro, Singh, \& Guestrin}]{ribeiroWhyShouldTrust2016}
Ribeiro, M.~T., Singh, S., \& Guestrin, C. 2016, in Proceedings of the 22nd {{ACM SIGKDD International Conference}} on {{Knowledge Discovery}} and {{Data Mining}}, 1135--1144, \dodoi{10.1145/2939672.2939778}

\bibitem[{Ross {et~al.}(2017)Ross, Hughes, \& {Doshi-Velez}}]{rossRightRightReasons2017}
Ross, A.~S., Hughes, M.~C., \& {Doshi-Velez}, F. 2017, in IJCAI.
\newblock \doarXiv{1703.03717}

\bibitem[{Ruane {et~al.}(2019)Ruane, Ngo, Mawet, Absil, Choquet, Cook, Gonzalez, Huby, Matthews, Meshkat, Reggiani, Serabyn, Wallack, \& Xuan}]{ruaneReferenceStarDifferential2019}
Ruane, G., Ngo, H., Mawet, D., {et~al.} 2019, AJ, 157, 118, \dodoi{10.3847/1538-3881/aafee2}

\bibitem[{Ruffio {et~al.}(2017)Ruffio, Macintosh, Wang, Pueyo, Nielsen, De~Rosa, Czekala, Marley, Arriaga, Bailey, Barman, Bulger, Chilcote, Cotten, Doyon, Duch{\^e}ne, Fitzgerald, Follette, Gerard, Goodsell, Graham, Greenbaum, Hibon, Hung, Ingraham, Kalas, Konopacky, Larkin, Maire, Marchis, Marois, Metchev, {Millar-Blanchaer}, Morzinski, Oppenheimer, Palmer, Patience, Perrin, Poyneer, Rajan, Rameau, Rantakyr{\"o}, Savransky, Schneider, Sivaramakrishnan, Song, Soummer, Thomas, Wallace, {Ward-Duong}, Wiktorowicz, \& Wolff}]{ruffioImprovingAssessingPlanet2017}
Ruffio, J.-B., Macintosh, B., Wang, J.~J., {et~al.} 2017, ApJ, 842, 14, \dodoi{10.3847/1538-4357/aa72dd}

\bibitem[{Samland {et~al.}(2021)Samland, Bouwman, Hogg, Brandner, Henning, \& Janson}]{samlandTRAPTemporalSystematics2021}
Samland, M., Bouwman, J., Hogg, D.~W., {et~al.} 2021, A\&A, 646, A24, \dodoi{10.1051/0004-6361/201937308}

\bibitem[{Schramowski {et~al.}(2020)Schramowski, Stammer, Teso, Brugger, Herbert, Shao, Luigs, Mahlein, \& Kersting}]{schramowskiMakingDeepNeural2020}
Schramowski, P., Stammer, W., Teso, S., {et~al.} 2020, Nature Machine Intelligence, 2, 476, \dodoi{10.1038/s42256-020-0212-3}

\bibitem[{Simonyan {et~al.}(2013)Simonyan, Vedaldi, \& Zisserman}]{simonyanDeepConvolutionalNetworks2013}
Simonyan, K., Vedaldi, A., \& Zisserman, A. 2013, arXiv preprint.
\newblock \doarXiv{1312.6034}

\bibitem[{Smilkov {et~al.}(2017)Smilkov, Thorat, Kim, Vi{\'e}gas, \& Wattenberg}]{smilkovSmoothGradRemovingNoise2017}
Smilkov, D., Thorat, N., Kim, B., Vi{\'e}gas, F., \& Wattenberg, M. 2017, arXiv e-prints.
\newblock \doarXiv{1706.03825}

\bibitem[{Snik {et~al.}(2012)Snik, Otten, Kenworthy, Miskiewicz, Escuti, Packham, \& Codona}]{snikVectorAPPBroadbandApodizing2012}
Snik, F., Otten, G., Kenworthy, M., {et~al.} 2012, in Modern {{Technologies}} in {{Space-}} and {{Ground-based Telescopes}} and {{Instrumentation II}}, Vol. 8450 (SPIE), 224--234, \dodoi{10.1117/12.926222}

\bibitem[{Soummer {et~al.}(2011)Soummer, Hagan, Pueyo, Thormann, Rajan, \& Marois}]{soummerORBITALMOTIONHR2011a}
Soummer, R., Hagan, J.~B., Pueyo, L., {et~al.} 2011, ApJ, 741, 55, \dodoi{10.1088/0004-637X/741/1/55}

\bibitem[{Soummer {et~al.}(2012)Soummer, Pueyo, \& Larkin}]{soummerDETECTIONCHARACTERIZATIONEXOPLANETS2012}
Soummer, R., Pueyo, L., \& Larkin, J. 2012, ApJ, 755, L28, \dodoi{10.1088/2041-8205/755/2/L28}

\bibitem[{Sparks \& Ford(2002)}]{sparksImagingSpectroscopyExtrasolar2002}
Sparks, W.~B., \& Ford, H.~C. 2002, ApJ, 578, 543, \dodoi{10.1086/342401}

\bibitem[{Stolker {et~al.}(2019)Stolker, Bonse, Quanz, Amara, Cugno, Bohn, \& Boehle}]{stolkerPynPointModularPipeline2019}
Stolker, T., Bonse, M.~J., Quanz, S.~P., {et~al.} 2019, A\&A, 621, A59, \dodoi{10.1051/0004-6361/201834136}

\bibitem[{Sundararajan {et~al.}(2017)Sundararajan, Taly, \& Yan}]{sundararajanAxiomaticAttributionDeep2017}
Sundararajan, M., Taly, A., \& Yan, Q. 2017, in ICML, 3319--3328.
\newblock \url{https://proceedings.mlr.press/v70/sundararajan17a.html}

\bibitem[{Sutskever {et~al.}(2013)Sutskever, Martens, Dahl, \& Hinton}]{sutskeverImportanceInitializationMomentum2013}
Sutskever, I., Martens, J., Dahl, G., \& Hinton, G. 2013, in ICML, 1139--1147.
\newblock \url{https://proceedings.mlr.press/v28/sutskever13.html}

\bibitem[{{The Astropy Collaboration} {et~al.}(2013){The Astropy Collaboration}, {Robitaille}, {Tollerud}, {Greenfield}, {Droettboom}, {Bray}, {Aldcroft}, {Davis}, {Ginsburg}, {Price-Whelan}, {Kerzendorf}, {Conley}, {Crighton}, {Barbary}, {Muna}, {Ferguson}, {Grollier}, {Parikh}, {Nair}, {Günther}, {Deil}, {Woillez}, {Conseil}, {Kramer}, {Turner}, {Singer}, {Fox}, {Weaver}, {Zabalza}, {Edwards}, {Azalee Bostroem}, {Burke}, {Casey}, {Crawford}, {Dencheva}, {Ely}, {Jenness}, {Labrie}, {Lim}, {Pierfederici}, {Pontzen}, {Ptak}, {Refsdal}, {Servillat}, \& {Streicher}}]{astropy_2013}
{The Astropy Collaboration}, {Robitaille}, T.~P., {Tollerud}, E.~J., {et~al.} 2013, \aap, 558, A33, \dodoi{10.1051/0004-6361/201322068}

\bibitem[{{The Astropy Collaboration} {et~al.}(2018){The Astropy Collaboration}, {Price-Whelan}, {Sip\H{o}cz}, {Günther}, {Lim}, {Crawford}, {Conseil}, {Shupe}, {Craig}, {Dencheva}, {Ginsburg}, {VanderPlas}, {Bradley}, {Pérez-Suárez}, {de Val-Borro}, {Aldcroft}, {Cruz}, {Robitaille}, {Tollerud}, {Ardelean}, {Babej}, {Bach}, {Bachetti}, {Bakanov}, {Bamford}, {Barentsen}, {Barmby}, {Baumbach}, {Berry}, {Biscani}, {Boquien}, {Bostroem}, {Bouma}, {Brammer}, {Bray}, {Breytenbach}, {Buddelmeijer}, {Burke}, {Calderone}, {Rodríguez}, {Cara}, {Cardoso}, {Cheedella}, {Copin}, {Corrales}, {Crichton}, {D’Avella}, {Deil}, {Depagne}, {Dietrich}, {Donath}, {Droettboom}, {Earl}, {Erben}, {Fabbro}, {Ferreira}, {Finethy}, {Fox}, {Garrison}, {Gibbons}, {Goldstein}, {Gommers}, {Greco}, {Greenfield}, {Groener}, {Grollier}, {Hagen}, {Hirst}, {Homeier}, {Horton}, {Hosseinzadeh}, {Hu}, {Hunkeler}, {Ivezić}, {Jain}, {Jenness}, {Kanarek}, {Kendrew}, {Kern}, {Kerzendorf}, {Khvalko}, {King}, {Kirkby}, {Kulkarni}, {Kumar}, {Lee},
  {Lenz}, {Littlefair}, {Ma}, {Macleod}, {Mastropietro}, {McCully}, {Montagnac}, {Morris}, {Mueller}, {Mumford}, {Muna}, {Murphy}, {Nelson}, {Nguyen}, {Ninan}, {Nöthe}, {Ogaz}, {Oh}, {Parejko}, {Parley}, {Pascual}, {Patil}, {Patil}, {Plunkett}, {Prochaska}, {Rastogi}, {Janga}, {Sabater}, {Sakurikar}, {Seifert}, {Sherbert}, {Sherwood-Taylor}, {Shih}, {Sick}, {Silbiger}, {Singanamalla}, {Singer}, {Sladen}, {Sooley}, {Sornarajah}, {Streicher}, {Teuben}, {Thomas}, {Tremblay}, {Turner}, {Terrón}, {Kerkwijk}, {de la Vega}, {Watkins}, {Weaver}, {Whitmore}, {Woillez}, \& {Zabalza}}]{astropy_2018}
{The Astropy Collaboration}, {Price-Whelan}, A.~M., {Sip\H{o}cz}, B.~M., {et~al.} 2018, \aj, 156, 123, \dodoi{10.3847/1538-3881/aabc4f}

\bibitem[{{The Astropy Collaboration} {et~al.}(2022){The Astropy Collaboration}, {Price-Whelan}, {Lim}, {Earl}, {Starkman}, {Bradley}, {Shupe}, {Patil}, {Corrales}, {Brasseur}, {Nöthe}, {Donath}, {Tollerud}, {Morris}, {Ginsburg}, {Vaher}, {Weaver}, {Tocknell}, {Jamieson}, {van Kerkwijk}, {Robitaille}, {Merry}, {Bachetti}, {Günther}, {Aldcroft}, {Alvarado-Montes}, {Archibald}, {Bódi}, {Bapat}, {Barentsen}, {Bazán}, {Biswas}, {Boquien}, {Burke}, {Cara}, {Cara}, {Conroy}, {Conseil}, {Craig}, {Cross}, {Cruz}, {D’Eugenio}, {Dencheva}, {Devillepoix}, {Dietrich}, {Eigenbrot}, {Erben}, {Ferreira}, {Foreman-Mackey}, {Fox}, {Freij}, {Garg}, {Geda}, {Glattly}, {Gondhalekar}, {Gordon}, {Grant}, {Greenfield}, {Groener}, {Guest}, {Gurovich}, {Handberg}, {Hart}, {Hatfield-Dodds}, {Homeier}, {Hosseinzadeh}, {Jenness}, {Jones}, {Joseph}, {Kalmbach}, {Karamehmetoglu}, {Kałuszyński}, {Kelley}, {Kern}, {Kerzendorf}, {Koch}, {Kulumani}, {Lee}, {Ly}, {Ma}, {MacBride}, {Maljaars}, {Muna}, {Murphy}, {Norman}, {O’Steen},
  {Oman}, {Pacifici}, {Pascual}, {Pascual-Granado}, {Patil}, {Perren}, {Pickering}, {Rastogi}, {Roulston}, {Ryan}, {Rykoff}, {Sabater}, {Sakurikar}, {Salgado}, {Sanghi}, {Saunders}, {Savchenko}, {Schwardt}, {Seifert-Eckert}, {Shih}, {Jain}, {Shukla}, {Sick}, {Simpson}, {Singanamalla}, {Singer}, {Singhal}, {Sinha}, {Sip\H{o}cz}, {Spitler}, {Stansby}, {Streicher}, {Šumak}, {Swinbank}, {Taranu}, {Tewary}, {Tremblay}, {Val-Borro}, {Van Kooten}, {Vasović}, {Verma}, {de Miranda Cardoso}, {Williams}, {Wilson}, {Winkel}, {Wood-Vasey}, {Xue}, {Yoachim}, {Zhang}, \& {Zonca}}]{astropy_2022}
{The Astropy Collaboration}, {Price-Whelan}, A.~M., {Lim}, P.~L., {et~al.} 2022, \apj, 935, 167, \dodoi{10.3847/1538-4357/ac7c74}

\bibitem[{{Thompson} \& {Marois}(2021)}]{thompsonImprovedContrastImages2021}
{Thompson}, W., \& {Marois}, C. 2021, \aj, 161, 236, \dodoi{10.3847/1538-3881/abee7d}

\bibitem[{Tobin {et~al.}(2024)Tobin, Currie, Li, Chilcote, Brandt, Lacy, Kuzuhara, Vincent, El~Morsy, Deo, Williams, Guyon, Lozi, Vievard, Skaf, Ahn, Groff, Kasdin, Uyama, Tamura, Gibbs, Lewis, {Bowens-Rubin}, Salama, An, \& Chen}]{tobinDirectimagingDiscoverySubstellar2024}
Tobin, T.~L., Currie, T., Li, Y., {et~al.} 2024, AJ, 167, 205, \dodoi{10.3847/1538-3881/ad3077}

\bibitem[{Vallenari {et~al.}(2023)Vallenari, Brown, Prusti, de~Bruijne, Arenou, Babusiaux, Biermann, Creevey, Ducourant, Evans, Eyer, Guerra, Hutton, Jordi, Klioner, Lammers, Lindegren, Luri, Mignard, Panem, Pourbaix, Randich, Sartoretti, Soubiran, Tanga, Walton, {Bailer-Jones}, Bastian, Drimmel, Jansen, Katz, Lattanzi, van Leeuwen, Bakker, Cacciari, Casta{\~n}eda, Angeli, Fabricius, Fouesneau, Fr{\'e}mat, Galluccio, Guerrier, Heiter, Masana, Messineo, Mowlavi, Nicolas, Nienartowicz, Pailler, Panuzzo, Riclet, Roux, Seabroke, Sordo, Th{\'e}venin, {Gracia-Abril}, Portell, Teyssier, Altmann, Andrae, Audard, {Bellas-Velidis}, Benson, Berthier, Blomme, Burgess, Busonero, Busso, C{\'a}novas, Carry, Cellino, Cheek, Clementini, Damerdji, Davidson, de~Teodoro, Campos, Delchambre, Dell'Oro, Esquej, {Fern{\'a}ndez-Hern{\'a}ndez}, Fraile, Garabato, {Garc{\'i}a-Lario}, Gosset, Haigron, Halbwachs, Hambly, Harrison, Hern{\'a}ndez, Hestroffer, Hodgkin, Holl, Jan{\ss}en, de~Fombelle, Jordan, {Krone-Martins}, Lanzafame,
  L{\"o}ffler, Marchal, Marrese, Moitinho, Muinonen, Osborne, Pancino, Pauwels, {Recio-Blanco}, Reyl{\'e}, Riello, Rimoldini, Roegiers, Rybizki, Sarro, Siopis, Smith, Sozzetti, Utrilla, van Leeuwen, Abbas, {\'A}brah{\'a}m, Aramburu, Aerts, Aguado, Ajaj, {Aldea-Montero}, Altavilla, {\'A}lvarez, Alves, Anders, Anderson, Varela, Antoja, Baines, Baker, {Balaguer-N{\'u}{\~n}ez}, Balbinot, Balog, Barache, Barbato, Barros, Barstow, Bartolom{\'e}, Bassilana, Bauchet, Becciani, Bellazzini, Berihuete, Bernet, Bertone, Bianchi, Binnenfeld, {Blanco-Cuaresma}, Blazere, Boch, Bombrun, Bossini, Bouquillon, Bragaglia, Bramante, Breedt, Bressan, Brouillet, Brugaletta, Bucciarelli, Burlacu, Butkevich, Buzzi, Caffau, Cancelliere, {Cantat-Gaudin}, Carballo, Carlucci, Carnerero, Carrasco, Casamiquela, Castellani, {Castro-Ginard}, Chaoul, Charlot, Chemin, Chiaramida, Chiavassa, Chornay, Comoretto, Contursi, Cooper, Cornez, Cowell, Crifo, Cropper, Crosta, Crowley, Dafonte, Dapergolas, David, David, de~Laverny, Luise, March, Ridder,
  de~Souza, de~Torres, del Peloso, del Pozo, Delbo, Delgado, Delisle, Demouchy, Dharmawardena, Matteo, Diakite, Diener, Distefano, Dolding, Edvardsson, Enke, Fabre, Fabrizio, Faigler, Fedorets, Fernique, Fienga, Figueras, Fournier, Fouron, Fragkoudi, Gai, {Garcia-Gutierrez}, {Garcia-Reinaldos}, {Garc{\'i}a-Torres}, Garofalo, Gavel, Gavras, Gerlach, Geyer, Giacobbe, Gilmore, Girona, Giuffrida, Gomel, Gomez, {Gonz{\'a}lez-N{\'u}{\~n}ez}, {Gonz{\'a}lez-Santamar{\'i}a}, {Gonz{\'a}lez-Vidal}, Granvik, Guillout, Guiraud, {Guti{\'e}rrez-S{\'a}nchez}, Guy, Hatzidimitriou, Hauser, Haywood, Helmer, Helmi, Sarmiento, Hidalgo, Hilger, H{\l}adczuk, Hobbs, Holland, Huckle, Jardine, Jasniewicz, Piccolo, {Jim{\'e}nez-Arranz}, Jorissen, Campillo, Julbe, Karbevska, Kervella, Khanna, Kontizas, Kordopatis, Korn, K{\'o}sp{\'a}l, {Kostrzewa-Rutkowska}, Kruszy{\'n}ska, Kun, Laizeau, Lambert, Lanza, Lasne, Campion, Lebreton, Lebzelter, Leccia, Leclerc, {Lecoeur-Taibi}, Liao, Licata, Lindstr{\o}m, Lister, Livanou, Lobel, Lorca, Loup,
  Pardo, Romeo, Managau, Mann, Manteiga, Marchant, Marconi, Marcos, Santos, Pina, Marinoni, Marocco, Marshall, Polo, {Mart{\'i}n-Fleitas}, Marton, Mary, Masip, Massari, {Mastrobuono-Battisti}, Mazeh, McMillan, Messina, Michalik, Millar, Mints, Molina, Molinaro, Moln{\'a}r, Monari, Mongui{\'o}, Montegriffo, Montero, Mor, Mora, Morbidelli, Morel, Morris, Muraveva, Murphy, Musella, Nagy, Noval, Oca{\~n}a, Ogden, Ordenovic, Osinde, Pagani, Pagano, Palaversa, Palicio, {Pallas-Quintela}, Panahi, {Payne-Wardenaar}, Esteller, Penttil{\"a}, Pichon, Piersimoni, Pineau, Plachy, Plum, Poggio, Pr{\v s}a, Pulone, Racero, Ragaini, Rainer, Raiteri, Rambaux, Ramos, {Ramos-Lerate}, Fiorentin, Regibo, Richards, Diaz, Ripepi, Riva, Rix, Rixon, Robichon, Robin, Robin, Roelens, Rogues, Rohrbasser, {Romero-G{\'o}mez}, Rowell, Royer, Mieres, Rybicki, Sadowski, N{\'u}{\~n}ez, Sell{\'e}s, Sahlmann, Salguero, Samaras, Gimenez, Sanna, Santove{\~n}a, Sarasso, Schultheis, Sciacca, Segol, Segovia, S{\'e}gransan, Semeux, Shahaf, Siddiqui,
  Siebert, Siltala, Silvelo, Slezak, Slezak, Smart, Snaith, Solano, Solitro, Souami, Souchay, Spagna, Spina, Spoto, Steele, Steidelm{\"u}ller, Stephenson, S{\"u}veges, Surdej, Szabados, {Szegedi-Elek}, Taris, Taylor, Teixeira, Tolomei, Tonello, Torra, Torra, Elipe, Trabucchi, Tsounis, Turon, Ulla, Unger, Vaillant, van Dillen, van Reeven, Vanel, Vecchiato, Viala, Vicente, Voutsinas, Weiler, Wevers, Wyrzykowski, Yoldas, Yvard, Zhao, Zorec, Zucker, \& Zwitter}]{vallenariGaiaDataRelease2023}
Vallenari, A., Brown, A. G.~A., Prusti, T., {et~al.} 2023, A\&A, 674, A1, \dodoi{10.1051/0004-6361/202243940}

\bibitem[{{van der Walt} {et~al.}(2014){van der Walt}, {Schönberger}, {Nunez-Iglesias}, {Boulogne}, {Warner}, {Yager}, {Gouillart}, \& {Yu}}]{skimage}
{van der Walt}, S., {Schönberger}, J.~L., {Nunez-Iglesias}, J., {et~al.} 2014, PeerJ, 2, e453, \dodoi{10.7717/peerj.453}

\bibitem[{van Leeuwen(2007)}]{leeuwenValidationNewHipparcos2007}
van Leeuwen, F. 2007, A\&A, 474, 653, \dodoi{10.1051/0004-6361:20078357}

\bibitem[{Vigan {et~al.}(2021)Vigan, Fontanive, Meyer, Biller, Bonavita, Feldt, Desidera, Marleau, Emsenhuber, Galicher, Rice, Forgan, Mordasini, Gratton, Coroller, Maire, Cantalloube, Chauvin, Cheetham, Hagelberg, Lagrange, Langlois, Bonnefoy, Beuzit, Boccaletti, D'Orazi, Delorme, Dominik, Henning, Janson, Lagadec, Lazzoni, Ligi, Menard, Mesa, Messina, Moutou, M{\"u}ller, Perrot, Samland, Schmid, Schmidt, Sissa, Turatto, Udry, Zurlo, Abe, Antichi, {Asensio-Torres}, Baruffolo, Baudoz, Baudrand, Bazzon, Blanchard, Bohn, Sevilla, Carbillet, Carle, Cascone, Charton, Claudi, Costille, Caprio, Delboulb{\'e}, Dohlen, Engler, Fantinel, Feautrier, Fusco, Gigan, Girard, Giro, Gisler, Gluck, Gry, Hubin, Hugot, Jaquet, Kasper, Mignant, Llored, Madec, Magnard, Martinez, Maurel, {M{\"o}ller-Nilsson}, Mouillet, Moulin, Orign{\'e}, Pavlov, Perret, Petit, Pragt, Puget, Rabou, Ramos, Rickman, Rigal, Rochat, Roelfsema, Rousset, Roux, Salasnich, Sauvage, Sevin, Soenke, Stadler, Suarez, Wahhaj, Weber, \&
  Wildi}]{viganSPHEREInfraredSurvey2021}
Vigan, A., Fontanive, C., Meyer, M., {et~al.} 2021, A\&A, 651, A72, \dodoi{10.1051/0004-6361/202038107}

\bibitem[{{Virtanen} {et~al.}(2020){Virtanen}, {Gommers}, {Oliphant}, {Haberland}, {Reddy}, {et~al.}}]{scipy}
{Virtanen}, P., {Gommers}, R., {Oliphant}, T.~E., {et~al.} 2020, Nature Methods, 17, 261, \dodoi{10.1038/s41592-019-0686-2}

\bibitem[{Wahhaj {et~al.}(2015)Wahhaj, Cieza, Mawet, Yang, Canovas, de~Boer, Casassus, M{\'e}nard, Schreiber, Liu, Biller, Nielsen, \& Hayward}]{wahhajImprovingSignaltonoiseDirect2015}
Wahhaj, Z., Cieza, L.~A., Mawet, D., {et~al.} 2015, A\&A, 581, A24, \dodoi{10.1051/0004-6361/201525837}

\bibitem[{Wang {et~al.}(2015)Wang, Ruffio, De~Rosa, Aguilar, Wolff, \& Pueyo}]{wangPyKLIPPSFSubtraction2015}
Wang, J.~J., Ruffio, J.-B., De~Rosa, R.~J., {et~al.} 2015, Astrophysics Source Code Library, ascl:1506.001

\bibitem[{{Waskom}(2021)}]{seaborn}
{Waskom}, M. 2021, JOSS, 6, 3021, \dodoi{10.21105/joss.03021}

\bibitem[{{Wolf} {et~al.}(2024){Wolf}, {Jones}, \& {Bowler}}]{wolfDirectExoplanetDetection2023}
{Wolf}, T.~N., {Jones}, B.~A., \& {Bowler}, B.~P. 2024, \aj, 167, 92, \dodoi{10.3847/1538-3881/ad11eb}

\bibitem[{{Z{\'u}{\~n}iga-Fern{\'a}ndez} {et~al.}(2021){Z{\'u}{\~n}iga-Fern{\'a}ndez}, Bayo, Elliott, Zamora, Corval{\'a}n, Haubois, {Corral-Santana}, Olofsson, Hu{\'e}lamo, Sterzik, Torres, Quast, \& Melo}]{zuniga-fernandezSearchAssociationsContaining2021}
{Z{\'u}{\~n}iga-Fern{\'a}ndez}, S., Bayo, A., Elliott, P., {et~al.} 2021, A\&A, 645, A30, \dodoi{10.1051/0004-6361/202037830}

\end{thebibliography}
    \bibliographystyle{aasjournal}
    
    \appendix

\section{Linear PSF Subtraction Compared}
\label{sec:masking_strategies}

The right reason mask of \fours introduced in \Cref{sec:right_reason_mask} has similarities with the masks proposed by other post-processing algorithms.
Already the PSF subtraction based on the LOCI algorithm used PSF-sized masks to protect the signal of the planet \citep{lafreniereNewAlgorithmPointSpread2007,maroisExoplanetImagingLOCI2010,soummerORBITALMOTIONHR2011a}. 
Within this framework, two types of masks can be distinguished: (1) \emph{temporal masks}, for example based on field rotation; and (2) \emph{spatial PSF-sized masks}, applied to the data.
Both masks share the same idea: if the planetary signal is excluded from the fit, only the speckle noise can be captured. The temporal exclusion ensures that the planet is not part of the reference frames, while the spatial mask ensures that it is excluded from the frame we want to fit.
The same masks are commonly used with PCA \citep{soummerDETECTIONCHARACTERIZATIONEXOPLANETS2012,absilSearchingCompanionsAU2013,gomezgonzalezVIPVortexImage2017a}.
While the shape of these masks is similar to the right reason mask, the way they work is different. Unlike previous work, the right reason mask of \fours is applied to the model \emph{parameters} during the optimization and not to the \emph{data}.
\looseness=-1

\begin{figure}[t]
	\includegraphics[width=\linewidth]{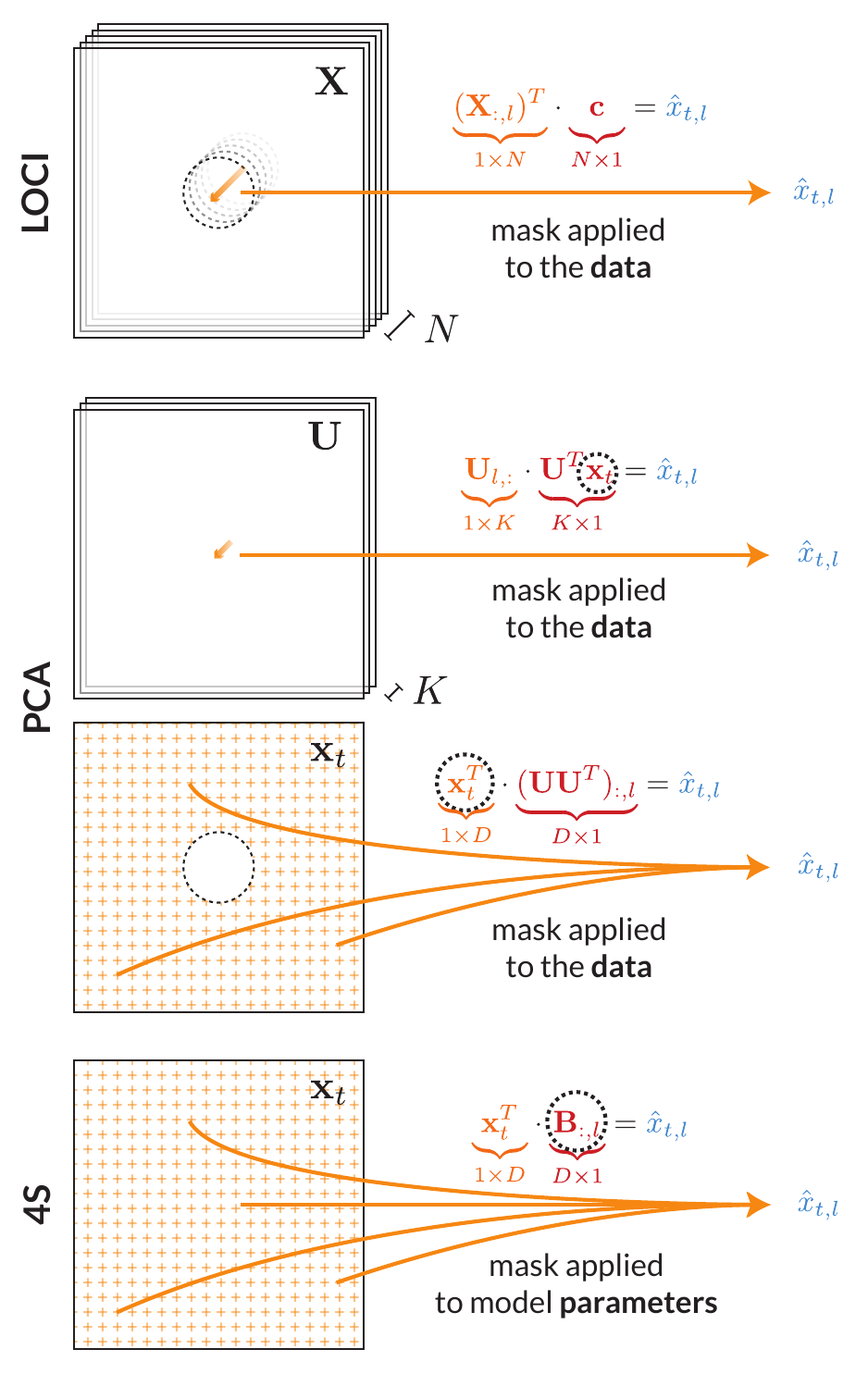}
	\caption{Illustration of the noise estimation implemented by the linear post-processing algorithms LOCI, PCA, and \fours. The noise estimate $\hat{x}_{t, l}$ at time $t$ and spatial position $l$ is a weighted sum of the values marked in orange, weighted by the parameters in dark red. The dotted circle marks the position of the mask. PCA can be interpreted in two ways, as explained in the text.}
	\label{fig:0a4_methods_compared}
\end{figure}

LOCI, PCA, and \fours are all linear PSF subtraction algorithms.
To better understand their differences and masking strategies, let us consider a single science frame $\mathbf{x}_t$ at time $t$. 
Using LOCI, PCA or \fours, we estimate the speckle noise for this particular science frame $\mathbf{\hat{x}}_t$.
For simplicity, let us now consider a single pixel position $l$ in this noise estimate $\hat{x}_{t, l}$. All three PSF subtraction algorithms have a different approach to calculating $\hat{x}_{t, l}$ (see \Cref{fig:0a4_methods_compared}).
\looseness=-1

\paragraph{LOCI}
This uses a linear combination of reference frames to fit and subtract the speckle noise \citep{lafreniereNewAlgorithmPointSpread2007}. In the case of ADI, these reference frames are other science frames from the same data set but excluding temporally adjacent frames \citep[see, e.g., Figure 2 in][]{maroisExoplanetImagingLOCI2010}. 
The noise estimate $\hat{x}_{t, l}$ for LOCI is a weighted sum of the reference frames at the position~$l$:
\begin{equation}
	\hat{x}_{t, l} \overset{\text{LOCI}}{=} \underbrace{(\mathbf{X}_{:,l})^T}_{1 \times N} \cdot \underbrace{\mathbf{c} \vphantom{(\mathbf{X}_{:,l})^T}}_{N \times 1} \, ,
\end{equation}
where $\mathbf{c} \in \mathbb{R}^{N}$ are the weights of the sum. Depending on the temporal exclusion criterion, some values in $\mathbf{c}$ can be zero. 
The weights $\mathbf{c}$ are fitted while masking a PSF-sized region around the position $l$. The region on which the fit is based is called the $\mathit{O}$-zone, while the region to which the model is applied to is called the $\mathit{S}$-zone. The $\mathit{S}$-zone can be a single pixel \citep{soummerORBITALMOTIONHR2011a}.

\paragraph{PCA first interpretation}
Instead of a linear combination of reference frames, PCA estimates the noise using a linear combination of the principal components $\mathbf{U}$. Since the number of components is smaller than the number of reference frames used by LOCI ($K \ll N$), PCA can be seen as a regularized version of LOCI. 
The noise estimate $\hat{x}_{t, l}$ for PCA is a weighted sum of the principal components at the position $l$:
\begin{equation}
	\label{eq:pca_int_1}
	\hat{x}_{t, l} \overset{\text{PCA}}{=} \underbrace{\mathbf{U}_{l, :}}_{1 \times K} \cdot \underbrace{\mathbf{U}^T \mathbf{x}_t \vphantom{\mathbf{U}_{l, :}}}_{K \times 1} \, .
\end{equation}
The weights of the sum are given by the projection of the science frame onto the component matrix $\mathbf{U}^T \mathbf{x}_t$. Similar to LOCI, it is possible to exclude a PSF-sized region in this projection by masking the data.

\paragraph{PCA second interpretation}
If we rearrange \Cref{eq:pca_int_1} we obtain the alternative interpretation of PCA:
\begin{equation}
	\hat{x}_{t, l} \overset{\text{PCA}}{=} \underbrace{\mathbf{x}_{t}^T}_{1 \times D} \cdot \underbrace{(\mathbf{U}\mathbf{U}^T)_{:,l} \vphantom{\mathbf{x}_{t}^T}}_{D \times 1} \, .
\end{equation}
The noise estimate $\hat{x}_{t, l}$ is not only a weighted sum of the principal components but also a weighted sum of the pixels in the science frame $\mathbf{x}_{t}$. The weights of this sum are given by the $l$-th column vector of the projection matrix $(\mathbf{U}\mathbf{U}^T)_{:,l}$. 
The saliency maps discussed in \Cref{sec:signal_loss_explained} are based on this interpretation of PCA. A mask can be applied directly to the data $\mathbf{x}_{t}$.

\paragraph{\fours}
Similar to the second interpretation of PCA, the noise estimate $\hat{x}_{t, l}$ in \fours is a weighted sum of the pixels in science frame $\mathbf{x}_{t}$:
\begin{equation}
	\hat{x}_{t, l} \overset{\text{4S}}{=} \underbrace{\mathbf{x}_{t}^T}_{1 \times D} \cdot \underbrace{\mathbf{B}_{:,l}}_{D \times 1} \, .
\end{equation}
The weights of the sum are given by the $l$-th column of the matrix $\mathbf{B}_{:,l}$. This means that, unlike LOCI, the noise estimate of \fours is not a weighted sum of the $N$ reference frames but of the $D$ pixels in the science frame $\mathbf{x}_{t}$.

\subsection{The Masks of PCA and \fours}
The principal components are not orthogonal to the signal of the planet. This means that a linear combination of the components can fit and subtract the signal of the planet. 
Similar to the masks used in LOCI, we can apply a mask in the projection step of PCA in \Cref{eq:pca_int_1}.
For comparison with \fours we use the same PSF-sized masks $\mathbf{m}_l$ as for the right reason mask:
\begin{align*}
	\hat{x}_{t, l} &= \mathbf{U}_{l, :} \cdot \mathbf{U}^T \left( \mathbf{x}_t \circ \mathbf{m}_l \right) \\
	&= \left(\mathbf{x}_t \circ \mathbf{m}_l \right)^T (\mathbf{U}\mathbf{U}^T)_{:,l} \\
	&= \mathbf{x}_t^T \left((\mathbf{U}\mathbf{U}^T)_{:,l} \circ \mathbf{m}_l\right)
\end{align*}
This means that instead of masking the data we can mask the $l$-th column vector of the projection matrix $(\mathbf{U}\mathbf{U}^T)_{:,l}$ and obtain the same result. 
The projection matrix of PCA ($\mathbf{U}\mathbf{U}^T$) and the model matrix of \fours ($\mathbf{B}$) are both square matrices with shape $D \times D$.
In both cases the mask $\mathbf{m}_l$ is applied to the diagonal of the matrix.
\begin{figure}[t]
	\includegraphics[width=\linewidth]{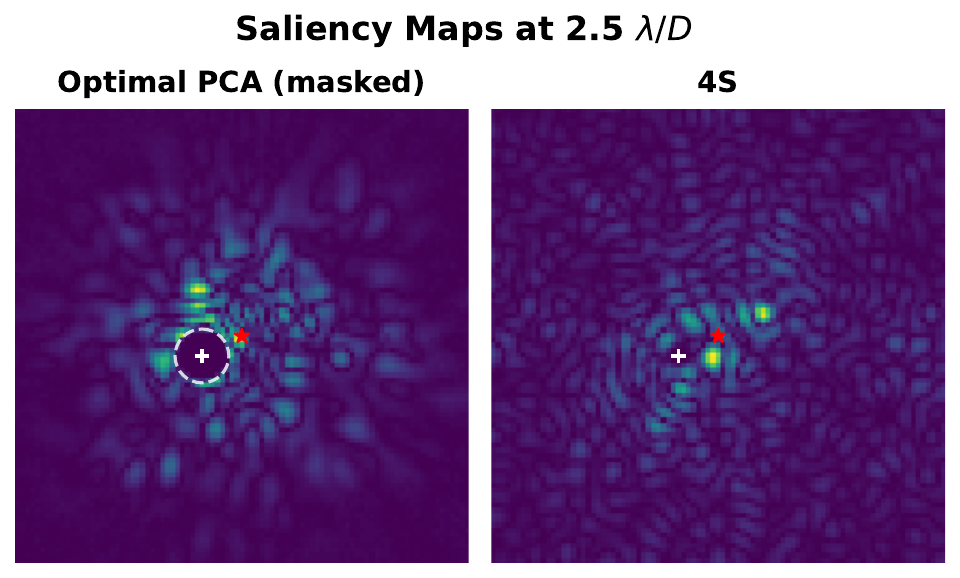}
	\caption{
	Comparison of the saliency maps of the masked version of PCA and \fours. The number of components for PCA was selected according to the best contrast performance at $2.5 \lambda/D$ which is reached for $K=90$. The white cross represents the position for which the saliency map is calculated. The red star marks the center of the frame.
    }
	\label{fig:0a5_masked_sailency_map}
\end{figure}
In \Cref{fig:0a5_masked_sailency_map} we compare the saliency map of the masked version of PCA with the saliency map of \fours. As shown in the figure, the information used by PCA is very different from the information used by \fours. 
While the point symmetries of the speckle noise are clearly identified by \fours, PCA uses information from the whole image.

\emph{If the masks are applied along the same dimension, what is the cause for this difference?}
\begin{itemize}
	\item The linear model of \fours is more general:
	the column vectors of the projection matrix in PCA $(\mathbf{U}\mathbf{U}^T)_{:,l}$ are linearly dependent. The rank of the matrix is set by the number of components:
	\begin{equation}
		\text{rank} \left( \mathbf{U}\mathbf{U}^T \right) = K \, .
	\end{equation}
	Further, the projection matrix is symmetric, $(\mathbf{U}\mathbf{U}^T)^T = \mathbf{U}\mathbf{U}^T$ and satisfies
	\begin{equation}
		\mathbf{U}\mathbf{U}^T \cdot \mathbf{U}\mathbf{U}^T = \mathbf{U}\mathbf{U}^T \, .
	\end{equation}
	In contrast, the column vectors $\mathbf{B}_{:,l}$ are generally not dependent, and the rank of the matrix is unconstrained. Further, $\mathbf{B}$ does not have to be symmetric, $\mathbf{B}^T \neq \mathbf{B}$ and is not a projection matrix, $\mathbf{B}^2 \neq \mathbf{B}$. The linear model of \fours has fewer constraints and can therefore be seen as a generalization of the PCA framework.
	\item For PCA, the component matrix $\mathbf{U}$ is the result of the eigenvalue decomposition (or SVD). The mask can be applied during the projection step. For large numbers of principal components, the information used by PCA converges to a \emph{de facto} identity (see \Cref{sec:right_for_wrong_reasons}). Even if we mask this identity in $\mathbf{U}\mathbf{U}^T$, there is no incentive for the model to explore other areas of the data. The right reason mask in \fours is embedded in $\mathbf{B}$. The parameters of \fours are optimized in an end-to-end fashion, taking into account the right reason mask. Since the mask is part of the optimization, the model is forced to search for additional information about the speckle noise.
	\item The loss function of PCA is a classical least-squares minimization (see \Cref{sec:Misguided_loss_function}). This loss function is replaced in \fours making it invariant to the signal of the planet (see \Cref{sec:4s_loss_function}). Temporal exclusion criteria as introduced for LOCI \citep{maroisExoplanetImagingLOCI2010} are not needed.
\end{itemize}

    \section{Fake planet experiments for AF Lep b}
\label{sec:af_lep_fake_planets}
\begin{figure}[!t]
	\includegraphics[width=\linewidth]{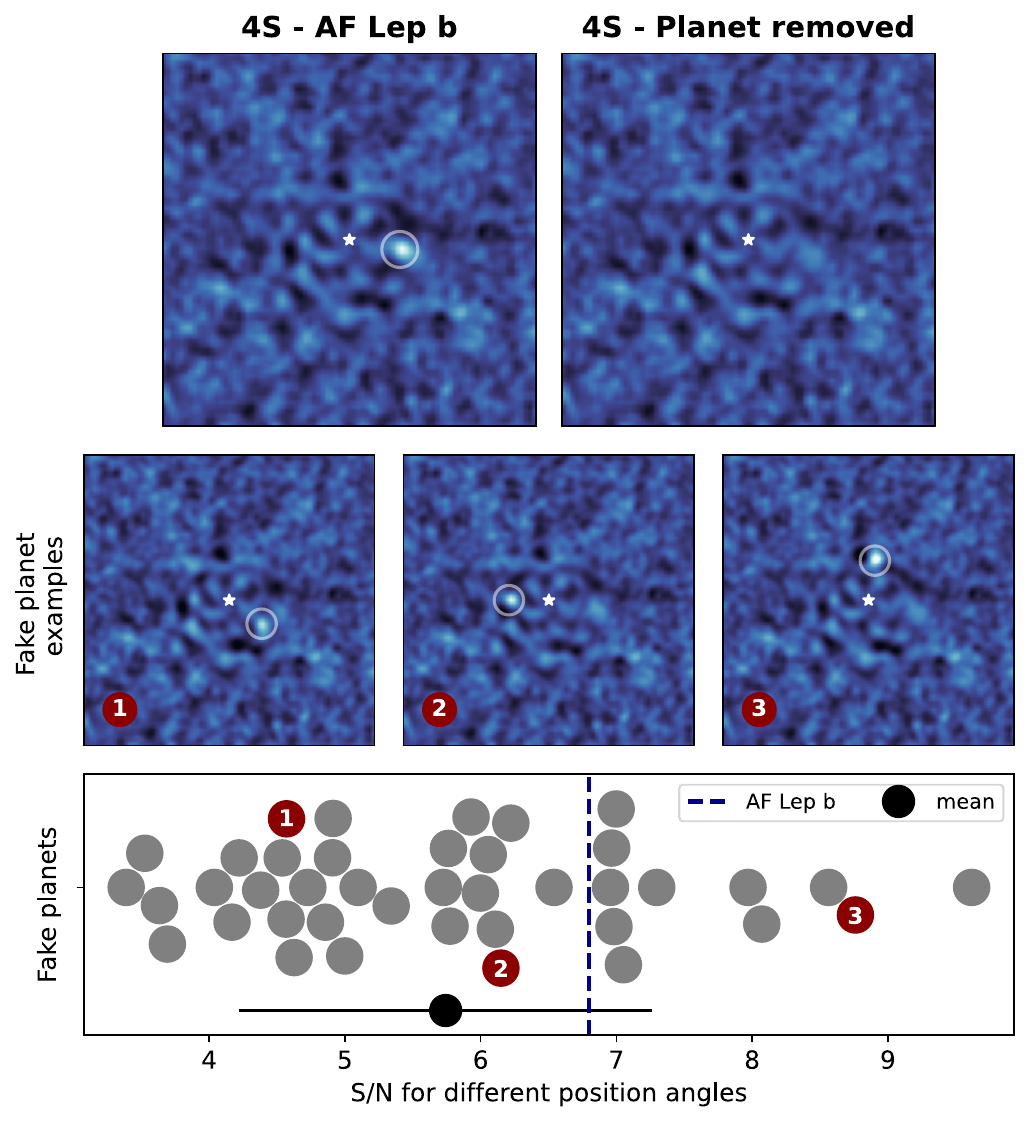}
	\caption{
		Verification of the astrometry and photometry estimated with the help of PCA. The top two panels show the residuals of \fours for AF~Lep~b (left) and after insertion of a negative fake planet (right) based on the results given in \Cref{tab:af_lep_prop}. In the middle row, fake planet residuals are shown that are inserted into the data at the same separation and contrast as AF~Lep~b, but at different position angles. The bottom panel summarizes the $S/N$ values of the 40 fake planet experiments described in the text. 
    }
	\label{fig:0a6_AF_Lep_fake_planets}
\end{figure}

In \Cref{sec:af_lep} we estimate the astrometry and photometry of AF~Lep~b using PCA. Ideally, we would like to use \fours to perform this analysis. Due to the high computational cost, this is currently not possible, a limitation we plan to overcome in future work using the ideas presented in \Cref{sec:future_work}. 
To verify that the astrometry and photometry extracted by PCA are consistent with the detection in \fours, we insert a negative fake planet with $\Delta L' = \qty{10.03}{\magnitude}$ at the position obtained with PCA (see~\Cref{tab:af_lep_prop}).
Afterward, we run \fours again on the data without the planet. As shown in the top right panel of \Cref{fig:0a6_AF_Lep_fake_planets}, the signal of AF~Lep~b is removed and the residual of \fours is flat. 
As an additional test, we insert fake planets with $\Delta L' = \qty{10.03}{\magnitude}$ at the same separation as AF~Lep~b but at 40 different position angles. 
For each fake planet, we rerun \fours and compute the $S/N$ of the artificial companion.
On average, the fake planets achieve $S/N = 5.7 \pm 1.5$ which is consistent with the real companion $S/N = 6.8$.
Examples of the fake planet residuals and the $S/N$ values computed for all 40 fake planets are shown in the middle and bottom rows of \Cref{fig:0a6_AF_Lep_fake_planets}.
The $S/N$ varies considerably for different position angles, an effect already reported for \naco data in \cite{bonseComparingApplesApples2023}.
Based on this experiment, we conclude that the photometry and position extracted using PCA are consistent with the detection in \fours.

    \section{Residual Noise Distribution}
\label{sec:a_residual_noise_dist}
\begin{figure*}[t]
	\includegraphics[width=\linewidth]{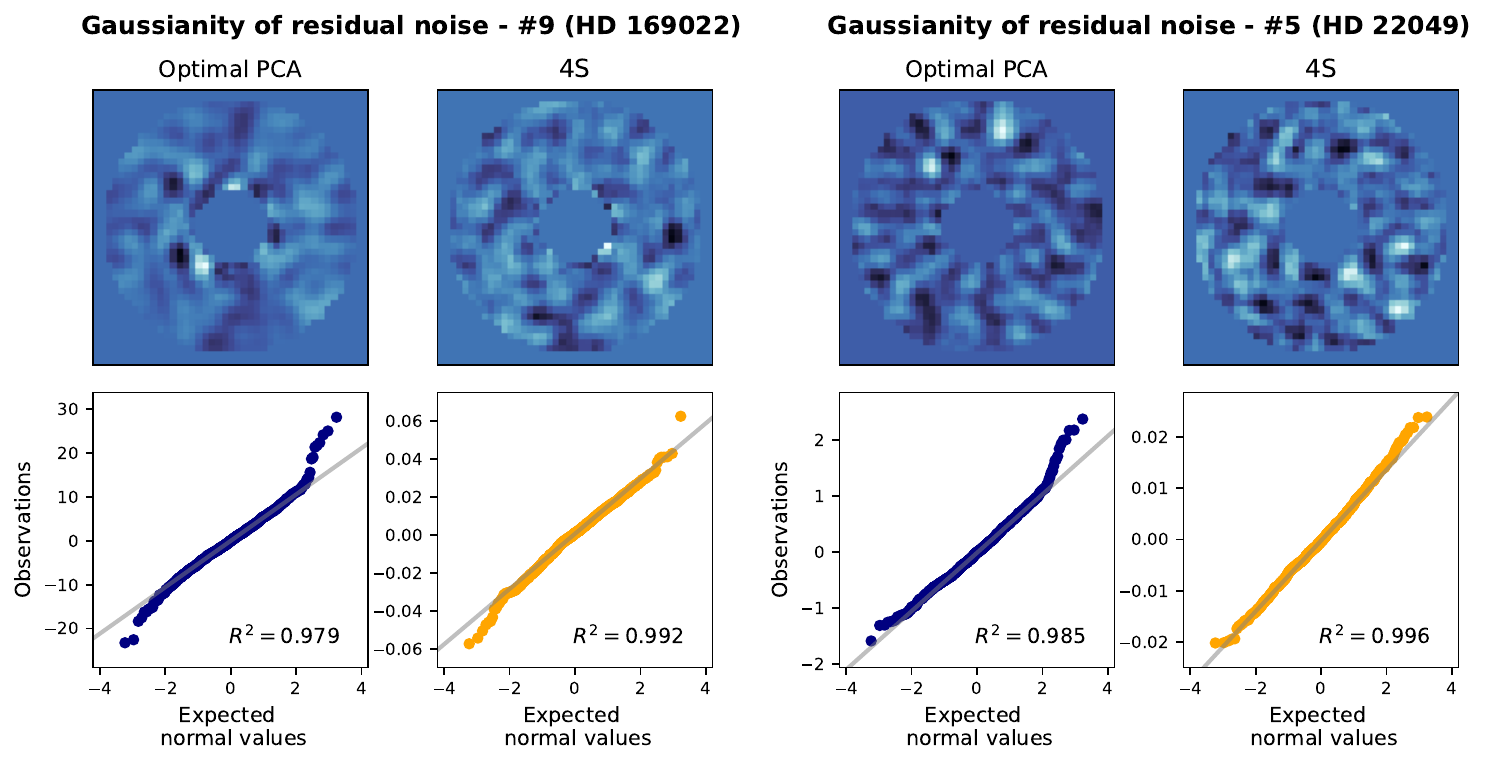}
	\caption{
        Two examples of the Gaussianity test performed in  \Cref{fig:0a2_gaussianity_residual_noise}. The top row shows examples of the residuals obtained with PCA and \fours. The pixels outside \qtyrange{2.5}{4.5}{\lod} were masked to extract the noise. Q--Q plots are shown in the bottom row. If the noise were perfectly Gaussian, we would expect all values to be on the gray diagonal line and the value of $R^2$ to be close to 1.
    }
	\label{fig:0a1_gaussianity_residual_noise}
\end{figure*}
\begin{figure*}[t]
    \centering
	\includegraphics[width=0.5\linewidth]{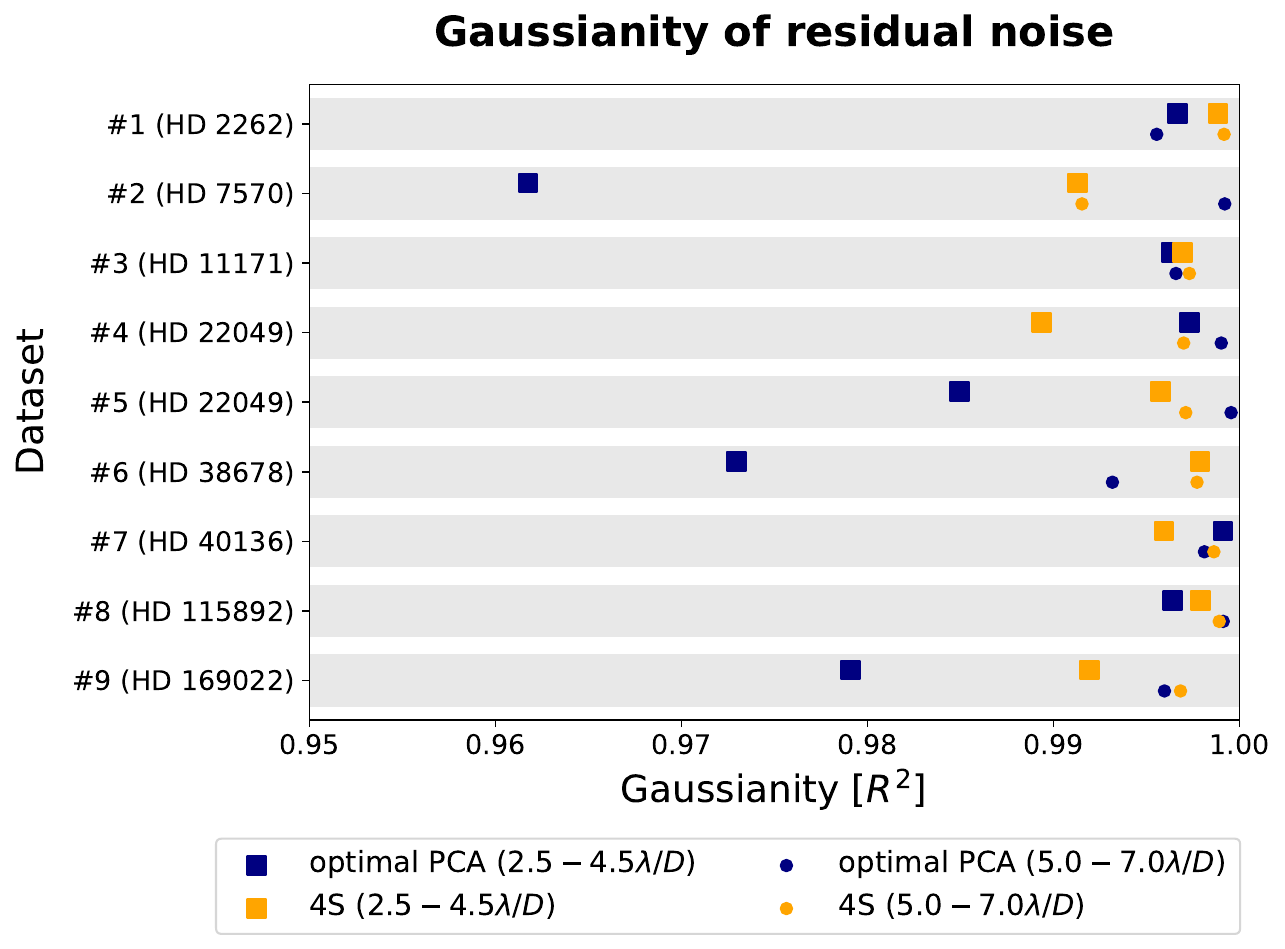}
	\caption{
        Comparison of the Gaussianity of the noise in the residuals of \fours and PCA for the data sets \#1--\#9. If the coefficient of determination $R^2$ is closer to 1, the noise is better explained by Gaussian noise. We choose the number of principal components $K$ and the regularization strength $\lambda$ of \fours that maximize the contrast (\emph{optimal} PCA).
    }
	\label{fig:0a2_gaussianity_residual_noise}
\end{figure*}

A common assumption for the quantification of detections and nondetections in HCI is that the residual noise after post-processing with PCA is Gaussian distributed \citep[see, e.g.,][]{mawetFUNDAMENTALLIMITATIONSHIGH2014}. 
This assumption is generally justified by the central limit theorem (we average many frames along time) and the whitening effect of PCA. Nevertheless, the residual noise after post-processing often deviates from Gaussian noise \citep{pairetSTIMMapDetection2019,bonseComparingApplesApples2023}. 
This is problematic because the risk of false positives (we claim a detection that is just noise) depends on the type of noise. 
More specifically, the probability of observing a value of $\snr = 5$ that is just noise is higher if the noise is heavy tailed, that is, the frequency of bright values such as speckles is higher than for Gaussian noise. 
\cite{bonseComparingApplesApples2023} found that if the residual noise follows a Laplacian distribution, the final contrast could be overestimated by up to one magnitude if the test of \cite{mawetFUNDAMENTALLIMITATIONSHIGH2014} is used.

In practice, the true noise distribution is usually unknown, making it difficult to compare limits across data sets. 
This is also true if we want to compare different post-processing techniques. 
We study the residuals of PCA and \fours for data sets \#1--\#9 (see \Cref{tab:datasets}) to get a better understanding of their noise distributions. 
Data sets \#10 and \#11 are excluded owing to strong saturation at close separations. 
We consider two regions: 1.~within \qtyrange{2.5}{4.5}{\lod}, and 2.~within \qtyrange{5.0}{7.0}{\lod} distance to the star. 
For each data set and region we plot Q--Q plots and calculate the coefficient of determination $R^2$ (see, e.g., \citealt{pairetSTIMMapDetection2019} for a detailed explanation). 
A comparison of the residual noise for data sets \#5 and \#9 is shown in \Cref{fig:0a1_gaussianity_residual_noise}. 
The Q--Q plots shown in the figure compare the pixel noise distribution with Gaussian noise. If the residual noise were perfectly Gaussian, we would expect all values to lie on the gray diagonal line. The better the observed noise can be explained by Gaussian noise, the closer the coefficient of determination $R^2$ is to 1.
For PCA, the large values are above the diagonal line, indicating that the noise is heavy-tailed. 
In the case of \fours, the residual noise is much closer to Gaussian, although it is still not perfectly Gaussian.
The quantification based on the $R^2$ values confirms this observation. A summary of the results for all data sets is shown in \Cref{fig:0a2_gaussianity_residual_noise}. 
In general, the residual noise at close separations tends to be more affected by speckle and is therefore often more non-Gaussian.
For most data sets, especially those for which PCA deviates the most from Gaussian noise, \fours improves the Gaussianity of the residuals. 
This means that \fours not only improves the contrast limits, as shown in \Cref{sec:ana}, but also reduces the risk of false positives. 

\onecolumngrid

\begin{figure*}[p]
	\centering
    \includegraphics[width=0.9\linewidth]{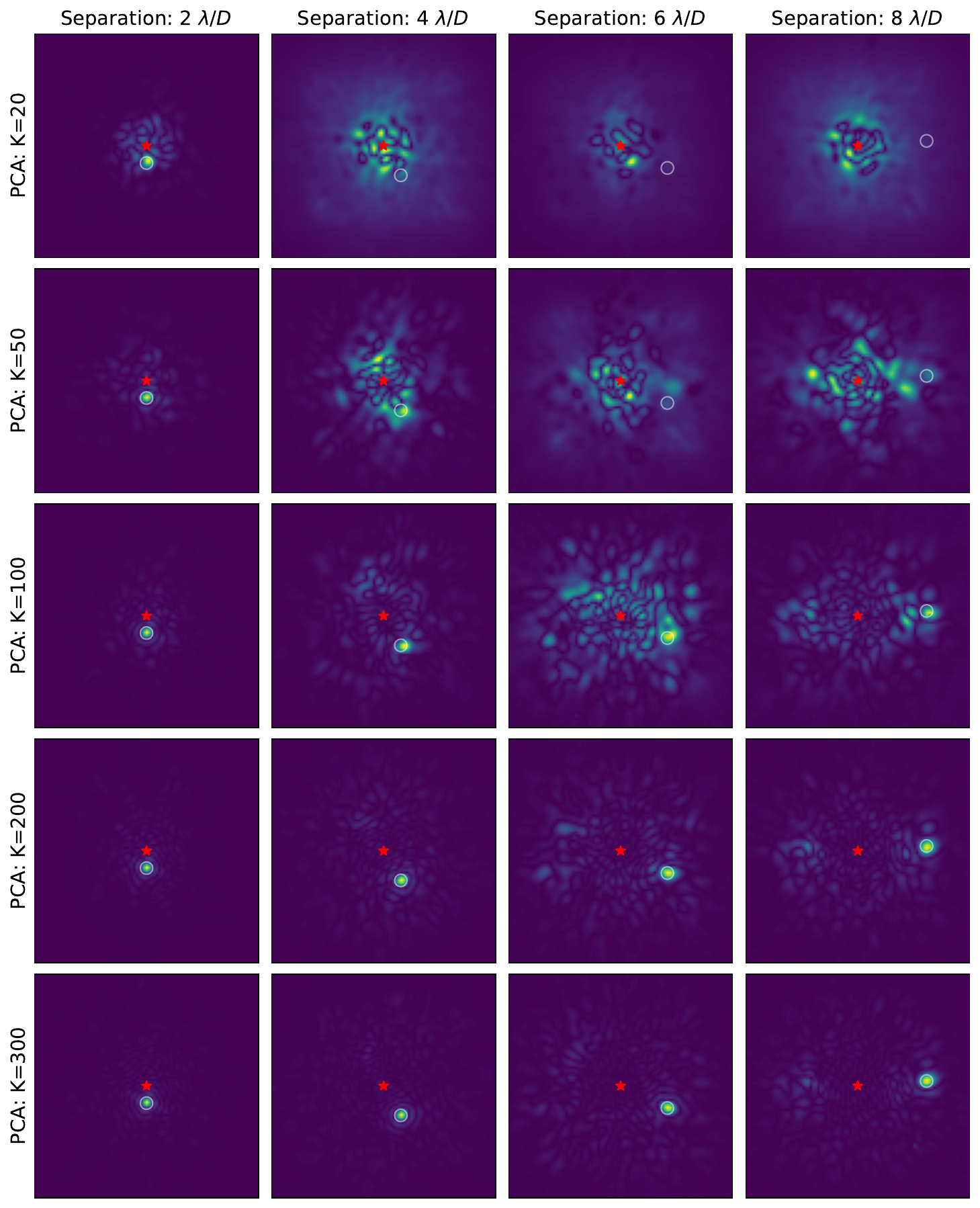}
	\caption{
        PCA saliency maps for different numbers of principal components $K$ and separations from the star. 
        The saliency maps show which information from the science frames is used by PCA to estimate the noise at the position marked by the white circle. 
        Especially for large $K$ and separations close to the star, only the information within the white circle is used, resulting in a substantial loss of the planet signal. 
        The red star marks the center of the frame. 
	}
	\label{fig:0a3_gallery_saliency}
\end{figure*}

\begin{figure*}[p]
    \centering
	\includegraphics[width=0.9\linewidth]{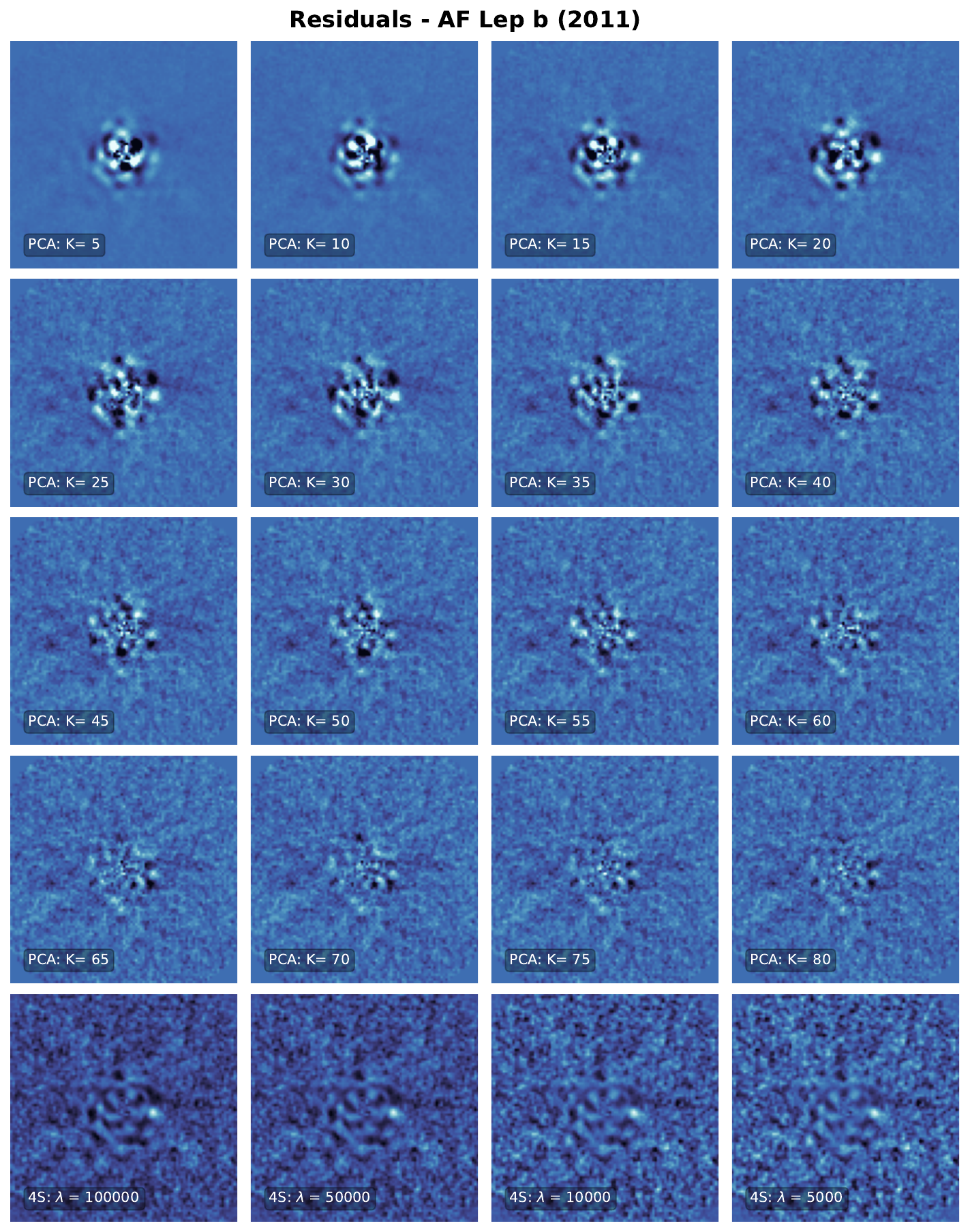}
	\caption{
        Residual images of the AF~Lep data set shown in \Cref{fig:05_af_lep} for several different settings of the algorithm's hyperparameters (principal components $K$ for PCA and regularization strength $\lambda$ for \fours). 
        The planet AF~Lep~b is barely visible in the PCA residuals. 
        For \fours, any choice of parameters yields a clear detection, highlighting the robustness of the method.
	}
	\label{fig:0a4_af_lep}
\end{figure*}

\end{document}